\documentclass[final, a4paper, 10pt,twocolumn]{IEEEtran}

\usepackage{import}
\usepackage{url}
\usepackage[T1]{fontenc}
\usepackage[latin1]{inputenc}
\usepackage{textcomp}
\usepackage{bm}
\usepackage{graphicx}
\usepackage{cite}
\usepackage{amsmath,amssymb,amsfonts}
\usepackage{balance}
\usepackage{xcolor}

\usepackage{hyperref}
\hypersetup{colorlinks=true,linkcolor=black,urlcolor=blue,}

\usepackage[acronym,style=super,nogroupskip,nonumberlist]{glossaries-extra}
\makeglossaries
\setabbreviationstyle[acronym]{long-short}
\newacronym{PLS}{PLS}{physical~layer~security}
\newacronym{RIS}{RIS}{reconfigurable~intelligent~surface}
\newacronym{IIoT}{IIoT}{industrial~internet~of~things}
\newacronym{LIS}{LIS}{large~intelligent~surface}
\newacronym{DQN}{DQN}{deep~Q-learning}
\newacronym{EM}{EM}{electromagnetic}
\newacronym{ARIS}{ARIS}{aerial~RIS}
\newacronym{DRL}{DRL}{deep~reinforcement~learning}
\newacronym{RL}{RL}{reinforcement~learning}
\newacronym{DDPG}{DDPG}{deep~deterministic~policy~gradient}
\newacronym{DPG}{DPG}{deterministic~policy~gradient}
\newacronym{QoS}{QoS}{quality~of~service}
\newacronym{MIMO}{MIMO}{multiple-input~multiple-output}
\newacronym{SIMO}{SIMO}{single-input~multiple-output}
\newacronym{SISO}{SISO}{single-input~single-output}
\newacronym{MISO}{MISO}{multiple-input~single-output}
\newacronym{RF}{RF}{radio~frequency}
\newacronym{ML}{ML}{machine~learning}
\newacronym{OWC}{OWC}{optical~wireless~communication}
\newacronym{KPIs}{KPIs}{key~performance~indicators}
\newacronym{5G}{5G}{fifth~generation}
\newacronym{6G}{6G}{sixth~generation}
\newacronym{EE}{EE}{energy~efficiency}
\newacronym{PNSC}{PNSC}{probability~of~non-zero~secrecy~capacity}
\newacronym{SCA}{SCA}{successive~convex~approximation}
\newacronym{BCD}{BCD}{block~coordinate~descent}
\newacronym{AO}{AO}{alternating~optimization}
\newacronym{LPWAN}{LPWAN}{low~power~wide~area~network}
\newacronym{SOP}{SOP}{secrecy~outage~probability}
\newacronym{SINR}{SINR}{signal-to-interference-plus-noise~ratio}
\newacronym{BS}{BS}{base~station}
\newacronym{Eav}{Eav}{eavesdropper}
\newacronym{mmWave}{mmWave}{millimeter~wave}
\newacronym{AN}{AN}{artificial~noise}
\newacronym{CSI}{CSI}{channel~state~information}
\newacronym{WPT}{WPT}{wireless~power~transfer}
\newacronym{WPC}{WPC}{wireless~power~communication}
\newacronym{SWIPT}{SWIPT}{simultaneous~wireless~information~and~power~transfer}
\newacronym{MEC}{MEC}{mobile~edge~computing}
\newacronym{SC}{SC}{secrecy~capacity}
\newacronym{SR}{SR}{secrecy~rate}
\newacronym{UAV}{UAV}{unmanned~aerial~vehicle}
\newacronym{DoF}{DoF}{degree~of~freedom}
\newacronym{BER}{BER}{bit~error~rate}
\newacronym{NOMA}{NOMA}{non-orthogonal~multiple~access}
\newacronym{LoS}{LoS}{line-of-sight}
\newacronym{NLoS}{NLoS}{non-line-of-sight}
\newacronym{THz}{THz}{terahertz}
\newacronym{SNR}{SNR}{signal-to-noise~ratio}
\newacronym{SDR}{SDR}{semidefinite~relaxation}
\newacronym{CCP}{CCP}{convex-concave~procedure}
\newacronym{IoE}{IoE}{internet-of-everything}
\newacronym{AR}{AR}{augmented~reality}
\newacronym{MR}{MR}{mixed~reality}
\newacronym{VR}{VR}{virtual~reality}
\newacronym{VLC}{VLC}{visible~light~communications}
\newacronym{MM}{MM}{majorization-minimization}
\newacronym{QCQP}{QCQP}{quadratically~constrained~quadratic~programming}
\newacronym{IoT}{IoT}{internet~of~things}
\newacronym{V2X}{V2X}{vehicle-to-everything}
\newacronym{V2V}{V2V}{vehicle-to-vehicle}
\newacronym{ASR}{ASR}{average~secrecy~rate}
\newacronym{ASC}{ASC}{average~secrecy~capacity}
\newacronym{SSR}{SSR}{sum~secrecy~rate}
\newacronym{SER}{SER}{symbol~error~rate}
\newacronym{ESC}{ESC}{ergodic~secrecy~capacity}
\newacronym{D2D}{D2D}{Device-to-Device}
\newacronym{RRS}{RRS}{reconfigurable~refractive~surface}
\newacronym{IOS}{IOS}{intelligent~omni-surface}
\newacronym{STAR}{STAR}{simultaneous~transmitting~and~reflecting}
\newacronym{SOCP}{SOCP}{second~order~cone~programming}
\newacronym{SIC}{SIC}{successive~interference~cancellation}
\newacronym{2D}{2D}{two-dimensional}
\newacronym{DC}{DC}{direct-current}
\newacronym{WBAN}{WBAN}{wireless~body~area~network}
\newacronym{TDD}{TDD}{time~division~duplexing}
\newacronym{CRN}{CRN}{cognitive~radio~network}
\newacronym{AP}{AP}{access~point}
\newacronym{ISAC}{ISAC}{integrated~sensing~and~communications}
\newacronym{MSE}{MSE}{mean~square~error}
\newacronym{IM/DD}{IM/DD}{intensity~modulation~and~direct~detection}
\newacronym{BF}{BF}{beamforming}
\newacronym{LED}{LED}{light~emitting~diode}
\newacronym{SPSC}{SPSC}{strictly~positive~secrecy~capacity}
\newacronym{FSO}{FSO}{free~space~optical}
\newacronym{RGB}{RGB}{red-green-blue}
\newacronym{IR}{IR}{infrared}
\newacronym{AI}{AI}{artificial~intelligence}
\newacronym{WPAN}{WPAN}{wireless~personal~area~network}
\newacronym{WSSCE}{WSSCE}{weighted~sum~secrecy~computation~efficiency}
\newacronym{AANET}{AANET}{aeronautical~Ad-hoc~networks}
\newacronym{DNN}{DNN}{deep~neural~network}
\newacronym{SDP}{SDP}{semidefinite~programming}
\newacronym{LC}{LC}{liquid~crystal}
\newacronym{COP}{COP}{connection~outage~probability}
\newacronym{ANT}{ANT}{average~number~of~transmission}
\newacronym{EST}{EST}{effective~secrecy~throughput}
\newacronym{IP}{IP}{intercept~probability}
\newacronym{PPSC}{PPSC}{probability~of~positive~secrecy~capacity}
\newacronym{HARQ}{HARQ}{hybrid~automatic~repeat~request}
\newacronym{IRIS}{RIS}{illegal~reconfigurable~intelligent~surface}
\newacronym{WSNs}{WSNs}{wireless~sensor~networks}
\newacronym{IoMT}{IoMT}{internet~of~medical~things}
\newacronym{CoMP}{CoMP}{coordinate~multi-point}
\newacronym{3D}{3D}{three-dimensional}
\newacronym{HAPS}{HAPS}{high-altitude~platform~systems}
\newacronym{PN}{PN}{primary~network}
\newacronym{SN}{SN}{secondary~network}
\newacronym{BackCom}{BackCom}{backscatter~communication}
\newacronym{EH}{EH}{energy~harvesting}
\newacronym{STAR-RIS}{STAR-RIS}{simultaneous~transmitting~and~reflecting~reconfigurable~intelligent~surface}
\newacronym{SymR}{SymR}{symbiotic~radio}
\newacronym{MA}{MA}{movable~antenna}
\newacronym{FPA}{FPA}{fixed~position~antenna}
\newacronym{RSMA}{RSMA}{rate~splitting~multiple~access}
\newacronym{MWN}{MWN}{multiband~wireless~network}
\newacronym{FA}{FA}{fluid~antenna}
\newacronym{FSS}{FSS}{frequency~selective~surface}
\newacronym{BD-RIS}{BD-RIS}{beyond-diagonal RIS} % file with acronyms
\glssetcategoryattribute{acronym}{glossdesc}{title}
\glsdisablehyper

\usepackage{enumitem}
\setlist{nosep}

\usepackage{subfloat} % For Subtable
\usepackage{multirow}
\usepackage{adjustbox}
\usepackage{pifont}
\begin{document}

\title{RIS-Assisted Physical Layer Security in Emerging RF and Optical Wireless Communication Systems: A Comprehensive Survey}
%
%
% author names and IEEE memberships
% note positions of commas and nonbreaking spaces ( ~ ) LaTeX will not break
% a structure at a ~ so this keeps an author's name from being broken across
% two lines.
% use \thanks{} to gain access to the fRISt footnote area
% a separate \thanks must be used for each paragraph as LaTeX2e's \thanks
% was not built to handle multiple paragraphs
%

\author{{Majid~H.~Khoshafa},~\IEEEmembership{Member,~IEEE,}  Omar~Maraqa,~Jules~M.~Moualeu,~\IEEEmembership{Senior Member,~IEEE,} Sylvester~Aboagye,~\IEEEmembership{Member,~IEEE,}~Telex~M.~N.~Ngatched,~\IEEEmembership{Senior Member,~IEEE,}~Mohamed~H.~Ahmed,\\~\IEEEmembership{Senior~Member,~IEEE,}~Yasser~Gadallah,~\IEEEmembership{Senior~Member,~IEEE,}~and~Marco~Di~Renzo,~\IEEEmembership{Fellow,~IEEE}% <-this % 
\thanks{
This work was supported in part by McMaster University and in part by the Natural Sciences and Engineering Research Council (NSERC) of Canada under its Discovery Grant (DG) program (RGPINs 2019-04626 and 2021-03323).  The work of J. M. Moualeu was supported in part by the Centre for Telecommunications Access Services (CeTAS) under the Telkom Centre of Excellence (CoE) Program. The work of S. Aboagye was supported in part by the University of Guelph. The work of M. Di Renzo was supported in part by the European Commission through the Horizon Europe project titled COVER under grant agreement number 101086228, the Horizon Europe project titled UNITE under grant agreement number 101129618, and the Horizon Europe project titled INSTINCT under grant agreement number 101139161, as well as by the Agence Nationale de la Recherche (ANR) through the France 2030 project titled ANR-PEPR Networks of the Future under grant agreement NF-PERSEUS, 22-PEFT-004, and the ANR-CHISTERA project titled PASSIONATE under grant agreement ANR-23-CHR4-0003-01.

Majid H. Khoshafa, Omar Maraqa, and Telex M. N. Ngatched are with the Department of Electrical and Computer Engineering, McMaster University, Hamilton, ON L8S 4L8, Canada (e-mail:\{khoshafm@mcmaster.ca, maraqao@mcmaster.ca, ngatchet@mcmaster.ca\}).

Jules M. Moualeu is with the School of Electrical and Information Engineering, University of the Witwatersrand, Johannesburg 2000, South Africa
(e-mail: jules.moualeu@wits.ac.za).

Sylvester Aboagye is with the School of Engineering, University of Guelph, Guelph, ON N1G 2W1, Canada (e-mail: saboagye@uoguelph.ca).

Mohamed H. Ahmed is with the School of Electrical Engineering and Computer Science, University of Ottawa, Ottawa, ON K1N 6N5, Canada (e-mail: mahme3@uottawa.ca). 

Yasser Gadallah is with the Department of Electronics and Communications Engineering, The American University in Cairo, New Cairo 11835, Egypt (e-mail: ygadallah@ieee.org).

Marco Di Renzo is with Universit\'e Paris-Saclay, CNRS, CentraleSup\'elec,
Laboratoire des Signaux et Syst\'emes, 91192 Gif-sur-Yvette, France (e-mail:
marco.di-renzo@universite-paris-saclay.fr).

This work has been submitted to the IEEE for possible publication. Copyright may be transferred without notice, after which this version may no longer be accessible.
} }

% note the % following the last \IEEEmembership and also \thanks - 
% these prevent an unwanted space from occurring between the last author name
% and the end of the author line. i.e., if you had this:
% 
% \author{....lastname \thanks{...} \thanks{...} }
%                     ^------------^------------^----Do not want these spaces!
%
% a space would be appended to the last name and could cause every name on that
% line to be shifted left slightly. This is one of those "LaTeX things". For
% instance, "\textbf{A} \textbf{B}" will typeset as "A B" not "AB". To get
% "AB" then you have to do: "\textbf{A}\textbf{B}"
% \thanks is no different in this regard, so shield the last } of each \thanks
% that ends a line with a % and do not let a space in before the next \thanks.
% Spaces after \IEEEmembership other than the last one are OK (and needed) as
% you are supposed to have spaces between the names. For what it is worth,
% this is a minor point as most people would not even notice if the said evil
% space somehow managed to creep in.

% The paper headers
\markboth{}%
{Shell \MakeLowercase{\textit{et al.}}: Bare Demo of IEEEtran.cls for IEEE Journals}

\maketitle

\begin{abstract}
Security and latency are crucial aspects in the design of future wireless networks. Physical layer security (PLS) has received a growing interest from the research community in recent years for its ability to safeguard data confidentiality without relying on key distribution or encryption/decryption, and for its latency advantage over bit-level cryptographic techniques. However, the evolution towards the fifth generation (5G) technology and beyond poses new security challenges that must be addressed in order to fulfill the unprecedented performance requirements of future wireless communication networks. Among the potential key-enabling technologies, reconfigurable intelligent surface (RIS) has attracted extensive attention due to its ability to proactively and intelligently reconfigure the wireless propagation environment to combat dynamic wireless channel impairments. Consequently, the RIS technology can be adopted to improve the information-theoretic security of both radio frequency (RF) and optical wireless communications (OWC) systems. It is worth noting that the configuration of RIS in RF communications is different from the one in optical systems at many levels (e.g., RIS materials, signal characteristics, and functionalities). This survey paper provides a comprehensive overview of the information-theoretic security of RIS-based RF and optical systems. The article first discusses the fundamental concepts of PLS and RIS technologies, followed by their combination in both RF and OWC systems. Subsequently, some optimization techniques are presented in the context of the underlying system model, followed by an assessment of the impact of RIS-assisted PLS through a comprehensive performance analysis. Given that the computational complexity of future communication systems that adopt RIS-assisted PLS is likely to increase rapidly as the number of interactions between the users and infrastructure grows, machine learning (ML) is seen as a promising approach to address this complexity issue while sustaining or improving the network performance. A discussion of recent research studies on RIS-assisted PLS-based systems embedded with ML is presented. Furthermore, some important open research challenges are proposed and discussed to provide insightful future research directions, with the aim of moving a step closer towards the development and implementation of the forthcoming sixth-generation (6G) wireless technology. 
\end{abstract}


\tableofcontents

\textcolor{black}{{\footnotesize{\printglossary[type=\acronymtype,title={List of Abbreviations}]}}}

%%%%%%%%%%%%%%%%%%%%%%%%%%%%%%%%%%%%%%%%%%%%%%%%%%%%%%%%%%%%%%%%%%%%%%

\section{Introduction}
\label{intro}
\begin{figure}[!t]
\centering
\includegraphics[width=3.4in]{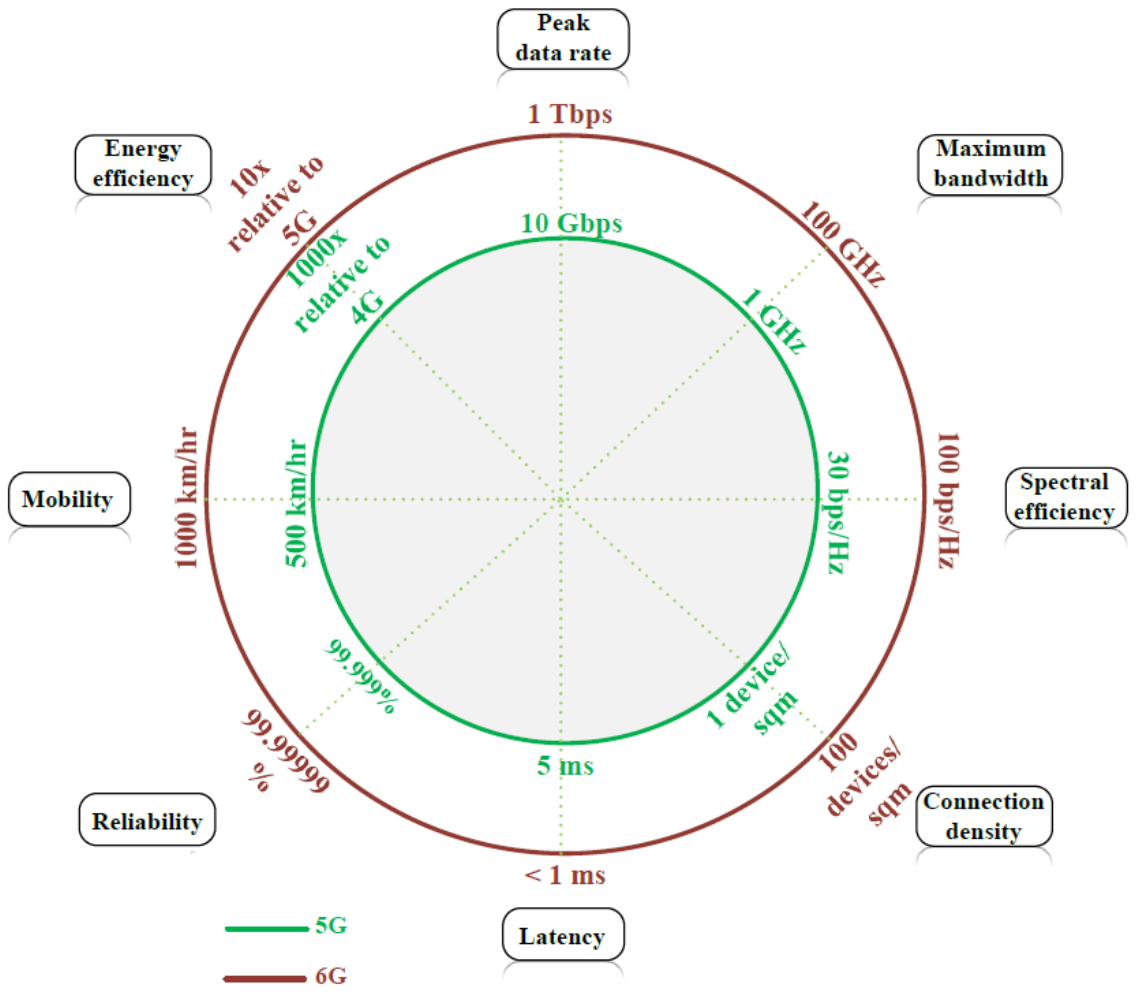}
\caption{\color{black}{Comparison of some \gls{KPIs} between \gls{5G} and \gls{6G} communication systems~\cite[Table IV]{wang2023road}.}}
\label{FigKPI}
\end{figure}

Despite the enormous potential of the \gls{5G} technology as a key enabler for the \gls{IoE}, it is anticipated that the rapid emergence of fully intelligent and automated systems such as tactile internet, industrial automation, \gls{AR}, \gls{MR}, \gls{VR}, telemedicine, haptics, flying vehicles, brain-computer interfaces, and connected autonomous systems, will overburden the capacity and limit the performance of \gls{5G} mobile networks in supporting the stringent requirements of next-generation networks such as extremely high-spectrum- and energy-efficiency, ultra-low latency, ultra-massive, and ubiquitous wireless connectivity, full dimensional network coverage, as well as connected intelligence~\cite{zhang2019sixG,saad2020vision,tataria2021sixG}. As a result, there have been intensive research efforts from both industry and academia devoted to the \gls{6G} technology 
to meet such technical requirements and demands as it is expected to provide much improved \gls{KPIs}~\cite{letaief2019roadmap,wang2023road}. {Fig.~\ref{FigKPI} shows a comparison between the \gls{5G} and \gls{6G} technologies in terms of some important \gls{KPIs}, including peak data rate, maximum bandwidth, energy efficiency, mobility, reliability, latency, connection density and spectral efficiency}.

Although the key requirements of the above-mentioned future wireless technologies are mainly characterized by low latency and high reliability, the leakage of critical and confidential information remains a challenge that must be addressed to fulfill its full potential. These security loopholes are due to the broadcast nature of the wireless environment, which makes it difficult to prevent information leakage by unauthorized receivers (e.g., an eavesdropper). To deal with security threats or attacks, cryptography-based tools have traditionally been adopted. However, such techniques may not be suitable for next-generation technologies for the following reasons: (a) cryptography algorithms cannot be adopted in resource-limited networks because they require a high computational complexity yielding a large amount of energy consumption~\cite{ mukherjee2014principles}; (b)  encryption-based techniques may not be able to sufficiently protect against information leakage due to the unlimited computing resources of the adversaries~\cite{ ahmed2018socially }; (c) processes involving key distribution and management in distributed networks (e.g., \gls{IoT} and vehicular networks) are challenging. To overcome these limitations, \gls{PLS} has emerged as a promising solution to achieve secrecy by exploiting the inherent randomness of wireless channels~\cite{wu2018survey,wang2018survey,shiu2011physical,moualeu2021physical,khoshafa2021performance}. Unlike encryption-based techniques, \gls{PLS} does not require key distribution/sharing that may violate the low-latency requirements of future wireless systems and is therefore, an attractive approach to provide security and privacy for such systems. Several studies reported in the literature have investigated the \gls{PLS} enhancement of wireless networks through \gls{AN}~\cite{feng2018user}, cooperative jamming~\cite{hu2018cooperative, khoshafa2020improving,khoshafa2022relay}, \gls{BF}~\cite{zhu2016improving}, directional modulation~\cite{kalantari2016directional}, \gls{MIMO}~\cite{khisti2010secure}, etc. However, these techniques cannot always guarantee secure communication under certain system configurations. Additionally, they do require extra power and/or hardware costs for their practical implementation.

%%%%%%%%%%%%%%%
%%%%%%%%%%%%%%%%%%%%%%%%%%

The advancement of metamaterial techniques has led to the emergence of a low-cost and energy-efficient wireless device termed \gls{RIS} that can control the phase shift of the reflecting units via a programmable surface~\cite{huang2019reconfigurable,tang2021wireless,direnzo2020smart}. These features enable the \gls{RIS} technology to intelligently reconfigure the wireless propagation environment. Unlike conventional techniques that only adapt to or with limited control over dynamic wireless channels, \gls{RIS} provides a new and cost-effective approach to combat channel impairments by guiding the reflected signal in a desirable direction for better reception reliability while suppressing the interference at unintended or unauthorized receivers~\cite{wu2020towards}. These appealing features have motivated the integration of \gls{RIS} into next-generation wireless networks for performance improvement. More specifically, \gls{RIS} has been adopted into future wireless networks since it has the potential to offer an effective way for \gls{PLS} improvements. In light of the above, \gls{RIS} has emerged as a viable solution to tackle wireless security concerns from an information-theoretic perspective. By incorporating the \gls{RIS} technology into \gls{PLS}, a new paradigm that mitigates eavesdropping threats 
is introduced, leading to more secure and reliable communications for future wireless networks~\cite{feng2021physical,yu2020robust,almohamad2020smart,ge2022robust}.  

To fully implement the \gls{IoE} concept, next-generation wireless networks will require large bandwidths due to their real-time demands. To this end, the exploration of high-frequency bands for \gls{RF} communications (e.g., \gls{mmWave} and \gls{THz}) and optical communications (e.g., \gls{FSO}, \gls{VLC}, \gls{IR}) is imperative. However, communications occurring at these frequency bands suffer from severe atmospheric attenuations - owing to the absorption by water vapor and oxygen molecules - yielding short transmission distances. Consequently, a dead-zone problem may arise in many scenarios (e.g., a long distance between the transmitter and the receiver or a blocked received signal due to an obstacle between the transmitter and the receiver). Recent efforts have adopted the \gls{RIS} technology as an effective approach to solve skip-zone situations as opposed to the cooperative relaying technology (see~\cite{mamaghani2022terahertz,ruiz2021analysis,khoshafa2019physical,pan2020secure,aboagye2021energy, osman2023secrecy} and the references therein) where the propagation of radio waves cannot be altered and/or controlled. Hence, \gls{RIS} can be deployed in \gls{RF} and optical communications in the high-frequency bands to mitigate privacy and security threats (e.g., jamming, eavesdropping, and pilot contamination attacks). 

\subsection{Related Work and Existing Surveys}
Few works have reported the progress made on the integration of \gls{RIS} in \gls{RF} and/or optical systems for the development of future wireless communications to provide a plethora of services to the users (see~\cite{naeem2022irs,pogaku2022uav,aboagye2022ris} and references therein). In~\cite{naeem2022irs}, a contemporary overview of the \gls{RIS} architecture and deployment strategies in the future \gls{6G} was presented. On the latter, the authors particularly focused on the integration of the \gls{RIS} technology with other \gls{6G}-enabling applications such as \gls{NOMA}, \gls{THz}/\gls{mmWave}, \gls{UAV}, \gls{MEC} and \gls{PLS} for the system performance improvement. 
The research work of~\cite{pogaku2022uav} provided an in-depth survey on the integration of \gls{RIS} and \gls{UAV} in the context of emerging communications, \gls{DRL}, optimization, secrecy performance, and \gls{IoT}. The authors of~\cite{aboagye2022ris} presented a comprehensive tutorial on the design and application of the \gls{RIS} technology in indoor \gls{VLC} systems. Moreover, an overview of optical \glspl{RIS} was provided and the differences with \gls{RF}-\gls{RIS} in terms of functionalities were highlighted. However, the security aspect of these systems was not investigated.

%\textcolor{black}{The focus of these surveys are either on \gls{RIS} as in~\cite{basar2019wireless, liang2019large, gong2020toward, direnzo2020smart, liu2021reconfigurable, munochiveyi2021reconfigurable, naeem2022irs, zhang2022intelligent, pogaku2022uav, aboagye2022ris} or on \gls{PLS} as in~\cite{illi2023physical, pakravan2023physical}. All aforementioned surveys briefly touch the integration of \gls{RIS}-assisted \gls{PLS} in both \gls{RF} and \gls{OWC} systems.}

Only a handful of research works has considered the integration of \gls{RIS}-assisted \gls{PLS} for future wireless systems~\cite{almohamad2020smart,naeem2023security,khalid2023reconfigurable,kaur2024survey}. In~\cite{almohamad2020smart}, the authors conducted a literature review on the information-theoretic security of \gls{RIS}-assisted wireless networks focusing on the classification of the \gls{RIS}-assisted \gls{PLS} applications, on multi-antenna configurations, and on optimization problems for the maximization of some performance metrics (e.g., \gls{SR}, \gls{SC}). In~\cite{naeem2023security}, the authors reviewed the security challenges impacting the integration of \gls{RIS} in future wireless networks. To this end, they outlined the security and privacy threats associated with the adoption of \gls{RIS} and \gls{mmWave}, \gls{THz}, \gls{D2D}, \gls{IoT} networks, \gls{MEC}, \gls{SWIPT}, \gls{ISAC}, and \glspl{UAV}. In~\cite{khalid2023reconfigurable}, the authors discussed the designs of \gls{RIS}-assisted \gls{PLS} for \gls{6G}-\gls{IoT} networks against security and privacy attacks. Moreover, simulation results are provided to illustrate the effectiveness of \gls{RIS} from an information-theoretic security viewpoint. \textcolor{black}{The authors in~\cite{kaur2024survey} discussed the various secrecy performance metrics of wireless networks and presented an overview of the \gls{RIS}-assisted \gls{PLS} for various antenna configurations including \gls{SISO}, \gls{MISO} and \gls{MIMO}}. Some common points of the existing research works are: (a) the reported studies of \gls{RIS}-assisted \gls{PLS} are mainly investigated in \gls{RF} systems; (b) the issue of security from an information-theoretic security perspective is partially investigated except for~\cite{kaur2024survey}. 

\textcolor{black}{However, considering the aforementioned antenna architectures, the work of~\cite{kaur2024survey} mainly focuses on exploring the \gls{RIS}-assisted \gls{PLS} in \gls{RF} systems \textendash with a brief discussion in \gls{OWC} systems. To the best of the authors' knowledge, there is no up-to-date survey work in the open literature that presents a comprehensive overview of the information-theoretic security aspects in both the \gls{RIS}-assisted \gls{RF} and \gls{OWC} systems. This is the knowledge gap our survey paper aims to partially fill by emphasizing on the following aspects:}
\begin{enumerate}
    \item[$\bullet$] \textit{Integration of \gls{RIS} technology with \gls{PLS} in \gls{RF} communication systems}: We first introduce some \gls{PLS} techniques such as \gls{AN}, \gls{BF} and cooperative jamming to mitigate the security threats in different scenarios for future wireless communication systems. Subsequently, we provide valuable insights into the adaptability and optimization of \gls{RIS} for various communication scenarios. We also highlight the scalability and adaptability of \gls{RIS}-assisted \gls{PLS} techniques when the network complexity increases. Moreover, we outline the integration of \gls{RIS} in emerging \gls{RF} communication systems, namely with multi-antenna communications, \gls{mmWave} communications, \gls{THz} communications, \gls{UAV} communications, \gls{D2D} communications, \glspl{CRN}, \gls{WPC}/\gls{SWIPT}, \gls{MEC}, satellite-enabled networks, multicast communications, cell-free networks, relay-aided networks, vehicular communications, \gls{WBAN}, Ad-hoc networks, \gls{ISAC}, radar communications, IoT networks, backscatter communications, wireless sensor networks, and \gls{HAPS}.  
   
    \item[$\bullet$] \textit{Integration of \gls{RIS} technology with \gls{PLS} in \gls{OWC} systems}: 
    We review the integration of \gls{RIS}-assisted \gls{PLS} in \gls{VLC} systems by discussing its evolution from single-user scenarios for static or mobile user to multi-user scenarios. Furthermore, an extension of the \gls{RIS}-assisted \gls{PLS} in hybrid \gls{VLC}-\gls{RF} and \gls{FSO}-\gls{RF} systems is discussed with emphasis on the motivation for such mixed systems and some useful secrecy performance measures.
   
    \item[$\bullet$] \textit{Optimization techniques}: We first review state-of-the-art optimization techniques for \gls{PLS} and then summarize existing optimization strategies for the \gls{SR} maximization and various secrecy metrics associated with \gls{RIS}-assisted \gls{PLS} systems such as: (a) \gls{AO} and \gls{BCD}) to decouple joint optimization variables into sub-problems; (b) \gls{SDR} to relax non-convex \gls{RIS} phase constraints; (c) \gls{MM}, \gls{SCA}, and quadratic transform for non-convex and nonlinear objective functions and constraint; (d) \gls{CCP} to solve the \gls{BS} \textcolor{black}{\gls{BF}} sub-optimization problem. A discussion on the optimization of the \gls{SR} and various secrecy metrics (e.g. \gls{SNR}, power consumption, computation and energy efficiencies, ergodic secrecy rate, channel gain, secrecy key rate, utility function and outage probability) ensue.
    
    \item[$\bullet$] \textit{Machine learning techniques}: \Gls{ML} has become popular in wireless communications due to its ability to solve complex optimization problems in wireless networks. We explore state-of-the-art \gls{ML} techniques for optimizing \gls{RIS}-assisted \gls{PLS} in \gls{RF} and \gls{OWC} systems, such as \gls{DRL} and unsupervised learning. Besides addressing the issues associated with the stringent requirements of future wireless communication networks, the adoption of \gls{ML} algorithms in improving the security of wireless networks results in some advantages, such as proactive detection and preventive threats.
    
    \item[$\bullet$] \textit{Performance analysis}: We explore the improvement of some performance metrics (\gls{SNR}, \gls{SC} and  \gls{SOP}) via the integration of the \gls{RIS} technology within the context of \gls{PLS} in \gls{RF} and \gls{OWC} systems. Furthermore, we highlight the important role of \gls{RIS} to mitigate the noise and increase the signal strength on one hand and to improve the \gls{SC} and reduce the \gls{SOP} on the other hand.

    \item[$\bullet$] \textit{Applications and challenges}: We give a general review of applications of \gls{RIS}-assisted \gls{PLS} towards future wireless networks including all 
    the aforementioned technologies mentioned in the first two bullets. Moreover, we discuss how the integration of \gls{RIS} and cutting-edge technologies can improve the security of future wireless networks. Finally, we highlight some key research challenges and future directions 
    for \gls{RIS}-assisted \gls{PLS} in upcoming wireless networks.
\end{enumerate}

\subsection{Paper Organization}

The remainder of this survey paper is organized as follows. Section~\ref{Section: PLS Fundamentals} presents the basic concepts of \gls{PLS} followed by a description of the type of attacks and performance metrics in \gls{PLS}. Section~\ref{Section: RIS: Architecture and Capabilities} introduces the \gls{RIS} technology and its fundamental principles that enable 
its basic operation. A description of the \gls{RIS} architectures and the integration of this emerging technology in optical systems is subsequently provided. Section~\ref{Section: RIS-Assisted PLS in RF Communication Systems} discusses \gls{PLS} techniques associated with \gls{RIS}-aided systems, 
and further explores the efficacy of integrating \gls{RIS} with \gls{RF}-based emerging technologies 
for security and privacy improvements of future 
wireless communication systems. Section~\ref{Section: RIS-Assisted PLS in OWC Systems} elaborates on \gls{PLS} in wireless systems that integrate \gls{RIS} with \gls{OWC}. Some optimization techniques tailored for \gls{RIS}-assisted \gls{PLS} are comprehensively reviewed in Section~\ref{Section: Optimization Techniques for RIS-Assisted PLS}. Section~\ref{Section: Machine Learning Techniques for RIS-Assisted PLS} explores various \gls{ML} techniques adopted for the optimization of \gls{RIS}-assisted \gls{PLS} for both \gls{RF} and optical systems. Section~\ref{Section: Performance Analysis for RIS-Assisted PLS} highlights the impact of \gls{RIS}-assisted \gls{PLS} from a performance analysis viewpoint. Section~\ref{Section: Open Research Challenges and Future Directions} highlights the key challenges and future research directions associated with the adoption of \gls{RIS}-assisted \gls{PLS} in future wireless networks. Further discussions on the integration of \gls{PLS} and cutting-edge technologies to improve wireless network security are presented. Concluding remarks are given in Section~\ref{Section: Conclusion}. %Fig.~\ref{Fig:0_Structure} presents the main topics covered in this survey paper.

\section{PLS Fundamentals}
\label{Section: PLS Fundamentals}
\textcolor{black}{This section provides some description on the information-theoretic security through some basic concepts, and in the context of attack classifications and performance metrics ranging from \gls{SR}, \gls{SC}, \gls{SOP}, \gls{PNSC}, \gls{IP}, \gls{ESC} to \gls{EST}. Moreover, we introduce some techniques for achieving information-theoretic security, including \gls{AN} and \gls{BF}.}

\subsection{Introduction to PLS}

Wireless communication is crucial in connecting individuals globally, regardless of location or time. While wireless technologies offer numerous advantages, it is essential to acknowledge the significant risk users face from potential attacks due to the broadcast nature of wireless signals. As a result, the attention to communication security has increased. The conventional approach to enhancing communication security involves employing encryption techniques in the upper layers of the communication stack to protect user data~\cite{wu2018survey}. However, these methods have limitations regarding adaptability and flexibility owing to computational tasks and system complexity. Nevertheless, the unique characteristics of wireless channels enable signals to be transmitted securely between transmitters and receivers at the physical layer. This concept has proven effective in improving communication security. Consequently, \gls{PLS} has emerged as a prominent topic of interest, given the advancements in wireless communication techniques~\cite{wang2018survey,yang2015safeguarding}. In this section, we will provide the fundamentals of \gls{PLS}, covering its concepts, technical evolution, and applications in wireless networks.

%In his seminal work published

In 1949, Shannon established the information-theoretic principles that form the basis of modern cryptography~\cite{shannon1949communication}. Shannon's model encompassed the assumption that a non-reusable private key, denoted as $\mathcal{K}$, is employed to encrypt a confidential message, referred to as $\mathcal{M}$, generating a cryptogram, denoted as $\mathcal{C}$. This cryptogram is subsequently transmitted across a channel that is assumed to be noise-free. \textcolor{black}{The eavesdropper is assumed to have unlimited computational power, full knowledge of the transmission coding scheme, and access to an identical copy of the signal intended for the receiver.} The concept of perfect secrecy was introduced, which necessitates that the \textit{a posteriori} probability of the eavesdropper correctly deducing the secret message based on the received signal is equal to the \textit{a priori} probability of that message. In essence, perfect secrecy implies that the eavesdropper gains no additional knowledge about the secret message beyond what was already known before intercepting the signal. From the information-theoretic perspective, perfect secrecy can be expressed as
\begin{equation}    I(\mathcal{M};\mathcal{C})=0,
    \label{001}
\end{equation}
where $I(\cdot; \cdot)$ represent mutual information. Equation~\eqref{001} states that the mutual information between the message $\mathcal{M}$ and the cryptogram $\mathcal{C}$ is zero, indicating their statistical independence. This absence of correlation implies that there exists no means for the adversary to obtain information about the original message. Thus, perfect secrecy can only be assured if and only if the secret 
key $\mathcal{K}$ possesses entropy that is equal to 
or greater than that of the message $\mathcal{M}$, i.e., $H(\mathcal{K})\geq H(\mathcal{M})$, where $H(\cdot)$ 
is the entropy.  

\begin{figure}[!t]
\centering
\includegraphics[width=3.4in]{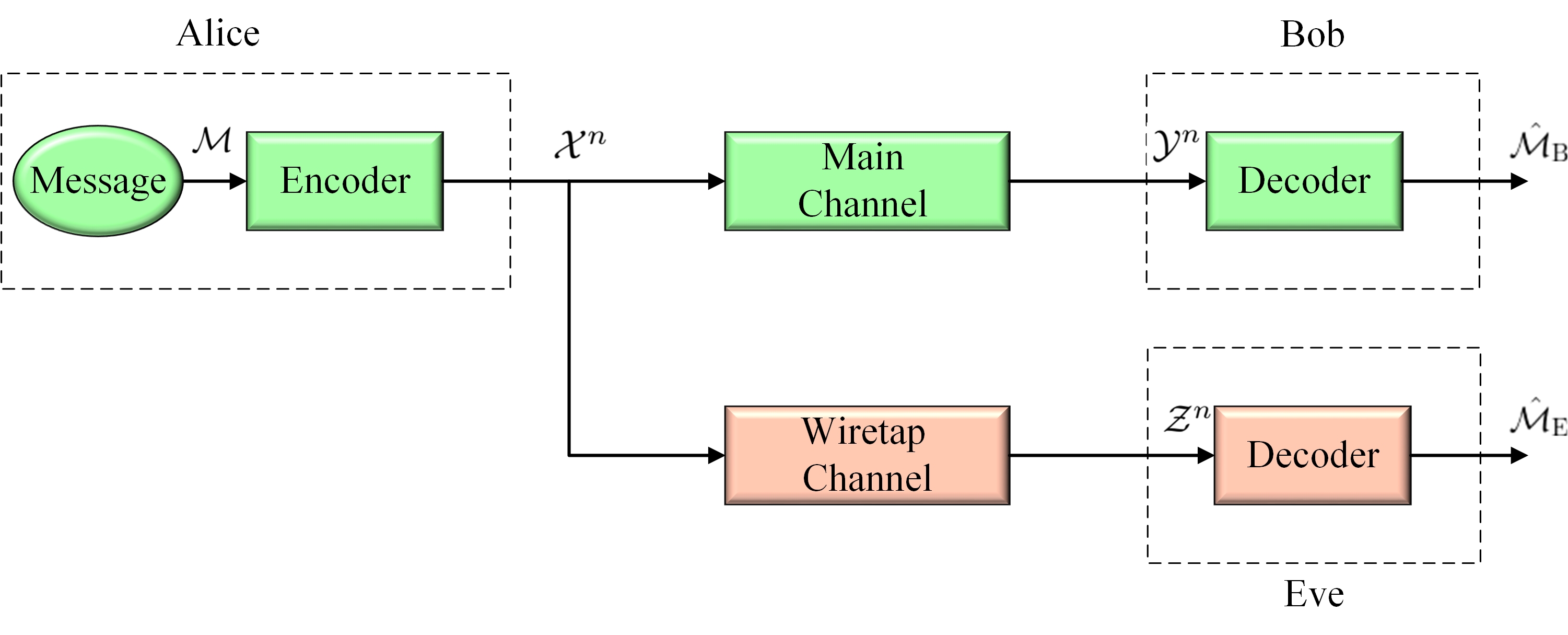}
\caption{{Wireless wiretap system model.}}
\label{Fig01}
\end{figure}

In $1975$, Wyner introduced a significant advancement in the area of information-theoretic security with the proposal of the wiretap channel~\cite{wyner1975wire}. Wyner's system, in contrast to Shannon's original secrecy system, incorporated a random noise channel, which is considered an inherent element of physical communications. As illustrated in Fig.~\ref{Fig01}, the system consists of a legitimate transmitter and receiver, which are commonly referred to as Alice and Bob, respectively; the general wiretap model of \gls{PLS} endeavors to establish communication while contending with the existence of an eavesdropper, commonly referred to as  Eve. Both the legitimate channel and the wiretap channel are modeled as discrete memoryless channels. The transmitter encodes the message $\mathcal{M}$ into a codeword $\mathcal{X}^n$ comprising $n$ symbols, where $\mathcal{X}^n=\left [ \mathcal{X}_{1},\cdots,\mathcal{X}_{n} \right ]$, and transmits it to the intended receiver as a degraded message $\mathcal{Y}^n$, where $\mathcal{Y}^n=\left [ \mathcal{Y}_{1},\cdots,\mathcal{Y}_{n} \right ]$, while simultaneously sending it via a wiretap channel to the adversary as $\mathcal{Z}^n$, where $\mathcal{Z}^n=\left [ \mathcal{Z}_{1},\cdots,\mathcal{Z}_{n} \right ]$, as illustrated in Fig. 2. Moreover, Wyner proposed an alternative definition for perfect secrecy instead of~\eqref{001}. A new secrecy condition was introduced, according to which the equivocation rate $\mathcal{R}_{e}=\frac{1}{n} H\left ( \mathcal{M}/\mathcal{Z}^n \right )
$ should approach the entropy rate $\mathcal{R}$ of the message $\mathcal{R}=\frac{1}{n} H\left ( \mathcal{M}\right )
$ as $n$ tends to infinity.  Consequently, as the value of $\mathcal{R}-\mathcal{R}_{e}=\frac{1}{n} I\left ( \mathcal{M};\mathcal{Z}^n \right )$ suggests, the message $\mathcal{M}$ gradually achieves perfect security against the adversary if $ \frac{1}{n} I\left ( \mathcal{M};\mathcal{Z}^n \right )\leq \varepsilon $, where $\varepsilon$ is an arbitrary small, positive number~\cite{liu2013new}. Furthermore, Wyner defined the \textcolor{black}{\gls{SC}} as the highest achievable transmission rate to the legitimate receiver that guarantees reliability and information-theoretic security against eavesdroppers.

%%%%%%%%%%%%%%%
%%%%%%%%%%%%%%%%%%%%%%%%%%

%%%%%%%%%%%%%%%%%%%%%%%%%%%%%%%%%%%%%%%%%%%%%%%%%%%%%
\subsection{Categorization of Attacks in PLS}
In \gls{PLS}, attacks can be classified into two categories based on the capabilities of the illegitimate nodes: passive attacks and active attacks~\cite{shiu2011physical}. Passive attacks involve illegitimate nodes assuming the role of eavesdroppers, discreetly intercepting transmitted information from legitimate wireless channels without actively transmitting any signals. By concealing their presence, these nodes do not disrupt network operations. Their primary objective is to clandestinely intercept and potentially analyze the information received from the legitimate source, Alice. Consequently, it becomes imperative to prevent eavesdroppers from successfully intercepting information by employing carefully designed signaling techniques~\cite{kapetanovic2015physical}.

On the other hand, active attacks involve illegitimate nodes with the capacity to withstand the risk of detection by legitimate nodes, enabling them to engage in powerful active attacks. These nodes transmit deceptive signals to confuse the intended recipient, Bob. They can intercept and forge messages, thus compromising the security performance of the communication system. Such attacks are also known as masquerade attacks~\cite{shiu2011physical}. Also, malicious nodes can act as jammers, transmitting noisy signals intending to interrupt communication~\cite{wang2016physical}. When Bob receives both the desired and jamming signals simultaneously, the legitimacy of the intended signal becomes less trustworthy. Consequently, the legitimate signal may fail to be decoded. Active attacks can significantly impact normal network operations since adversaries seek to manipulate network data. In the event of an attack, legitimate users must identify the presence of such attacks and subsequently implement appropriate protective measures to safeguard the transmitted signal accordingly.

\subsection{Secrecy Performance Metrics}
\subsubsection{Secrecy Rate} 
The \gls{SR} of the Gaussian noise wiretap channel is determined by calculating the difference between the achievable rates of the main channel and the wiretap channel when utilizing a Gaussian codebook~\cite{wang2016physical}. Mathematically, the \gls{SR}, $\mathcal{R}_{s}$, can be represented as 
\begin{equation}    
\mathcal{R}_{s}=\left [ \mathcal{R}_{b}-\mathcal{R}_{e},0 \right ]^{+},
\end{equation}
where $\mathcal{R}_{b}$ is the achievable rate of the legitimate link,  $\mathcal{R}_{e}$ is the achievable rate of the eavesdropping link, and $\left [x,0 \right ]^{+}=\max\left ( x,0 \right )$.
\subsubsection{Secrecy Capacity}
 The \gls{SC}, $\mathcal{C}_{s}$, can be defined as the highest achievable rate $\mathcal{R}_{s}$ that ensures perfect secrecy rate, i.e., the maximum rate at which secure information can be transmitted over a wireless channel while maintaining confidentiality~\cite{gopala2008secrecy}. \textcolor{black}{\Gls{SC}} indicates the amount of secure communication that can be achieved in the presence of eavesdroppers. In this regard, $\mathcal{C}_{\textup{s}}$ can be obtained by
\begin{align}   \begin{split}       \mathcal{C}_{s}= \big [\mathcal{C}_{b}-\mathcal{C}_{e} ,0 \big ]^{+},\label{c61}    \end{split}\end{align}
where $\mathcal{C}_{b}$ and $\mathcal{C}_{e}$ are the legitimate link and the eavesdropping link capacities, respectively. The primary requirement in this context is that $\mathcal{C}_{b}$ should be greater than $\mathcal{C}_{e}$, highlighting the crucial aspect that the quality of the main channel must surpass that of the wiretap channel, regardless of the computational capabilities possessed by the eavesdropper.
\subsubsection{Secrecy Outage Probability}
In specific scenarios, Alice may not possess perfect \gls{CSI} regarding Bob and Eve. Consequently, the \textcolor{black}{\gls{SOP}} is employed as a performance metric. The secrecy outage occurs when the current secrecy rate $\mathcal{R}_s$ falls below a predefined threshold, indicating an inability to meet the security requirement. The \gls{SOP} corresponds to the likelihood of observing a secrecy outage given a specific fading distribution~\cite{barros2006secrecy}. Mathematically, it is represented as: 
\begin{align}   \begin{split}       \textup{SOP}&=\textrm{Pr}\left ( \mathcal{C}_{s} < \mathcal{R}_s \right ).\label{c6}    \end{split}\end{align}
\subsubsection{Probability of Non-Zero Secrecy Capacity}
The \gls{PNSC}, which can be called the \gls{PPSC} or \gls{SPSC}, is a performance metric used in \gls{PLS} to evaluate the probability that a secure communication link can be established, i.e., the probability that the \textcolor{black}{\gls{SC}} is greater than zero. It indicates the likelihood of achieving positive, secure communication rates and provides insights into the effectiveness of \gls{PLS} mechanisms~\cite{bloch2008wireless, yang2012transmit}, and is given by
\begin{align}   \begin{split}       \textup{PNSC}&=\textrm{Pr}\left ( \mathcal{C}_{s} > 0  \right ).\label{c66}    \end{split}\end{align}
%%%%%%%%%%%%%%%%%%%%
\subsubsection{Intercept Probability}
The \gls{IP}, $P_{\textnormal{int}}$, is a performance metric used to assess the vulnerability of wireless communication systems to eavesdropping attacks. The \textcolor{black}{\gls{IP}} is the likelihood that  $\mathcal{C}_{s}$ of a wireless communication system is below zero, indicating a situation where the system fails to achieve secure communication~\cite{zou2015improving, zou2013optimal}, and is given by
\begin{align}   \begin{split}       P_{\textnormal{int}}&=\textrm{Pr}\left ( \mathcal{C}_{s} < 0  \right ).\label{c666}    \end{split}\end{align} This probability quantifies the vulnerability of the system to unauthorized interception and provides insight into the probability of unsuccessful secrecy establishment.
A lower \textcolor{black}{\gls{IP}} indicates a more secure system, while a higher \textcolor{black}{\gls{IP}} indicates a higher vulnerability to eavesdropping. It serves as a crucial metric for evaluating system security and guiding the implementation of \gls{PLS} techniques to mitigate interception risks and enhance the confidentiality of communication.
%%%%%%%%%%%%%%%%%%%%
\subsubsection{Ergodic Secrecy Capacity}
The \gls{ESC} refers to the statistical mean of the secrecy rate across fading channels. This metric provides insights into the system's ability to maintain confidentiality over time. Mathematically, the ESC, $\mathcal{R}_{s}$, can be represented as
\begin{align} \begin{split}   \mathcal{R}_{s}= \mathbb{E}\big [\mathcal{C}_{b}-\mathcal{C}_{e} ,0 \big ]^{+}.\label{c6666}    \end{split}\end{align}
%%%%%%%%%%%%%%%%%%%%

\subsubsection{\textcolor{black}{Effective Secrecy Throughput}}
{\textcolor{black}{the \gls{EST} is a metric introduced in~\cite{yan2015optimization} to explicitly account for the reliability and secrecy constraints inherent in wiretap channels. This metric quantifies the average rate at which confidential information is transmitted from the transmitter to the receiver without being wiretapped. In this regard, the EST can be obtained by~\cite{yan2015optimization}
\begin{align}   \begin{split}       \textrm{EST}=(\mathcal{R}_b-\mathcal{R}_e)\left[1-\mathcal{O}_{r}(\mathcal{R}_b)\right]\left[1-\mathcal{O}_{s}(\mathcal{R}_e)\right],\label{c67} \end{split}\end{align}
where $\mathcal{O}_{r}(\mathcal{R}_b)=\Pr(\mathcal{R}_b > \mathcal{C}_{b})$ and $\mathcal{O}_s(\mathcal{R}_e) = \Pr(\mathcal{R}_e < \mathcal{C}_{e})$ are the reliability outage probability and the SOP, respectively.}}

%\subsection{Other Secrecy Performance Metrics}
%On top of the mentioned secrecy performance metrics above, other metrics can be utilized to evaluate the secrecy level, such as the \gls{ESC} and \gls{EST}. The \gls{ESC} refers to the statistical mean of the secrecy rate across fading channels. This metric provides insights into the system's ability to maintain confidentiality over time. In addition, the \gls{EST} is a metric introduced in~\cite{yan2015optimization} to explicitly account for the reliability and secrecy constraints inherent in wiretap channels. This metric quantifies the average rate at which confidential information is transmitted from the transmitter to the receiver without being wiretapped.

%\subsubsection{\textcolor{black}{Connection Outage Probability (COP)}}

%\subsubsection{\textcolor{black}{Average number of transmission (ANT)}}

%\subsubsection{\textcolor{black}{Probability of Positive Secrecy Capacity (PPSC) or Strictly Positive Secrecy Capacity (SPSC)}}
%\textcolor{red}{If these metrics are the same as `Probability of Non-Zero Secrecy Capacity'. Please mention there that researchers use those terms interchangeably and then remove this sub-subsection.}

\subsection{Techniques for Achieving PLS}
\gls{PLS} techniques have been extensively studied and developed to enhance the security of wireless communications at the physical layer. These techniques exploit the unique characteristics of the wireless channel to protect the confidentiality and integrity of transmitted data. One well-known technique is \textcolor{black}{\gls{AN}} generation~\cite{goel2008guaranteeing}, which involves deliberately introducing carefully designed random noise to confuse eavesdroppers and make it difficult for them to decode the desired signal accurately. \textcolor{black}{\Gls{BF}} is another effective technique~\cite{wang2012distributed} that focuses the transmitted signal toward the intended receiver while minimizing signal leakage to unintended eavesdroppers, thereby improving communication security. Secure coding schemes, incorporating error correction and channel coding techniques, add redundancy and error correction capabilities to thwart eavesdroppers' decoding attempts~\cite{harrison2013coding}. Physical layer key generation leverages the channel's randomness to establish shared secret keys between communicating parties~\cite{zeng2015physical}. Interference alignment aligns interference caused by eavesdroppers to minimize its impact on the desired signal~\cite{hu2023interference}. Cooperative jamming involves the coordinated transmission of jamming signals to disrupt eavesdroppers' reception~\cite{khoshafa2020physical, khoshafa2020secure}. These techniques, among others, have demonstrated their effectiveness in enhancing the security of wireless communications at the physical layer (see~\cite{jameel2018comprehensive} and the references therein). Their selection and adaptation depend on system requirements, channel conditions, and potential eavesdropper capabilities, and combining multiple techniques can further bolster communication security. The significance of \gls{PLS} approaches within relay networks is highlighted by the heightened susceptibility of intermediate relay nodes to potential wiretapping compared to other network terminals~\cite{khoshafa2020enhancing, khoshafa2020secure40}. 
%%%%%%%%%%%%%%%%%%%%%%%%%%%%%%%%%%%%%%%%%%%%%%%%%%%%%%%%%%%%%%%%%
\section{RIS: Architecture and Capabilities}
\label{Section: RIS: Architecture and Capabilities}

\textcolor{black}{An important aspect of this survey paper is the \gls{RIS} technology. In that regard, this section first gives a brief introduction of the \gls{RIS} technology followed by a review of its basic architecture and fundamental working principles that enable both its mode of operation and deployment techniques. Moreover, the versatility of the \gls{RIS} technology is presented by highlighting its integration into standalone \gls{RF} and \gls{OWC} systems.}

\subsection{Introduction to RISs}
\glspl{RIS} are man-made sheets of \gls{EM} material with programmable macroscopic physical characteristics that intelligently reconfigure the wireless propagation environment by guiding or modifying the impinged radio waves in a desirable direction (e.g., through reflection, refraction, diffraction properties) to boost the signal power at intended receivers and/or to suppress interference at unintended  receivers~\cite{wu2020towards, basar2019wireless,direnzo2020smart, liaskos2019new,xu2023reconfiguring}. \textcolor{black}{Furthermore, they consist of a large number of low-cost and low-power scattering elements that can individually adjust the wireless channel with amplitude and/or phase shift control of the incident signals and act as mere supportive technology for communication (passive \glspl{RIS}). They can also integrate power amplifiers in some elements to further improve the tuning mechanism of the reflected signals (active \glspl{RIS}), making the \gls{RIS} technology attractive to fulfill one of the requirements for future wireless communication systems.} In general, \glspl{RIS} can be considered as nearly-passive devices as they require power only to ensure their reconfigurability. Recently, hybrid \glspl{RIS} have been proposed where some elements may be active~\cite{fotock2024energy}. Unlike similar key-enabling technologies for \gls{5G} and beyond (e.g., active relays, massive \gls{MIMO}, ultra-dense networks and \gls{mmWave} communications~\cite{boccardi2014five}) that can compensate or adapt to limited control over dynamic wireless channels, \glspl{RIS} provide an innovative and 
cost-effective approach to achieve the \gls{KPIs} of \gls{5G} and beyond (see Fig.~\ref{FigKPI}) without the need for power amplifiers, RF chains, and information encoding/decoding algorithms~\cite{saad2020vision}. 

In light of the potential benefits, the \gls{RIS} technology has emerged as a promising solution to mitigate a range of practical challenges inherent to future wireless systems, such as ever-increasing energy consumption, hardware cost, and intra-/inter-system interference. Besides controlling the wireless environment to a certain extent, \glspl{RIS} are economical and environmentally friendly due to their low power consumption and carbon footprint. Moreover, they yield improvement of key performance metrics such as network coverage, spectral efficiency, throughput, and \textcolor{black}{\gls{EE}}, especially in deep-fade and \gls{NLoS} environments wherein the transmitted signal cannot reach the end user with enough power. Moreover, they can readily and seamlessly be integrated into existing wireless networks by mounting them on various structures such as building facades, walls, ceilings, roadside billboards, clothes as well as mobile vehicles due to their flexibility~\cite{liu2021reconfigurable}. %Fig.~\ref{Fig:0_RF_Chart} illustrates examples of some \glspl{RIS} deployments that have been widely investigated in the existing literature. 

\subsection{Working Principles and Basic Architecture}
In what follows, we provide an overview of the working principles of the \gls{RIS} technology as well as its architecture from a hardware architecture and operating principles viewpoint.
%Survey paper: On Channel Estimation and Practical Passive BF Design
\subsubsection{Working Principles}
An \gls{RIS} is a \gls{2D} planar metasurface consisting of a very large number of nearly-passive elements that reflect the impinging \gls{EM} in a desired manner by adjusting the amplitude and/or the phase of the incident signals. In practice, an \gls{RIS} can be implemented by using different technologies, such as \textcolor{black}{\glspl{LC}}, microelectromechanical systems,
doped semiconductors, and electromechanical
switches~\cite{huang2020holographic}. Broadly speaking, to operate an \gls{RIS} needs a smart controller, and it often consists of three layers:
\begin{enumerate}
    \item[$\bullet$] A meta-atom layer that consists of a large number of passive scattering elements interacting directly with the incident signals.
    \item[$\bullet$] A control layer that aims at adjusting the reflection amplitude and phase shift of each meta-atom element, triggered by the smart controller.
    \item[$\bullet$] A communication layer that serves as a gateway to communicate between the control layer and other network components (e.g. \glspl{BS}, \glspl{AP}, etc.).
\end{enumerate}
As mentioned, the most studied implementation for \gls{RIS} assumes that the \gls{RIS} elements are nearly passive, i.e., they do not amplify the incident signals (see~\cite{huang2019reconfigurable,wu2020towards,pan2022overview,direnzo2020smart,pan2020multicell, basar2019wireless,wang2020intelligent} and references therein). Due to the absence of power amplifiers, a nearly-passive \gls{RIS} needs to be sufficiently large in size in order to achieve the desired \textcolor{black}{\gls{BF}} gain in the far field, since in this regime the path-loss scales with the product of the transmitter-\gls{RIS} and \gls{RIS}-receiver links~\cite{danufane2021pathloss,direnzo2020reconfigurable}. 
To increase the coverage of an \gls{RIS}-aided link while keeping the size of the \gls{RIS} small, a possible solution is to use active or hybrid \glspl{RIS}~\cite{khoshafa2021active,xu2021resource,zhang2023active,long2021active}. The reflecting elements of an active \gls{RIS} consist of active \gls{RF} components, which are capable of amplifying the incident signals. Although the basic operation principle is the same in both the passive and active \gls{RIS}, the latter necessitates additional power consumption during the amplification process to support the active elements.

\subsubsection{RIS Architecture}
\begin{figure*}[!ht]
\centering
\includegraphics[width=1.0\textwidth]{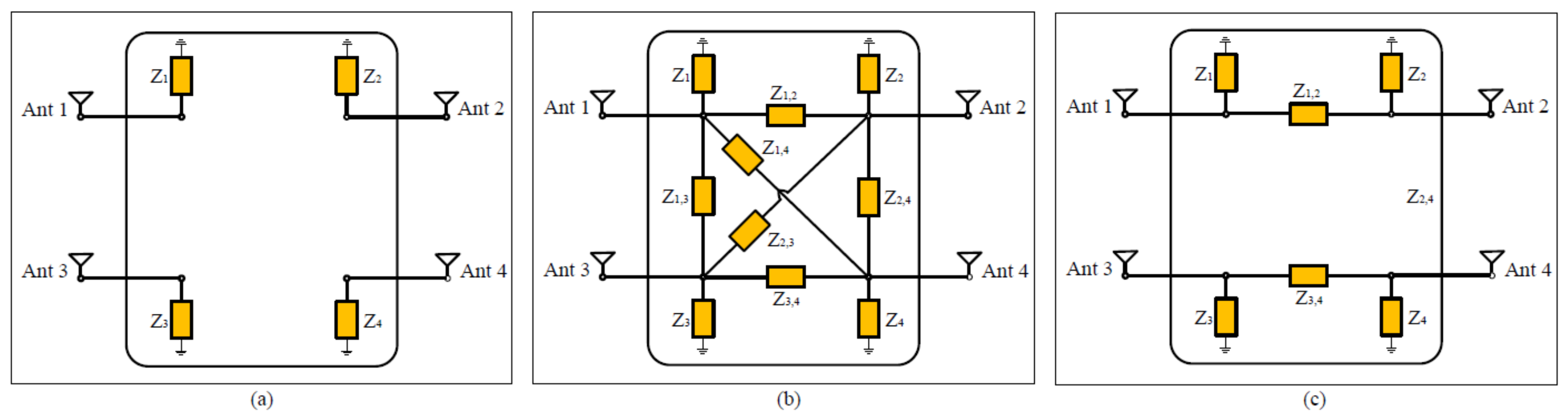}
\caption{\textcolor{black}{4-element \gls{RIS} with (a) single connected reconfigurable impedance network, (b) fully connected reconfigurable impedance network, and (c) group connected (2 groups and group size of 2) reconfigurable impedance network.}}
\label{Fig_RISarchitectures}
\end{figure*}

Each \gls{RIS} reflecting element is conventionally controlled by a tunable circuit, which can be modeled as a tunable impedance connected to the ground~\cite{abeywickrama2020intelligent}. In~\cite{shen2022modeling}, two new circuit topologies have been proposed, wherein all or a subset of the \gls{RIS} reflecting elements are connected via a reconfigurable impedance network. Based on the connection configuration of the reflecting elements, an \gls{RIS} can be classified into three types of architecture (see Fig.~\ref{Fig_RISarchitectures}):
\begin{enumerate}
    \item[$\bullet$] Single connected \gls{RIS}: This is the conventional architecture widely adopted in the literature in which each \gls{RIS} reflecting element is independently controlled by a reconfigurable impedance connected to the ground. It is the simplest of all the \gls{RIS} architectures with limited performance.
    \item[$\bullet$] Fully connected \gls{RIS}: In this setup, each \gls{RIS} element is connected to all the other reflecting elements through a reconfigurable impedance network. This architecture enables an improvement of the \gls{RIS} received signal power due to the additional degrees of freedom~\cite{shen2022modeling}. However, this performance gain comes at the cost of a complex circuit topology.
   \item[$\bullet$] Group or partially connected \gls{RIS}: As the number of reconfigurable impedance components in the network becomes increasingly large, the practical use of the fully connected \gls{RIS} is limited. Consequently, the group/partially \gls{RIS} architecture represents a good trade-off between complexity and performance as it combines both the single and fully connected \gls{RIS} architectures. 
\end{enumerate}

\textcolor{black}{A novel type of \gls{RIS} referred to as \gls{BD-RIS} which goes beyond the conventional \gls{RIS} with inter-element circuit connections and diagonal phase-shift matrices at the expense of increasing circuit complexity, has recently been proposed in~\cite{shen2022modeling}. Recent results have shown that \glspl{BD-RIS} are especially useful in the presence of mutual coupling among the \gls{RIS} elements, as they enable to better control the \textcolor{black}{\gls{EM}} coupling among the elements~\cite{li2024beyond}.}

\subsection{Reflection Modes and Deployment Techniques}
\subsubsection{Reflection Modes}
\begin{figure}[!ht]
\centering
\includegraphics[width=3.5in]{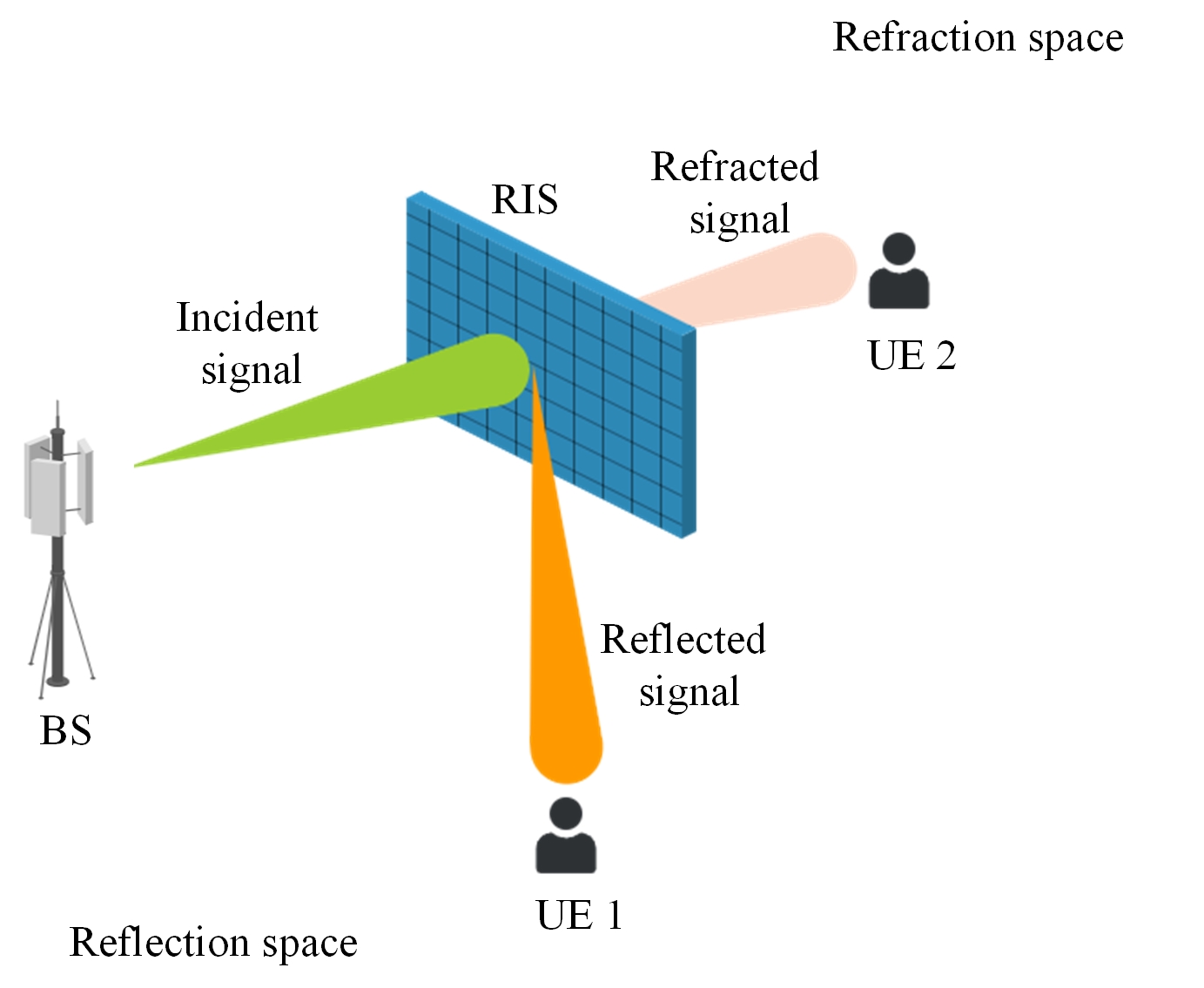}
\caption{\textcolor{black}{Illustration of signal propagation on an \gls{RIS}, where UE$1$ and UE$2$ denote User $1$ and User 2, respectively.}}
\label{Fig_RISSignalPropagation}
\end{figure}

Fig.~\ref{Fig_RISSignalPropagation} depicts the reflected and refracted signals from the incident signal of a generic \gls{RIS} element. With an appropriate design of the geometric parameters and arrangement of the meta-atoms, the incident signal on the meta-surface can be controlled in three possible modes~\cite{xu2022simultaneously,zhang2020beyond}: reflective, refractive, and reflective-refractive. 
%%%%%%%%%%%%%%%%%%%%%%%%%%%%
%%%%%%%%%%%%%%%%%%%%%%%%%%%%

\begin{enumerate}
    \item[$\bullet$] In the reflective-refractive mode, the surface can simultaneously reflect and refract the incident signal to serve users on either side of the surface. This surface is also called \gls{IOS}~\cite{zhang2022intelligent-commag} or \gls{STAR}-\gls{RIS}~\cite{liu2021star,zhang2023star}, and has been proposed to address the half-space limitation of reflecting-only \glspl{RIS}~\cite{direnzo2020smart,nguyen2023design,alghamdi2020intelligent}. Here, the amplitude response of both the reflection and transmission coefficients, denoted by $\Gamma_{l}$ and $\Gamma_{r}$, respectively, are non-negligible.
    \item[$\bullet$] In the reflective mode, the surface fully reflects the incident signal in the direction of the users located on the same side of the transmitter. This is possible through an \gls{RIS}~\cite{yu2021smart} whose transmission coefficient is close to zero to ensure that the energy of the incident signal can be optimally reflected by the surface. 
    \item[$\bullet$] In the refractive mode, the surface also known as \gls{RRS} (see~\cite{zeng2022reconfigurable}) fully refracts the incident \gls{EM} wave to serve users located on the opposite side of the transmitter with respect to the surface. To ensure that the incident signal fully penetrates the surface, it is required that the transmission coefficient be much larger than the reflection coefficient, which should be ideally equal to zero.  
\end{enumerate}
As illustrated in Table~\ref{tab: RISmodes_table}, an  \gls{IOS} encompasses reflecting and refracting \glspl{RIS}, provided that the reflection and transmission coefficients can be appropriately 
optimized.

%%%%%%%%%%%%%%%%%%%%%%%%%%%%
%\begin{figure}[!ht]
%\centering
%\includegraphics[width=3.0in]{Figures/Jules/RIS_modes.pdf}
%\caption{Different {\gls{RIS} modes.}}
%\label{Fig_RISmodes_chart}
%\end{figure}
%%%%%%%%%%%%%%%%%%%%%%%%%%%%

\begin{table}[t] %!h
\centering
\caption{Various RIS modes}
\label{tab: RISmodes_table}
\resizebox{0.3\textwidth}{!}{%
\begin{tabular}{|c|cc|}
\hline
\multirow{3}{*}{\begin{tabular}[c]{@{}c@{}}RIS\\ Modes\end{tabular}} & \multicolumn{2}{c|}{\begin{tabular}[c]{@{}c@{}}IOS\\ Reflective-Refractive\end{tabular}} \\ \cline{2-3} 

 & \multicolumn{1}{c|}{\begin{tabular}[c]{@{}c@{}}IRS\\ Reflective\end{tabular}} & \begin{tabular}[c]{@{}c@{}}RRS\\ Refractive\end{tabular} \\ \hline
 
\end{tabular}%
}
\end{table}

\subsubsection{Deployment Techniques}
\begin{figure*}[!ht]
\centering
\includegraphics[width=0.95\textwidth]{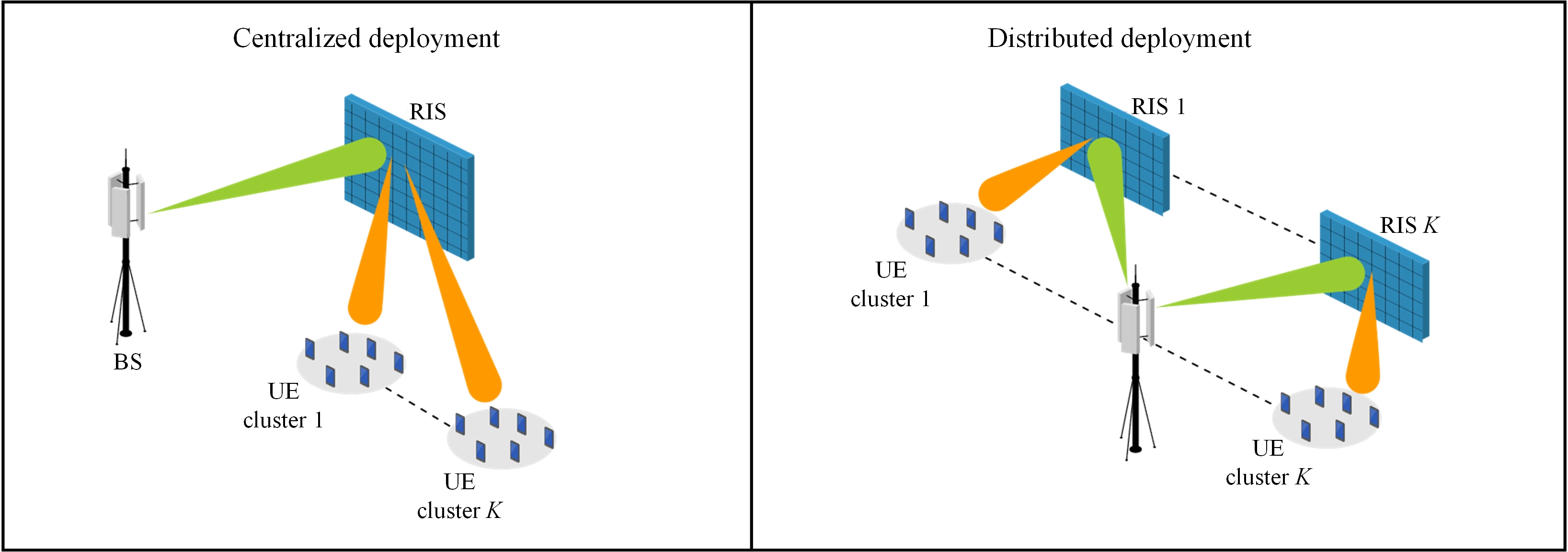}
\caption{
An \gls{RIS}-aided multi-user downlink communication system with different \gls{RIS} deployment strategies.}
\label{Fig_RISdeployentstrategies}\end{figure*}

To further improve the performance gains of \gls{RIS}-assisted wireless networks, it is crucial to appropriately design the reflection pattern of the \gls{RIS} and optimize its deployment in various system setups (i.e., single/multi-antenna, single/multi-user~\cite{he2020cascaded,wu2019intelligent,huang2019reconfigurable}). However, the design of the reflection coefficients is an intricate task since the location of the \gls{RIS}, which is dependent on the path-loss function, is different from that of a relay. To this end, the pioneer work of~\cite{wu2021intelligent} has devoted great effort to address this issue for a single-user scenario. For a more general multi-user cluster scenario,~\cite{ you2022howtodeploy} reviews two \gls{RIS} deployment architectures in an effort to reduce the resultant double path loss of the \gls{RIS}-enabled link. From the perspective of \gls{RIS} deployment, there are two main strategies~\cite{cheng2021downlink,yang2021reconfigurable,cheng2021NOMA,ding2020simple}: distributed \gls{RIS} and centralized \gls{RIS}. Fig.~\ref{Fig_RISdeployentstrategies} illustrates a downlink communication between a \gls{BS} and a single and multiple user clusters in a centralized and distributed configurations, respectively.

%%%%%%%%%%%%%%%%%%%%%%%%%%%%
\begin{figure}[!ht]
\centering
\includegraphics[width=3.5in]{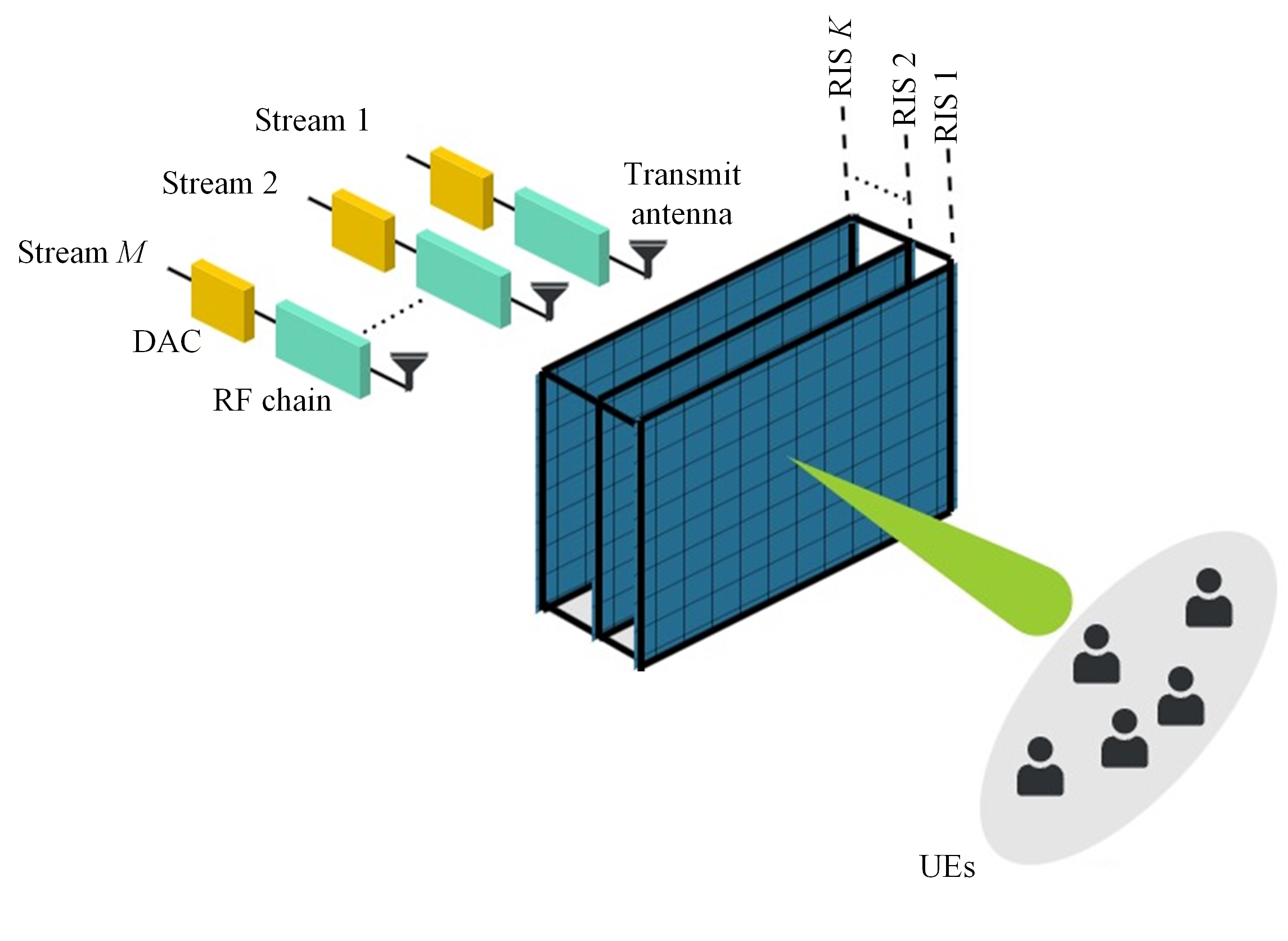}
\caption{An stacked intelligent metasurface-assisted multi-user \gls{MISO} downlink system.}
\label{Fig_SIM-assisted}
\end{figure}
%%%%%%%%%%%%%%%%%%%%%%%%%%%%

\begin{itemize}

    \item[$\bullet$] \textbf{Centralized RIS:} For the centralized \gls{RIS} setup, multiple \gls{RIS} elements are co-located or are grouped as a single large \gls{RIS} and placed in the vicinity of an \gls{AP} as shown on the left-hand side of Fig.~\ref{Fig_RISdeployentstrategies}.

    \item[$\bullet$] \textbf{Distributed RIS:} For the distributed \gls{RIS} strategy as illustrated on the right-hand side of Fig.~\ref{Fig_RISdeployentstrategies}, all the \gls{RIS} elements can be partitioned into multiple \glspl{RIS}, each of them placed near a user cluster to serve the latter. Due to the multiple \glspl{RIS}, there is a greater probability of establishing \gls{LoS} between the \gls{AP} or \gls{BS} and the \glspl{RIS} than the one in the centralized \gls{RIS} scenario.  However, the communication between the \gls{AP} or \gls{BS} and the \glspl{RIS} requires some level of coordination leading to an increase in the signaling overhead. 
\end{itemize}

From a system performance viewpoint, both strategies yield different user achievable rates. This is because their respective effective channels between the user and the \gls{AP} are different from one another. Although the users or user clusters in a central deployment scenario enjoy strong passive \textcolor{black}{\gls{BF}}, they suffer from a reduction in the gain. However, the users in the distributed deployment strategy enjoy a better performance gain albeit a weaker \textcolor{black}{\gls{BF}} gain. This weak gain is due to the fact that the reflected signals by other \glspl{RIS} placed far apart from the \gls{RIS} of interest, are very weak as a result of the significant path loss.

\subsubsection{Multi-layer RIS (or Stacked Intelligent Meta-surface)}
Beyond-diagonal \glspl{RIS} offer more \textcolor{black}{\gls{DoF}} for optimization in the presence of mutual coupling; however, they increase the complexity and power consumption due to the connections among the \gls{RIS} elements and diagonal phase-shift matrices. An alternative way to realize a non-diagonal scattering matrix\footnote{\textcolor{black}{A scattering matrix denoted by $\Phi\in\mathbb{C}^{N\times N}$ provides the complete description of the scattering characteristics of an $N$-port reconfigurable impedance network regardless of specific circuit designs~\cite{shen2022modeling}. These characteristics depend on the circuit topology of the $N$-port reconfigurable impedance network. From a \gls{RIS} perspective, a conventional \gls{RIS} 
where each port is only connected to its own reconfigurable impedance, leads to a diagonal scattering matrix. However, a \gls{BD-RIS} yields a scattering matrix not limited to diagonal as all the ports are connected to each other.}} is to leverage the multi-layer \gls{RIS} architecture that is presented in~\cite{an2023stacked} and~\cite{hassan2024efficient}, and is illustrated in Fig.~\ref{Fig_SIM-assisted}. A multi-layer \gls{RIS} is constituted by closely spaced refracting metasurfaces. The incident signal illuminates the first layer, which then refracts the signal towards the second layer, etc..., until the signal reaches the last layer, where it is finally radiated towards the receiver. 
As the wave propagates through the layers, it is appropriately shaped by the tunable meta-atoms in each of the layers, resulting in an end-to-end scattering matrix that is not diagonal anymore thanks to the broadcast nature of wireless signals when propagating from one layer to another~\cite[Eq. 7 and Fig. 2]{hassan2024efficient}. In other words, a non-diagonal scattering matrix is obtained by utilizing dynamic refracting metasurfaces each equipped with single-connected electronics.

\subsection{Optical RIS}
The \gls{RIS} technology discussed thus far in this section is based on its integration for \gls{RF} communications. Unlike \gls{RF} systems where \glspl{RIS} can manipulate the phase
of the reflected signal, the integration of \gls{RIS} in
optical wireless systems\footnote{\textcolor{black}{\Glspl{OWC}} have gained significant interest in recent years due to (a) their ability to offer large bandwidth required for the next generation of wireless applications and beyond~\cite{khalighi2014survey,al2018multi,maraqa2021achievable} and (b) their low cost and ease of implementation with cheap transceivers.} is possible using either a mirror arrays-based \glspl{RIS} or a meta-surface based \glspl{RIS} designed to steer the light~\cite{abdelhady2021visible,najafi2019intelligent,abumarshoud2021lifi}.

In \gls{RIS}-aided \gls{VLC} systems, phase control is not directly applicable, as these systems preferably operate with \gls{IM/DD}, focusing solely on the non-negative and real part of the signal~\cite{abumarshoud2022intelligent}. There are three primary approaches to incorporating \gls{RIS} into \gls{VLC} systems, namely, mirror array-based \gls{RIS}~\cite{sun2022optimization,saifaldeen2022dr,qian2021secure,abumarshoud2022intelligent,soderi2022vlc}, meta-surface-based \gls{RIS}~\cite{zhao2022secure,zhang2022physical,zhang2023security-Phys,wang2023uplink,rahman2023ris}, and \gls{LC}-based \gls{RIS}~\cite{ndjiongue2022digital,maraqa2023optical,aboagye2023liquid,maraqa2023optimized}.
Both the mirror array-based and meta-surface-based optical \glspl{RIS} consist of nearly-passive elements in the form of mirror arrays or intelligent metasurfaces used to control incident lights based on the light reflection or refraction principle and the manipulation of the \gls{EM} waves, while the latter uses \glspl{LC} as front-row materials embedded in a layered structure to control light reflection and/or refraction.

\subsubsection{Mirror Array-based RIS}
The mirror array-based \gls{RIS} is typically made 
of glass with a flat or curvy surface covered by a reflective coat. Moreover, it relies on geometric 
optics principles such as the generalized Snell's 
law of reflection and refraction. Each mirror within 
the array is controlled by a mechanical unit which enables independent rotation around two axes that 
are both orthogonal.

\subsubsection{Meta-surface-based RIS}
In contrast, the meta-surface-based \gls{RIS} relies 
on a range of meta-materials to control the light propagation behavior~\cite{sun2022joint}. These materials are metallic nanospheres in a dielectric structure, thin metallic rods isotropically distributed in a dielectric medium, splitting metallic elements in 
a dielectric structure, negative index meta-materials, and hyperbolic metamaterials among others. 
Unlike its mirror-array counterpart, which does not operate in refractive mode as it only reflects the incident light, the meta-surface-based \gls{RIS} can yield all light phenomena related to the impact of photon on the unit surface, i.e., reflection, refraction, scattering, and absorption. However, this versatility comes at the expense of cost and complexity. Furthermore, a mirror array-based \gls{RIS} exhibits better performance gains than its meta-surface-based \gls{RIS} counterpart both in \gls{VLC} systems according to the findings in~\cite{abdelhady2020visible}. 

\subsubsection{Liquid Crystal-based RIS}
This type of \gls{RIS} primarily consists of \gls{LC} materials due to their duplex transparency which can be used as reflectors and/or refractors~\cite{ndjiongue2022digital,maraqa2023optical}, since an \gls{RIS} element can smoothly steer and amplify the light signal owing to its electrically controllable birefingence property. Furthermore, \gls{RIS}-based \gls{LC} materials can be placed (a) in front of the \gls{LED} arrays of the \gls{VLC} transmitter for high data rates and adequate illumination performance~\cite{aboagye2023liquid}; (b) in the \gls{VLC} receiver for light amplification and beam steering~\cite{maraqa2023optimized}. The integration of \gls{LC} in \gls{RIS}-aided systems can greatly contribute to satisfying the joint illumination and communication needs of indoor \gls{VLC} technology.

\section{RIS-Assisted PLS in RF Communication Systems}
\label{Section: RIS-Assisted PLS in RF Communication Systems}

\gls{RIS}-assisted \gls{PLS} has emerged as a promising strategy for enhancing the security of \gls{RF} communication systems. By incorporating the \gls{RIS}, this approach introduces a new paradigm for mitigating eavesdropping risks and achieving secure wireless communication. \Gls{RIS} elements act as passive reflecting surfaces that can intelligently modify the wireless channel characteristics. Through careful adjustment of the reflection coefficients, transmitted signals can be manipulated to optimize signal strength at legitimate receivers while simultaneously degrading signal quality at potential eavesdroppers. Achieving this dual optimization involves joint consideration of \textcolor{black}{\gls{BF}}, power allocation, and \gls{RIS} phase shifts. \Gls{RIS}-assisted \gls{PLS} not only enhances \gls{SC} but also enhances the performance of \gls{RF} communication systems against eavesdropping attacks. Various aspects of \gls{RIS} design and deployment have been explored, including \gls{RIS} placement, channel estimation, resource allocation, and signal processing algorithms, to optimize the efficacy of \gls{RIS}-assisted \gls{PLS} techniques. Integration of \gls{RIS} technology with \gls{PLS} in \gls{RF} communication systems holds substantial potential for addressing security challenges in wireless networks and enabling secure and reliable communication in applications such as \gls{IoE}, \gls{5G}/\gls{6G}, and future wireless systems. \textcolor{black}{This section discusses the integration of RIS into RF-based emerging technologies and further explores its efficacy from a security/privacy perspective.} 

%%%%%%%%%%%%%%%%%%%%
\subsection{RIS-Assisted PLS Techniques}
%%%%%%%%%%%%%%%%%%%%%%%%%%%%%%%%%%%%%%%%%%%%%%%%%%%%%%%%%%%%%%%%%
\gls{RIS} offers a unique opportunity to enhance the \gls{PLS} in wireless communication systems through their ability to manipulate the wireless channel characteristics intelligently. Here are some techniques that can be used in conjunction with \gls{RIS} to improve \gls{PLS}.
%%%%%%%%%%%%%%%%%%%%
%%%%%%%%%%%%%%%%%%%%%
%%%%%%%%%%%%%%%%
%%%%%%%%%%%%%%%%%%%%%%%%%%%%%%%%%%%%%%%%%%
\subsubsection{Artificial Noise}
%%%%%%%%%%%%%%%%%%%%%%%%%%%%%%%%%%%%%%%%%%%%%%%%%%%%%%%%%%%%%%%%%
%%%%%%%%%%%%%%%%%%%%%%%%%%%%%%%%
To enhance the security of communication, \gls{AN} is deliberately injected into the transmitted signal with the purpose of confounding potential eavesdroppers~\cite{goel2008guaranteeing}. A critical aspect of \gls{AN} design lies in its ability to avoid causing interference to the intended receiver while simultaneously degrading the intercepted signal quality for the eavesdropper. To achieve this objective, \gls{AN} is integrated with multiple-antenna techniques. By leveraging the spatial degrees of freedom provided by multiple transmit antennas, spatial \textcolor{black}{\gls{BF}} allows for the joint adjustment of \gls{AN} and the transmit signal's directions, thereby optimizing the secrecy performance~\cite{yang2015artificial,
lin2013secrecy}. The effectiveness of \gls{AN} depends on the transmitter's \gls{CSI} accuracy. When the transmitter possesses perfect \gls{CSI}, it gains access to the maximum spatial degrees of freedom for designing the beamformer. However, in practical scenarios, the \gls{CSI} at the eavesdropper's end is often imperfect or entirely unavailable, introducing challenges and limitations to the \gls{AN}'s overall performance.

Integrating \gls{AN} into \gls{RIS} presents a promising avenue to enhance \gls{PLS} in wireless communication systems. By strategically incorporating AN during transmission, \gls{RIS} can introduce controlled random noise alongside the primary signal, confounding potential eavesdroppers and enhancing the confidentiality of transmitted information. This approach benefits from \gls{RIS}'s unique channel manipulation capabilities, enabling it to create constructive interference at the intended receiver and destructive interference at unauthorized recipients. Combining \gls{AN} and \gls{RIS} can strengthen security measures, reducing the risk of unauthorized data interception and providing additional protection beyond conventional encryption techniques. The potential benefits of integrating \gls{AN} to enhance the \gls{SR} in a wireless communication system assisted by \gls{RIS} were investigated in~\cite{guan2020intelligent}.  To optimize the achievable \gls{SR}, a joint design problem was formulated for optimizing transmit \textcolor{black}{\gls{BF}} with \gls{AN} or jamming alongside \gls{RIS} reflect \textcolor{black}{\gls{BF}}. The complexity of the problem arises from its non-convex nature and coupled variables, leading to the proposal of an efficient \textcolor{black}{\gls{AO}} algorithm as a suboptimal solution. Simulation results demonstrate the benefits of incorporating \gls{AN} in transmit \textcolor{black}{\gls{BF}}, particularly in the context of \gls{RIS} reflect \textcolor{black}{\gls{BF}}. Notably, the findings reveal that the \gls{RIS}-aided design without \gls{AN} performs worse than the AN-aided design without \gls{RIS}, particularly when a higher number of eavesdroppers are in proximity to the \gls{RIS}. The authors in~\cite{arzykulov2023artificial} proposed an approach based on the virtual division of \gls{RIS} into two distinct segments. This division aims to strategically configure the phase shifts for each partition, enhance the achievable rate for legitimate users, and strengthen the impact of \gls{AN} on illegitimate users. Two optimization problems were formulated, one focusing on maximizing \gls{SC} and the other on minimizing power consumption. Closed-form solutions were derived by jointly optimizing the partitioning ratio and signal or noise power levels while considering rate constraints for both legitimate and illegitimate users' effectiveness of the proposed \gls{RIS}-partitioning strategy in significantly improving the \gls{SC}. In~\cite{zhang2021improving}, the \gls{PLS} for a \gls{RIS}-aided \gls{NOMA} was investigated, considering scenarios involving both internal and external eavesdropping. A sub-optimal scheme, combining joint \textcolor{black}{\gls{BF}} and power allocation, was proposed to enhance the system's \gls{PLS} against internal eavesdropping, while \gls{AN} scheme was introduced to counter external eavesdropping effectively.

Using \gls{AN} for a STAR-\gls{RIS}-assisted \gls{NOMA} transmission system was investigated in~\cite{han2022artificial} to maximize the \gls{SSR}. To address the inherent non-convexity of the optimization problem, a decoupling strategy was proposed, wherein the optimization of active and passive \textcolor{black}{\gls{BF}} vectors at the \gls{BS} and STAR-\gls{RIS}, respectively, were separated. It is found that the proposed algorithm provides better secrecy performance with less \gls{AN} power compared with the other schemes by using more \gls{RIS} elements to reduce the \gls{AN} power. Increasing the number of transmit antennas at the \gls{BS} reduces the \gls{AN} power if the eavesdropper is quite close to the transmitter while improving it when the eavesdropper is far away. The authors in~\cite{arzykulov2023artificial} utilized the \gls{AN} and \gls{RIS}-partitioning to optimize secure communication. The proposed technique maximizes the legitimate \gls{SC} while constraining Eve's achievable rate by simultaneously reflecting the signal towards the legitimate user and jamming Eve with \gls{AN}. The proposed scheme showed improved \gls{SC} performance compared to traditional \gls{AN}-only and \gls{RIS}-only scenarios. An \gls{AN}-aided secure \gls{MIMO} wireless communication system was investigated in~\cite{hong2020artificial}, where an advanced \gls{RIS} technology was employed, and multiple antennas were deployed at the \gls{BS}, legitimate receiver, and Eve. The main objective is to maximize the \gls{SR} while considering transmit power constraints and unit modulus conditions on \gls{RIS} phase shifts. The efficacy of the proposed approach was demonstrated through simulation results, affirming the effectiveness of the \gls{RIS} in enhancing system security. 

The authors in~\cite{li2021reconfigurable} proposed a secure aerial-ground communication system integrated with \gls{RIS} to counter potential eavesdropping threats via \gls{AN}. The proposed system involves confidential data transmission from a \gls{UAV} operating at a fixed altitude to a legitimate user, while an \gls{RIS} strategically placed on a building's facade enhances secrecy communication performance. The focus is jointly optimizing the phase shifts, trajectory, and transmit power of the \gls{UAV} to improve the secrecy performance of the downlink communication. A prototype of the aerial-ground communication system was implemented, revealing the superiority of the proposed scheme and showing significant performance improvement. Given the imperfect \gls{CSI} possessed by potential eavesdroppers, the authors in~\cite{yu2020robust} introduced a strategy for optimizing the system sum rate while ensuring that information leakage to eavesdroppers remains within acceptable bounds. However, acquiring the \gls{CSI} of potential eavesdroppers is challenging, particularly in scenarios where a passive eavesdropper aims to remain inconspicuous within the network. Furthermore, increasing the transmitted power corresponds to an increased susceptibility of the eavesdropper to extract valuable information, thereby decreasing the secrecy performance. The utilization of the \gls{RIS} characterized by multiplicative randomness to enhance security against passive eavesdropping in wireless communication systems was investigated in~\cite{luo2021reconfigurable}. Enhanced security was achieved without prior knowledge of specific wiretap channels by applying the dynamic adjustment of reflection coefficients during data transmission while maintaining the reliability of the main channel. Performance assessment employs \gls{DoF}, spectral efficiency, and \gls{BER}. The investigation highlighted the potential of \gls{RIS}-enabled multiplicative randomness to enhance wireless network security substantially.

\subsubsection{Beamforming}
%\vspace{2pc}
%%%%%%%%%%%%%%%%%%%%%%%%%%%%%%%%%
%%%%%%%%%%%%%%%%%%%%%%%%%%%%%%%%%%%%%%%%%%%%%%%%%%%%%%%%%%%%%%%
%%%%%%%%%%%%%%%%%%%%%%%%%%%%%%%%%%%%%%%%%%%%%%%%%%%%%%%%%%%%%%%
Employing \textcolor{black}{\gls{BF}} techniques together with \gls{RIS} presents a significant advancement in enhancing PLS within wireless communication systems~\cite{jeong2011joint}. \textcolor{black}{\gls{BF}}, a signal processing method, holds the potential to profoundly influence the signal propagation characteristics by focusing the radiated energy in a specific direction or spatial region, thereby shaping the wireless channel for improved communication quality and performance~\cite{chen2016survey}. \textcolor{black}{Two main approaches exist: passive \gls{BF} at the \gls{RIS} and active \gls{BF} at the transmitter~\cite{wu2019intelligent}. Passive \gls{BF} employs a RIS to reflect incoming signals, directing the desired signal toward the intended receiver and creating null zones around eavesdroppers. It is an energy-efficient solution for resource-constrained applications like the \gls{IoT} and wireless sensor networks~\cite{li2024privacy}. On the other hand, active \gls{BF} involves controlling the phase and amplitude of an antenna array at the transmitter to direct the beam toward the intended receiver. Although this approach necessitates complex signal processing, it provides higher gain and adaptability. Active \gls{BF} is particularly suitable for high-security and dynamic environments, such as massive \gls{MIMO} systems. The choice between passive and active \gls{BF} depends on the unique needs of the specific application.} When integrated with \gls{RIS}s, \textcolor{black}{\gls{BF}} confers several substantial benefits that enhance \gls{PLS}. Firstly, strategically manipulating the reflecting elements' phase shifts through \textcolor{black}{\gls{BF}} enables precise control over signal propagation paths~\cite{pei2021ris}. This control extends to actively steering the transmitted signal towards intended legitimate users while simultaneously attenuating or redirecting signals intended for potential eavesdroppers. Such targeted control over signal distribution significantly mitigates eavesdropping, thereby improving the security of transmitted information. Secondly, the adaptability of \gls{RIS}s in real-time facilitates dynamic adjustments to the reflecting elements' configurations, responding to changing communication scenarios~\cite{he2022ris}. Integrating \textcolor{black}{\gls{BF}} with \gls{RIS}s allows for real-time adjustments in signal directionality and spatial focus, thereby aiding in establishing secure communication links amidst dynamic and potentially insecure environments. This adaptability also empowers the system to counteract eavesdropping attempts through instantaneous reconfiguration of the reflecting elements, making identifying secure transmission paths challenging for potential adversaries~\cite{huang2021multi-JSAC}. Moreover, the integration of \textcolor{black}{\gls{BF}} with \gls{RIS}s augments the overall link quality by mitigating signal fading and shadowing effects. By intelligently redistributing and focusing energy, the combined approach enhances \gls{SNR}s and minimizes the impact of interference, thereby contributing to improved communication reliability and strength against unauthorized access.

In~\cite{hu2021robust}, a self-sustainable \gls{RIS} was employed to enhance the security at the \gls{PLS} within a \gls{MISO} broadcast configuration featuring multiple eavesdroppers. The unique characteristic of the \gls{RIS}s is their ability to manage the activation status of their reflecting elements, directing them either for signal reflection or energy-harvesting purposes.
The authors in~\cite{li2021robust} proposed a secure communication system between \glspl{UAV} and ground stations, where the operation of an \gls{RIS} was incorporated. A joint optimization endeavor concerning the optimal trajectory of the \gls{UAV} and the passive \textcolor{black}{\gls{BF}} configuration of the \gls{RIS}, where imperfect \gls{CSI} on the \gls{RIS}-eavesdropper and \gls{UAV}-eavesdropper connections were considered. In addition, enhancing the ergodic \gls{SR} within a secure communication paradigm employing \gls{MIMO} architectures was investigated in~\cite{liu2021secrecy}, where direct links between the BS and authorized users, and potential eavesdroppers were blocked, with statistical \gls{CSI} for the \gls{RIS}. A deterministic approximation for this scenario was derived using random matrix theory. The resource allocation in multi-user communication networks was investigated in~\cite{yang2020beamforming}, focusing on a scenario where a \gls{RIS} enhances a wireless transmitter. The 
aim was to minimize the total network transmit power by optimizing \gls{RIS} phase \textcolor{black}{\gls{BF}} and \gls{BS} transmit power, subject  to user \gls{SINR} constraints. A dual technique was proposed, transforming the problem into a \textcolor{black}{\gls{SDP}} form. A closed-form \gls{RIS} phase \textcolor{black}{\gls{BF}} was provided, and an optimal transmit power was obtained based on standard interference functions. Simulation results highlighted the proposed method's effectiveness, achieving substantial reductions in total transmit power compared to maximal ratio transmission and zero-forcing \textcolor{black}{\gls{BF}} techniques.

In~\cite{sun2022ris}, the \gls{RIS} was used to reduce power consumption and enhance information security in a multi-user cellular network, particularly with imperfect angular \gls{CSI}. The main objective was to maximize worst-case sum rates by designing the system to optimize the receive decoders, digital precoding, and \gls{AN} at the transmitter and analog precoders at the \gls{RIS}. The optimization was done while meeting the constraints, including minimum achievable rate, maximum wiretap rate, and maximum power allocation. The authors in~\cite{cheng2023ris} investigated enhancing \gls{PLS} in a multi-antenna communication system using an \gls{RIS} by jointly optimizing the active and passive \textcolor{black}{\gls{BF}} to maximize \gls{SR}s and minimize power consumption. An algorithm was proposed to utilize the mathematical structure of optimal active \textcolor{black}{\gls{BF}} vectors, in which the iterative updates between transmit and \gls{RIS}-reflecting beamformers were not required. Robust and secure \textcolor{black}{\gls{BF}} in a \gls{RIS}-assisted \gls{mmWave} \gls{MISO} system was investigated in~\cite{lu2020robust} in the presence of multiple single-antenna eavesdroppers near the legitimate receiver, and the \gls{CSI} of cascaded wiretap channels was imperfectly known to the legitimate transmitter. The target was formulating optimization design problems to maximize worst-case achievable \gls{SR} while considering constraints on total transmission power and unit modulus. Maximizing \gls{SR}s in a \gls{mmWave} network containing a \gls{BS}, multiple \gls{RIS}s, multiple users, and a single eavesdropper was introduced in~\cite{rafieifar2023secure}. The objective was to ensure fairness in \gls{SR}s among users, which gives rise to a mixed integer problem under a max-min fairness criterion. The proposed algorithm converges to a Karush-Kuhn-Tucker point for the original problem, and its overall convergence and computational complexity were analyzed. 

The authors in~\cite{ye2022secure} investigated integrating secure directional modulation in \gls{RIS}-assisted communication networks. The technique involves selecting specific antenna subsets to generate randomized signals, enhancing \gls{PLS}. However, introducing \gls{RIS} poses the challenge of aligning an additional beam, which can lead to high sidelobe effects due to discrete optimization in antenna subset selection. To address this, a cross-entropy iterative method was proposed to achieve low-sidelobe hybrid \textcolor{black}{\gls{BF}} for secure directional modulation in \gls{RIS}-aided networks. The approach minimizes maximum sidelobe energy and employs the Kullback-Leibler divergence to select suitable antenna subsets. The \gls{RIS}-assisted \gls{SIMO} system was utilized in~\cite{tang2021securing} to improve the \gls{PLS}, considering the presence of a multi-antenna eavesdropper. To do so, the legitimate receiver employs active full-duplex jamming while simultaneously coordinating the passive \textcolor{black}{\gls{BF}} capabilities of the \gls{RIS} with legitimate reception and jamming. This coordination leads to a joint optimization approach encompassing receive \textcolor{black}{\gls{BF}}, active jamming, and passive \textcolor{black}{\gls{BF}}. The complexity of this optimization was managed through a \textcolor{black}{\gls{BCD}} framework, utilizing techniques such as the generalized eigenvalue decomposition and \textcolor{black}{\gls{SDP}}. In~\cite{huang2023smart}, using an \gls{RIS} from the standpoint of wireless attackers to degrade communication performance in a time-division duplex system was examined, considering a \gls{TDD} massive \gls{MIMO} system. The \gls{RIS} was strategically employed to disrupt channel estimates obtained by the \gls{BS} during the training phase and distort signals received by users during the transmission phase. The performance assessment utilized \gls{MSE} of users, and an efficient method for optimizing the \gls{RIS}'s reflection pattern was proposed based on theoretical findings. 
In~\cite{wang2023zero}, considering active and passive eavesdroppers, the authors investigated a multi-antenna secure transmission system with \gls{RIS} enhancement. A zero-forcing \textcolor{black}{\gls{BF}} strategy was proposed to nullify the transmit beam towards active eavesdroppers' channels and simultaneously enhance \gls{SNR}s for legitimate users and passive eavesdroppers, even without perfect \gls{CSI}. The goal was to optimize the user's \gls{SNR} while respecting constraints on transmit power, \gls{SNR}s of passive eavesdroppers, and \gls{RIS} reflection.

%%%%%%%%%%%%%%%%%%%%%%%%%%%%%%%%%%%%%%%
\subsubsection{Cooperative Jamming}
\begin{figure*}[!th]
\centering
\includegraphics[width=0.95\textwidth]{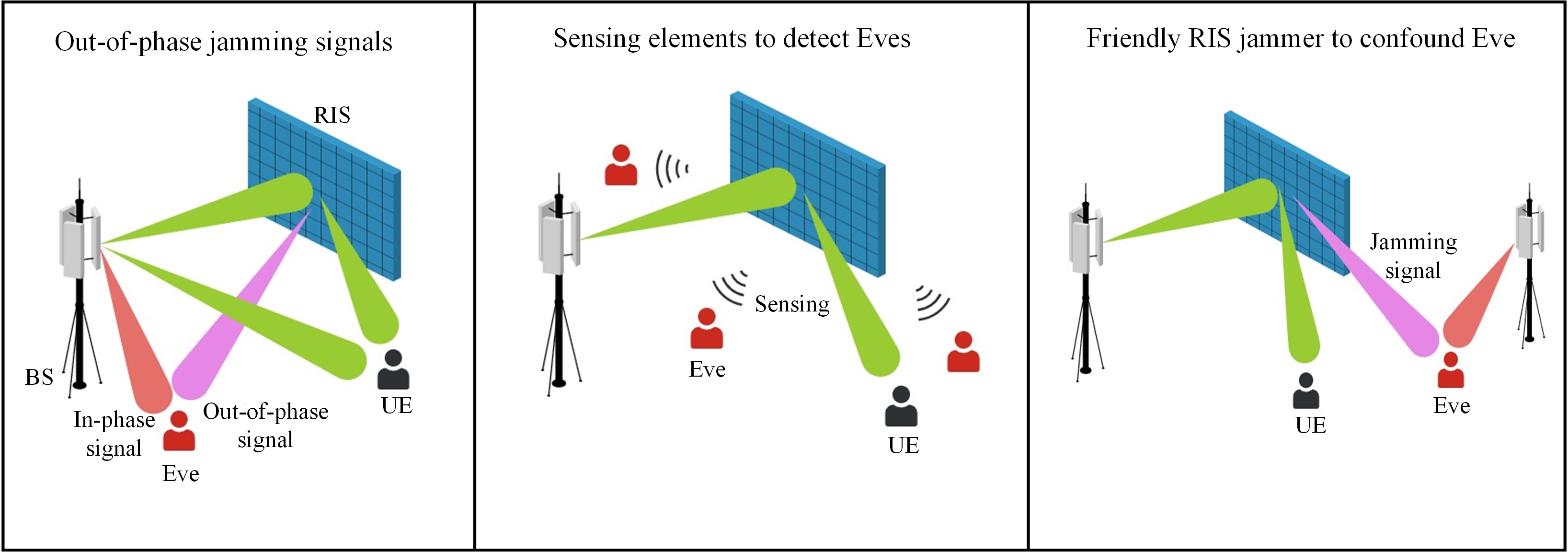}
\caption{{ \textcolor{black}{\gls{RIS}-enabled techniques for enhancing \gls{PLS}: Destructive interference, active sensing, and friendly jamming.}}}
\label{Fig:RIS_Scenarios}
\end{figure*}

%%%%%%%%%%%%%%%%%%%%%%%%%%
In wireless communication systems, ensuring robust security while maintaining efficient data transmission has become an increasingly critical challenge. The emergence of the \gls{RIS} technology has introduced a novel dimension to enhancing the \gls{PLS}. This subsection investigates the promising avenue of employing cooperative jamming methods to increase the security of systems featuring \gls{RIS} integration. As an advanced signal processing strategy, cooperative jamming offers the potential to prevent eavesdropping attempts and optimizes the use of \gls{RIS}-enabled \textcolor{black}{\gls{BF}} to enhance the confidentiality of information transmission~\cite{khoshafa2022physical}.  \textcolor{black}{As shown in Fig.~\ref{Fig:RIS_Scenarios}, RIS presents a novel approach to strengthening communication security through intelligent manipulation of the wireless propagation channel. RIS can induce destructive interference by strategically reflecting signals with a phase shift towards potential eavesdroppers, significantly degrading the intercepted signal. Additionally, active sensing elements embedded within the RIS can gather real-time information regarding the eavesdropper's location. Friendly jamming signals can also be reflected by the RIS to disrupt the eavesdropper's reception further.}

%%%%%%%%%%%%%%%%%%%%
%%%%%%%%%%%%%%%%%%%%

Enhancing secure communications through a \gls{RIS}-assisted \gls{MISO} wireless communication network with independently cooperative jamming was investigated in~\cite{wang2020energy}. The main objective of maximizing \textcolor{black}{\gls{EE}} was achieved through the joint optimization of \textcolor{black}{\gls{BF}}, jamming precode vectors, and the \gls{RIS} phase shift matrix under perfect and imperfect \gls{CSI} conditions. Additionally, a trade-off between \gls{SR} and \textcolor{black}{\gls{EE}} was observed, highlighting the capacity of the \gls{RIS} to enhance \textcolor{black}{\gls{EE}}, even when faced with cases of imperfect \gls{CSI}. The authors in~\cite{tang2021jamming} investigated the potential of utilizing the \gls{RIS}, focusing on \gls{ARIS}, for anti-jamming communications. A joint \gls{ARIS} deployment and passive \textcolor{black}{\gls{BF}} optimization through an \textcolor{black}{\gls{AO}} framework was explored, effectively mitigating jamming attacks and ensuring the security of legitimate data transmissions. In~\cite{ma2021interference}, the \gls{RIS} was utilized to suppress interference and jamming in radio wireless communication systems. The goal was to enhance the \gls{QoS}. The results demonstrated the significance of employing \gls{RIS}, yielding significant anti-interference and jamming benefits. An approach to enhance the resistance of wireless communication systems against jamming interference using \gls{RIS} was presented in~\cite{yang2020intelligent} by improving signal reception for legitimate users while mitigating jamming signals. An optimization problem was formulated to jointly optimize transmit power allocation at the \gls{BS} and reflecting \textcolor{black}{\gls{BF}} at the \gls{RIS}, resulting in anti-jamming performance enhancement. A passive secure communication scheme using an \gls{RIS} was investigated in~\cite{wu2022passive}, where the \gls{RIS} allocated power from incoming signals to transmit confidential information and generate passive jamming signals simultaneously. Each \gls{RIS}'s reflection coefficient served communication and jamming functions, and their joint optimization aimed to maximize the \gls{SR}. The authors of~\cite{tang2023robust} investigated the secure communication techniques utilizing a fixed \gls{RIS} in conjunction with an aerial platform equipped with another \gls{RIS} and a friendly jamming device to enhance security. The \gls{ARIS} improved the legitimate signal, and the fixed \gls{RIS} reinforced cooperative jamming, considering imperfect \gls{CSI}. The problem of configuring \gls{RIS}s to counter jamming attacks in a multi-user OFDMA system was discussed in~\cite{yuan2023joint},  with the uncertainties posed by an uncooperative jammer and limited \gls{CSI} availability. To tackle this problem, a \gls{DRL}-based approach was proposed. The \gls{DRL} framework efficiently learned the \gls{RIS} configuration, expedited learning through this strategy, and achieved rapid convergence in simulations.

%%%%%%%%%%%%%%%%%%%%%%%%%%%%%%%%%%%%%%%%%%%%%

%%%%%%%%%%%%%%%%%%%%%%%%%%%%
\begin{figure*}[!th]
\centering
\includegraphics[width=0.75\textwidth]{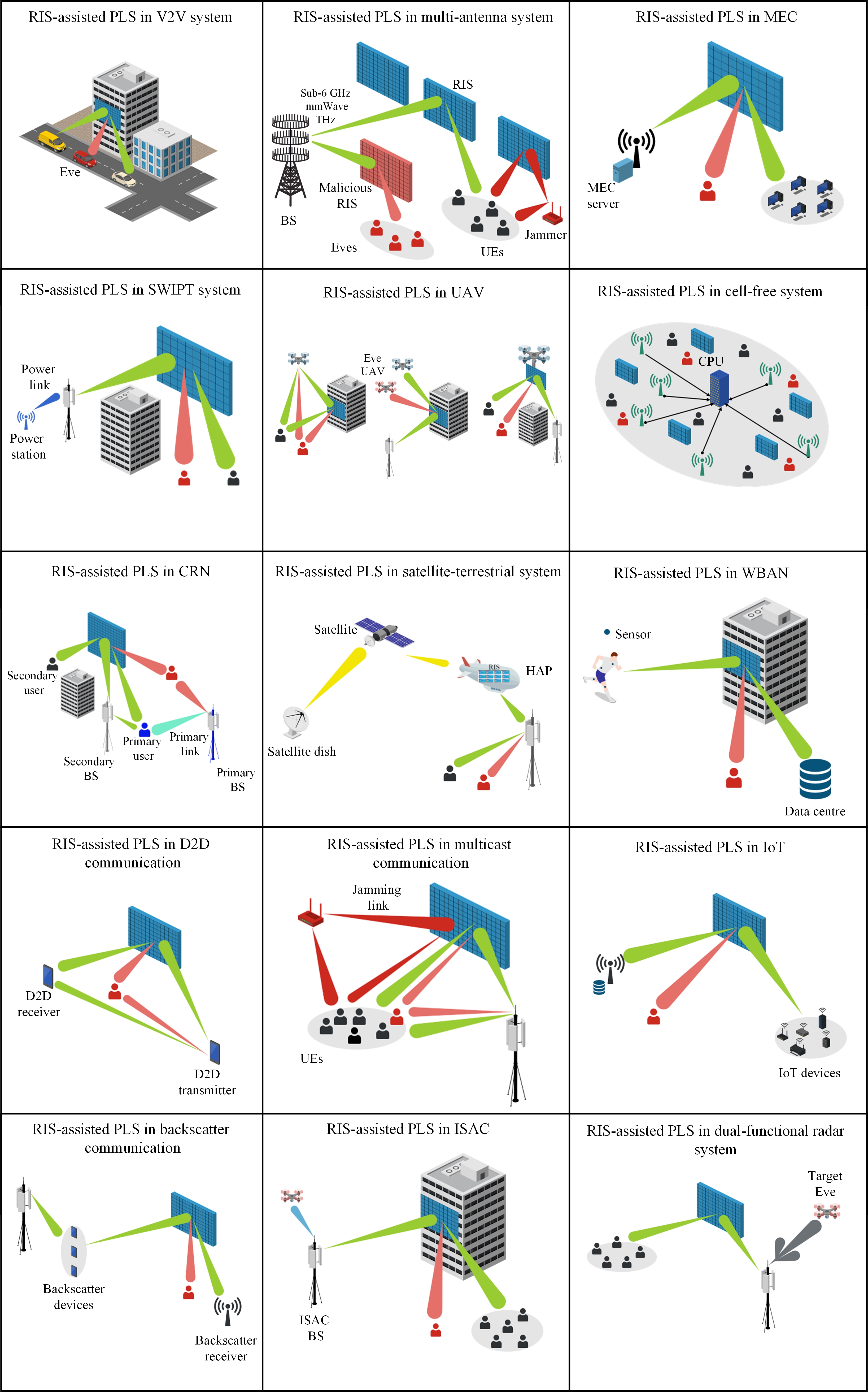}
\caption{{The integration of \gls{RIS} in \gls{PLS}-aided systems with \gls{RF} enabling technologies for secure communications.}}
\label{Fig:0_RF_Chart}
\end{figure*}
%%%%%%%%%%%%%

\begin{subtables}

\begin{table*}[!t]
\centering
\vspace{-2em} 
\caption{Summary of \gls{RIS}-Assisted \gls{PLS} adopted system models in \gls{RF} Communication Systems.} %\textcolor{red}{Please check the text highlighted in red}
\vspace{-0.7em} 
\label{tab:RF-table-1}
\resizebox{0.95\textwidth}{!}{%
\begin{tabular}{|c|cccc|cc|cc|c|}
\hline
\multirow{2}{*}{\textbf{{[}\#{]}}} & \multicolumn{4}{c|}{\textbf{System related content}} & \multicolumn{2}{c|}{\textbf{RIS related content}} & \multicolumn{2}{c|}{\textbf{Security related content}} & \multirow{2}{*}{\textbf{\begin{tabular}[c]{@{}c@{}}Performance\\ Metric(s)\end{tabular}}} \\ \cline{2-9}

& \multicolumn{1}{c|}{\textbf{Technology}} & \multicolumn{1}{c|}{\textbf{System Model}} & \multicolumn{1}{c|}{\textbf{\begin{tabular}[c]{@{}c@{}}CSI\\ Condition\end{tabular}}} & \textbf{\begin{tabular}[c]{@{}c@{}}Channel\\ Model\end{tabular}} & \multicolumn{1}{c|}{\textbf{\begin{tabular}[c]{@{}c@{}}RIS\\ Type\end{tabular}}} & \textbf{RIS(s)} & \multicolumn{1}{c|}{\textbf{\begin{tabular}[c]{@{}c@{}}PLS\\ Technique\end{tabular}}} & \textbf{Eve(s)} &  \\ \hline

%Multi-antenna
\cite{wang2020energy} & \multicolumn{1}{c|}{\multirow{50}{*}{\begin{tabular}[c]{@{}c@{}}Multi-\\antenna \\ commun. \end{tabular}}} & \multicolumn{1}{c|}{SU-MISO-DL} & \multicolumn{1}{c|}{\begin{tabular}[c]{@{}c@{}} Perfect/\\Imperfect \end{tabular}} & Rayleigh & \multicolumn{1}{c|}{Passive} & Single & \multicolumn{1}{c|}{Jamming} & Multiple & EE \\ \cline{1-1} \cline{3-10}

\cite{cui2019secure} & \multicolumn{1}{c|}{} & \multicolumn{1}{c|}{SU-MISO-DL} & \multicolumn{1}{c|}{Perfect} & Rician & \multicolumn{1}{c|}{Passive} & Single & \multicolumn{1}{c|}{BF} & Single & ASR \\ \cline{1-1} \cline{3-10} 

\cite{ren2022performance} & \multicolumn{1}{c|}{} & \multicolumn{1}{c|}{MU-MISO-DL} & \multicolumn{1}{c|}{Statistical} & Rician & \multicolumn{1}{c|}{Passive} & Single & \multicolumn{1}{c|}{BF} & Single & SSR \\ \cline{1-1} \cline{3-10}

\cite{li2021secure} & \multicolumn{1}{c|}{} & \multicolumn{1}{c|}{MU-MISO-DL} & \multicolumn{1}{c|}{Perfect} & Rayleigh & \multicolumn{1}{c|}{Passive} & Multiple & \multicolumn{1}{c|}{BF} & Single & SR \\ \cline{1-1} \cline{3-10}

\cite{niu2022efficient} & \multicolumn{1}{c|}{} & \multicolumn{1}{c|}{SU-MISO-DL} & \multicolumn{1}{c|}{Perfect} & Rayleigh & \multicolumn{1}{c|}{Passive} & Single & \multicolumn{1}{c|}{AN} & Single & SR \\ \cline{1-1} \cline{3-10}

\cite{liu2022secure} & \multicolumn{1}{c|}{} & \multicolumn{1}{c|}{MU-MISO-DL} & \multicolumn{1}{c|}{Perfect} & Rayleigh & \multicolumn{1}{c|}{Passive} & Multiple & \multicolumn{1}{c|}{BF} & Multiple & SR \\ \cline{1-1} \cline{3-10}

\cite{zhou2021secure} & \multicolumn{1}{c|}{} & \multicolumn{1}{c|}{SU-MISO-DL} & \multicolumn{1}{c|}{Perfect} & Rayleigh & \multicolumn{1}{c|}{Passive} & Single & \multicolumn{1}{c|}{BF} & Single & SR \\ \cline{1-1} \cline{3-10}

\cite{feng2020physical} & \multicolumn{1}{c|}{} & \multicolumn{1}{c|}{SU-MISO-DL} & \multicolumn{1}{c|}{Perfect} & Rician & \multicolumn{1}{c|}{Passive} & Single & \multicolumn{1}{c|}{BF} & Single & SR \\ \cline{1-1} \cline{3-10}

\cite{shen2019secrecy} & \multicolumn{1}{c|}{} & \multicolumn{1}{c|}{SU-MISO-DL} & \multicolumn{1}{c|}{Perfect} & Rayleigh & \multicolumn{1}{c|}{Passive} & Single & \multicolumn{1}{c|}{BF} & Single & SR \\ \cline{1-1} \cline{3-10}

\cite{chen2019intelligent} & \multicolumn{1}{c|}{} & \multicolumn{1}{c|}{MU-MISO-DL} & \multicolumn{1}{c|}{Perfect} & Rician & \multicolumn{1}{c|}{Passive} & Single & \multicolumn{1}{c|}{BF} & Multiple & SR \\ \cline{1-1} \cline{3-10}

\cite{chu2019intelligent} & \multicolumn{1}{c|}{} & \multicolumn{1}{c|}{MU-MISO-DL} & \multicolumn{1}{c|}{Perfect} & Rayleigh & \multicolumn{1}{c|}{Passive} & Single & \multicolumn{1}{c|}{BF} & Single & Transmit power \\ \cline{1-1} \cline{3-10}

\cite{alexandropoulos2023counteracting} & \multicolumn{1}{c|}{} & \multicolumn{1}{c|}{MU-MIMO-DL} & \multicolumn{1}{c|}{Perfect} & Rayleigh & \multicolumn{1}{c|}{Passive} & Single & \multicolumn{1}{c|}{AN} & Single & SR \\ \cline{1-1} \cline{3-10}

\cite{sun2023joint} & \multicolumn{1}{c|}{} & \multicolumn{1}{c|}{MU-MIMO-DL} & \multicolumn{1}{c|}{Perfect} & Rayleigh & \multicolumn{1}{c|}{Passive} & Multiple & \multicolumn{1}{c|}{Jamming} & Single & Sum rate \\ \cline{1-1} \cline{3-10}

\cite{chai2023secure} & \multicolumn{1}{c|}{} & \multicolumn{1}{c|}{MU-MISO-DL} & \multicolumn{1}{c|}{Perfect} & Rayleigh & \multicolumn{1}{c|}{Active} & Single & \multicolumn{1}{c|}{BF} & Multiple & SSR \\ \cline{1-1} \cline{3-10}

\cite{zhang2021physical} & \multicolumn{1}{c|}{} & \multicolumn{1}{c|}{MU-MISO-DL} & \multicolumn{1}{c|}{Perfect} & Nakagami-$m$ & \multicolumn{1}{c|}{Passive} & Single & \multicolumn{1}{c|}{BF} & Single & \begin{tabular}[c]{@{}c@{}} SOP, PNSC, \\and ASR \end{tabular} \\ \cline{1-1} \cline{3-10}

\cite{jiang2020intelligent} & \multicolumn{1}{c|}{} & \multicolumn{1}{c|}{SU-MIMO-DL} & \multicolumn{1}{c|}{Perfect} & \begin{tabular}[c]{@{}c@{}} Rayleigh/\\Rician \end{tabular} & \multicolumn{1}{c|}{Passive} & Single & \multicolumn{1}{c|}{BF} & Single & ASR \\ \cline{1-1} \cline{3-10}

\cite{li2023enhancing}  & \multicolumn{1}{c|}{} & \multicolumn{1}{c|}{MU-MIMO-DL} & \multicolumn{1}{c|}{Imperfect} & Rayleigh & \multicolumn{1}{c|}{Passive} & Single & \multicolumn{1}{c|}{-} & Multiple & Secrecy EE, SR \\ \cline{1-1} \cline{3-10}

\cite{shi2023secrecy}  & \multicolumn{1}{c|}{} & \multicolumn{1}{c|}{SU-MISO-DL} & \multicolumn{1}{c|}{Perfect} & Rician & \multicolumn{1}{c|}{Passive} & Single & \multicolumn{1}{c|}{BF} & Multiple & SOP and ESC\\ \cline{1-1} \cline{3-10}

\cite{liu2022minimization} & \multicolumn{1}{c|}{} & \multicolumn{1}{c|}{SU-MISO-DL} & \multicolumn{1}{c|}{Statistical} & Rayleigh & \multicolumn{1}{c|}{Passive} & Single & \multicolumn{1}{c|}{BF} & Single & SOP \\ \cline{1-1} \cline{3-10}

\cite{ji2023secure} & \multicolumn{1}{c|}{} & \multicolumn{1}{c|}{MU-MIMO-DL} & \multicolumn{1}{c|}{\begin{tabular}[c]{@{}c@{}} Perfect/\\Imperfect \end{tabular}} & \begin{tabular}[c]{@{}c@{}} Rayleigh/\\Rician \end{tabular} & \multicolumn{1}{c|}{Passive} & Single & \multicolumn{1}{c|}{Jamming} & Single & Sum rate \\ \cline{1-1} \cline{3-10}

\cite{dong2020secure} & \multicolumn{1}{c|}{} & \multicolumn{1}{c|}{SU-MIMO-DL} & \multicolumn{1}{c|}{Perfect} & \begin{tabular}[c]{@{}c@{}} Rayleigh/\\Rician \end{tabular} & \multicolumn{1}{c|}{Passive} & Single & \multicolumn{1}{c|}{BF} & Single & SR \\  \cline{1-1} \cline{3-10}

\cite{yang2023secure} & \multicolumn{1}{c|}{} & \multicolumn{1}{c|}{MU-MIMO-DL} & \multicolumn{1}{c|}{Statistical} & \begin{tabular}[c]{@{}c@{}} Correlated\\ Rayleigh \end{tabular} & \multicolumn{1}{c|}{Passive} & Single & \multicolumn{1}{c|}{AN} & Single & Ergodic SR \\ \cline{1-1} \cline{3-10}

\cite{song2022secrecy} & \multicolumn{1}{c|}{} & \multicolumn{1}{c|}{SU-MISO-DL} & \multicolumn{1}{c|}{Perfect} & \begin{tabular}[c]{@{}c@{}} Rayleigh/\\Rician  \end{tabular} & \multicolumn{1}{c|}{Passive} & Multiple & \multicolumn{1}{c|}{BF} & Multiple & Secrecy EE \\  \cline{1-1} \cline{3-10}

\cite{ge2021robust} & \multicolumn{1}{c|}{} & \multicolumn{1}{c|}{SU-MISO-FD} & \multicolumn{1}{c|}{Imperfect} & \begin{tabular}[c]{@{}c@{}} Rayleigh/\\Rician  \end{tabular} & \multicolumn{1}{c|}{Passive} & Single & \multicolumn{1}{c|}{BF} & Single & Worst-case SSR \\ \cline{1-1} \cline{3-10}

\cite{shi2023optimization} & \multicolumn{1}{c|}{} & \multicolumn{1}{c|}{SU-MISO-DL} & \multicolumn{1}{c|}{Statistical} & Rician & \multicolumn{1}{c|}{Passive} & Single & \multicolumn{1}{c|}{BF} & Single & Ergodic SR \\  \cline{1-1} \cline{3-10}

\cite{jiang2021joint} & \multicolumn{1}{c|}{} & \multicolumn{1}{c|}{SU-MIMO-DL} & \multicolumn{1}{c|}{Perfect} & \begin{tabular}[c]{@{}c@{}} Rayleigh/\\Rician \end{tabular} & \multicolumn{1}{c|}{Passive} & Single & \multicolumn{1}{c|}{BF} & Single & SSR \\  \cline{1-1} \cline{3-10}

\cite{fang2022intelligent} & \multicolumn{1}{c|}{} & \multicolumn{1}{c|}{SU-MIMO-DL} & \multicolumn{1}{c|}{Perfect} & \begin{tabular}[c]{@{}c@{}} Rayleigh/\\Rician \end{tabular} & \multicolumn{1}{c|}{Passive} & Single & \multicolumn{1}{c|}{AN} & Single & SR \\ \cline{1-1} \cline{3-10}

\textcolor{black}{\cite{de2024malicious}} & \multicolumn{1}{c|}{} & \multicolumn{1}{c|}{\textcolor{black}{MU-MIMO-DL}} & \multicolumn{1}{c|}{\textcolor{black}{Perfect}} & \textcolor{black}{Rayleigh} & \multicolumn{1}{c|}{\textcolor{black}{Passive}} & \textcolor{black}{Single} & \multicolumn{1}{c|}{\textcolor{black}{-}} & \textcolor{black}{Single}  & \textcolor{black}{ASR} \\ \cline{1-1} \cline{3-10}

\textcolor{black}{\cite{wang2024channel}} & \multicolumn{1}{c|}{} & \multicolumn{1}{c|}{\textcolor{black}{MU-MISO-DL}} & \multicolumn{1}{c|}{\textcolor{black}{Unavailable}} & \begin{tabular}[c]{@{}c@{}} \textcolor{black}{Rayleigh/}\\\textcolor{black}{Rician} \end{tabular} & \multicolumn{1}{c|}{\textcolor{black}{Passive}} & \textcolor{black}{Single} & \multicolumn{1}{c|}{\textcolor{black}{BF}} & \textcolor{black}{Single} & \textcolor{black}{Ergodic SR} \\ \cline{1-1} \cline{3-10}

\textcolor{black}{\cite{hong2023outage}} & \multicolumn{1}{c|}{} & \multicolumn{1}{c|}{\textcolor{black}{SU-MISO-DL}} & \multicolumn{1}{c|}{\textcolor{black}{Imperfect}} & \begin{tabular}[c]{@{}c@{}}  \textcolor{black}{Rayleigh/}\\\textcolor{black}{Rician} \end{tabular} & \multicolumn{1}{c|}{\textcolor{black}{Passive}} & \textcolor{black}{Single} & \multicolumn{1}{c|}{\textcolor{black}{AN}} & \textcolor{black}{Multiple} & \textcolor{black}{Transmit power} \\ \cline{1-1} \cline{3-10}

\textcolor{black}{\cite{huang2023smart}} & \multicolumn{1}{c|}{} & \multicolumn{1}{c|}{\textcolor{black}{MU-MISO-DL}} & \multicolumn{1}{c|}{\textcolor{black}{Statistical}} & \textcolor{black}{Rayleigh} & \multicolumn{1}{c|}{\textcolor{black}{Passive}} & \textcolor{black}{Single} & \multicolumn{1}{c|}{\textcolor{black}{-}} & \textcolor{black}{Single} & \textcolor{black}{MSE} \\ \cline{1-1} \cline{3-10}

\textcolor{black}{\cite{ma2023physical}} & \multicolumn{1}{c|}{} & \multicolumn{1}{c|}{\textcolor{black}{SU-MIMO-UL}} & \multicolumn{1}{c|}{\textcolor{black}{Perfect}} & \textcolor{black}{Rayleigh} & \multicolumn{1}{c|}{\textcolor{black}{Passive}} & \textcolor{black}{Single}  & \multicolumn{1}{c|}{\textcolor{black}{-}} & \textcolor{black}{Single} & \textcolor{black}{User SINR} \\ \cline{1-1} \cline{3-10}

\textcolor{black}{\cite{zhang2023secrecy}} & \multicolumn{1}{c|}{} & \multicolumn{1}{c|}{\textcolor{black}{SU-MIMO-DL}} & \multicolumn{1}{c|}{\textcolor{black}{Statistical}} & \textcolor{black}{Rayleigh} & \multicolumn{1}{c|}{\textcolor{black}{Passive}} & \textcolor{black}{Single} & \multicolumn{1}{c|}{\textcolor{black}{AN}} & \textcolor{black}{Single} & \textcolor{black}{Ergodic SR, SOP} \\ \cline{1-1} \cline{3-10}

\textcolor{black}{\cite{bao2023secrecy}} & \multicolumn{1}{c|}{} & \multicolumn{1}{c|}{\textcolor{black}{SU-MISO-DL}} & \multicolumn{1}{c|}{\textcolor{black}{Perfect}} & \begin{tabular}[c]{@{}c@{}}  \textcolor{black}{Rayleigh/}\\\textcolor{black}{Rician} \end{tabular} & \multicolumn{1}{c|}{\textcolor{black}{Active}} & \textcolor{black}{Single} & \multicolumn{1}{c|}{\textcolor{black}{BF}} & \textcolor{black}{Multiple} & \textcolor{black}{Secrecy EE} \\ \cline{1-1} \cline{3-10}

\textcolor{black}{\cite{dassanayake2023secrecy}} & \multicolumn{1}{c|}{} & \multicolumn{1}{c|}{\textcolor{black}{MU-MIMO-UL}} & \multicolumn{1}{c|}{\textcolor{black}{Imperfect}} & \textcolor{black}{Rayleigh} & \multicolumn{1}{c|}{\textcolor{black}{Passive}} & \textcolor{black}{Single} & \multicolumn{1}{c|}{\textcolor{black}{AN}} & \textcolor{black}{Multiple} & \textcolor{black}{SR} \\ \cline{1-1} \cline{3-10}

\textcolor{black}{\cite{yang2024spatially}} & \multicolumn{1}{c|}{} & \multicolumn{1}{c|}{\textcolor{black}{MU-MIMO-DL}} & \multicolumn{1}{c|}{\textcolor{black}{Imperfect}} & \textcolor{black}{Rayleigh} & \multicolumn{1}{c|}{\textcolor{black}{Passive}} & \textcolor{black}{Single} & \multicolumn{1}{c|}{\textcolor{black}{AN}} & \textcolor{black}{Single} & \textcolor{black}{Ergodic SR} \\ \cline{1-1} \cline{3-10}

\textcolor{black}{\cite{li2023sum}} & \multicolumn{1}{c|}{} & \multicolumn{1}{c|}{\textcolor{black}{MU-SIMO-UL}} & \multicolumn{1}{c|}{\textcolor{black}{Perfect}} & \textcolor{black}{Rician} & \multicolumn{1}{c|}{\textcolor{black}{Active}} & \textcolor{black}{Single} & \multicolumn{1}{c|}{\textcolor{black}{BF}} & \textcolor{black}{Single} & \textcolor{black}{Sum SR} \\ \cline{1-1} \cline{3-10}

\cite{dong2020enhancing} & \multicolumn{1}{c|}{} & \multicolumn{1}{c|}{SU-MIMO-DL} & \multicolumn{1}{c|}{\begin{tabular}[c]{@{}c@{}} Perfect/\\Unavailable \end{tabular}} & \begin{tabular}[c]{@{}c@{}} Rayleigh/\\Rician \end{tabular} & \multicolumn{1}{c|}{Passive} & Single & \multicolumn{1}{c|}{AN} & Single & SR \\ \hline

% mmWave Communications
\cite{xiu2020secure} & \multicolumn{1}{c|}{\multirow{12}{*}{\begin{tabular}[c]{@{}c@{}} mmWave \\ commun.\end{tabular}}} & \multicolumn{1}{c|}{SU-MISO-DL} & \multicolumn{1}{c|}{Perfect} & - & \multicolumn{1}{c|}{Passive} & Single & \multicolumn{1}{c|}{AN} & Single & ASR\\ \cline{1-1} \cline{3-10}

\cite{sun2021intelligent} & \multicolumn{1}{c|}{} & \multicolumn{1}{c|}{SU-MISO-DL} & \multicolumn{1}{c|}{Perfect} & Rayleigh & \multicolumn{1}{c|}{Passive} & Single & \multicolumn{1}{c|}{BF} & Single & SR \\ \cline{1-1} \cline{3-10}

\cite{guo2021learning} & \multicolumn{1}{c|}{} & \multicolumn{1}{c|}{MU-MISO-DL} & \multicolumn{1}{c|}{Imperfect} & Rayleigh & \multicolumn{1}{c|}{Passive} & Single & \multicolumn{1}{c|}{BF} & Multiple & SSR \\ \cline{1-1} \cline{3-10}

\cite{lu2020robust} & \multicolumn{1}{c|}{} & \multicolumn{1}{c|}{SU-MISO-DL} & \multicolumn{1}{c|}{Imperfect} & \begin{tabular}[c]{@{}c@{}} Rayleigh/\\Rician \end{tabular} & \multicolumn{1}{c|}{Passive} & Single & \multicolumn{1}{c|}{BF} & Multiple & SR \\ \cline{1-1} \cline{3-10}

\cite{rafieifar2023secure} & \multicolumn{1}{c|}{} & \multicolumn{1}{c|}{MU-MISO-DL} & \multicolumn{1}{c|}{Perfect} & Rayleigh & \multicolumn{1}{c|}{Passive} & Multiple & \multicolumn{1}{c|}{BF} & Single & SR \\ \cline{1-1} \cline{3-10}

\cite{yang2023secure-HB} & \multicolumn{1}{c|}{} & \multicolumn{1}{c|}{SU-MIMO-DL} & \multicolumn{1}{c|}{Perfect} & \begin{tabular}[c]{@{}c@{}} Rayleigh/\\Rician \end{tabular} & \multicolumn{1}{c|}{Passive} & Single & \multicolumn{1}{c|}{BF} & Single & SC \\ \cline{1-1} \cline{3-10}

\textcolor{black}{\cite{tu2024physical}} & \multicolumn{1}{c|}{} & \multicolumn{1}{c|}{\textcolor{black}{SU-MISO-DL}} & \multicolumn{1}{c|}{\textcolor{black}{Imperfect}} & \begin{tabular}[c]{@{}c@{}} \textcolor{black}{Rayleigh/}\\\textcolor{black}{Rician}  \end{tabular} & \multicolumn{1}{c|}{\textcolor{black}{Passive}} & \textcolor{black}{Multiple} & \multicolumn{1}{c|}{\textcolor{black}{BF}} & \textcolor{black}{Single} & \textcolor{black}{SR} \\ \cline{1-1} \cline{3-10}

\cite{ragheb2023ris} & \multicolumn{1}{c|}{} & \multicolumn{1}{c|}{SU-MISO-DL} & \multicolumn{1}{c|}{Perfect} & - & \multicolumn{1}{c|}{Passive} & Single & \multicolumn{1}{c|}{AN} & Single & Ergodic SR \\ \hline

%THz Communications 
\cite{xu2023sum} & \multicolumn{1}{c|}{\multirow{6}{*}{\begin{tabular}[c]{@{}c@{}}THz \\ commun.\end{tabular}}} & \multicolumn{1}{c|}{MU-MISO-DL} & \multicolumn{1}{c|}{Perfect} & - & \multicolumn{1}{c|}{Passive} & Single & \multicolumn{1}{c|}{BF} & Single & SSR\\ \cline{1-1} \cline{3-10} 

\cite{yuan2022secure} & \multicolumn{1}{c|}{} & \multicolumn{1}{c|}{SU-SISO-DL} & \multicolumn{1}{c|}{\begin{tabular}[c]{@{}c@{}}Perfect/ \\ Statistical\end{tabular}} & - & \multicolumn{1}{c|}{Passive} & Single & \multicolumn{1}{c|}{-} & Single &  Ergodic SR\\ \cline{1-1} \cline{3-10} 

\cite{qiao2022securing} & \multicolumn{1}{c|}{} & \multicolumn{1}{c|}{SU-MISO-DL} & \multicolumn{1}{c|}{Imperfect} & - & \multicolumn{1}{c|}{Passive} & Multiple & \multicolumn{1}{c|}{BF} & Single & Worst-case SR\\ \cline{1-1} \cline{3-10} 

\cite{qiao2020secure} & \multicolumn{1}{c|}{} & \multicolumn{1}{c|}{SU-MISO-DL} & \multicolumn{1}{c|}{Perfect} & - & \multicolumn{1}{c|}{Passive} & Single & \multicolumn{1}{c|}{BF} & Single & SR \\ \cline{1-1} \cline{3-10} 

\cite{khel2023secrecy} & \multicolumn{1}{c|}{} & \multicolumn{1}{c|}{SU-SISO-DL} & \multicolumn{1}{c|}{Statistical} & - & \multicolumn{1}{c|}{Passive} & Single & \multicolumn{1}{c|}{-} & Multiple & SC \\ \hline

\end{tabular}%
}

\end{table*} 

\begin{table*}[!t]
\centering
\vspace{-2em} 
\caption{Summary of \gls{RIS}-Assisted \gls{PLS} adopted system models in \gls{RF} Communication Systems}
\vspace{-0.7em} 
\label{tab:RF-table-2}
\resizebox{0.95\textwidth}{!}{%
\begin{tabular}{|c|cccc|cc|cc|c|}
\hline
\multirow{2}{*}{\textbf{{[}\#{]}}} & \multicolumn{4}{c|}{\textbf{System related content}} & \multicolumn{2}{c|}{\textbf{RIS related content}} & \multicolumn{2}{c|}{\textbf{Security related content}} & \multirow{2}{*}{\textbf{\begin{tabular}[c]{@{}c@{}}Performance\\ Metric(s)\end{tabular}}} \\ \cline{2-9}

& \multicolumn{1}{c|}{\textbf{Technology}} & \multicolumn{1}{c|}{\textbf{System Model}} & \multicolumn{1}{c|}{\textbf{\begin{tabular}[c]{@{}c@{}}CSI\\ Condition\end{tabular}}} & \textbf{\begin{tabular}[c]{@{}c@{}}Channel\\ Model\end{tabular}} & \multicolumn{1}{c|}{\textbf{\begin{tabular}[c]{@{}c@{}}RIS\\ Type\end{tabular}}} & \textbf{RIS(s)} & \multicolumn{1}{c|}{\textbf{\begin{tabular}[c]{@{}c@{}}PLS\\ Technique\end{tabular}}} & \textbf{Eve(s)} &  \\ \hline

% UAV        
\cite{jiang2021aerial} & \multicolumn{1}{c|}{\multirow{30}{*}{\begin{tabular}[c]{@{}c@{}}UAV \\ commun.\end{tabular}}} & \multicolumn{1}{c|}{SU-SISO-DL} & \multicolumn{1}{c|}{Perfect} & Rayleigh & \multicolumn{1}{c|}{Passive} & Single & \multicolumn{1}{c|}{BF} & Single & SR \\ \cline{1-1} \cline{3-10} 
        
\cite{li2021reconfigurable}  & \multicolumn{1}{c|}{} & \multicolumn{1}{c|}{SU-SISO-DL} & \multicolumn{1}{c|}{Perfect} & Rician & \multicolumn{1}{c|}{Passive} & Single & \multicolumn{1}{c|}{AN} & Single & ASR \\ \cline{1-1} \cline{3-10} 

\cite{wen2024ris}  & \multicolumn{1}{c|}{} & \multicolumn{1}{c|}{MU-SISO-DL} & \multicolumn{1}{c|}{Perfect} & Rician & \multicolumn{1}{c|}{Passive} & Single & \multicolumn{1}{c|}{AN} & Single & ASR \\ \cline{1-1} \cline{3-10} 

\cite{tang2023secure}   & \multicolumn{1}{c|}{} & \multicolumn{1}{c|}{SU-SISO-DL} & \multicolumn{1}{c|}{Perfect} &  Rayleigh & \multicolumn{1}{c|}{Passive} & Multiple & \multicolumn{1}{c|}{BF} & Multiple & ASR \\ \cline{1-1} \cline{3-10} 
        
\cite{li2021robust} & \multicolumn{1}{c|}{} & \multicolumn{1}{c|}{SU-SISO-UL/DL} & \multicolumn{1}{c|}{Perfect} & Rician & \multicolumn{1}{c|}{Passive} & Single & \multicolumn{1}{c|}{BF} & Single & ASR \\ \cline{1-1} \cline{3-10} 

%\cite{liu2022ris} & \multicolumn{1}{c|}{} & \multicolumn{1}{c|}{SU-SISO-UL} & \multicolumn{1}{c|}{Perfect} & Rician & \multicolumn{1}{c|}{Passive} & Single & \multicolumn{1}{c|}{BF} & Single & ASR \\ \cline{1-1} \cline{3-10} 

\cite{tang2023robust} & \multicolumn{1}{c|}{} & \multicolumn{1}{c|}{SU-SISO-DL} & \multicolumn{1}{c|}{Perfect} & Rician & \multicolumn{1}{c|}{Passive} &  Multiple & \multicolumn{1}{c|}{Jamming} & Multiple & SR \\ \cline{1-1} \cline{3-10}

\cite{wei2022secrecy} & \multicolumn{1}{c|}{} & \multicolumn{1}{c|}{SU-SISO-DL} & \multicolumn{1}{c|}{Perfect} & Rayleigh & \multicolumn{1}{c|}{Passive} & Single & \multicolumn{1}{c|}{BF} & Multiple & SC \\ \cline{1-1} \cline{3-10} 

\cite{wang2023uav} & \multicolumn{1}{c|}{} & \multicolumn{1}{c|}{SU-MISO-DL} & \multicolumn{1}{c|}{Perfect} & Rician & \multicolumn{1}{c|}{Passive} & Single & \multicolumn{1}{c|}{BF} & Multiple & SR \\ \cline{1-1} \cline{3-10}

\cite{pang2021irs} & \multicolumn{1}{c|}{} & \multicolumn{1}{c|}{SU-SISO-DL} & \multicolumn{1}{c|}{Perfect} & Rician & \multicolumn{1}{c|}{Passive} & Single & \multicolumn{1}{c|}{BF} & Single & ASR \\ \cline{1-1} \cline{3-10}

\cite{wei2023secure} & \multicolumn{1}{c|}{} & \multicolumn{1}{c|}{MU-MISO-DL} & \multicolumn{1}{c|}{Perfect} & Rayleigh & \multicolumn{1}{c|}{Passive} & Single & \multicolumn{1}{c|}{BF} & Multiple & SSR\\  \cline{1-1} \cline{3-10}

\cite{cheng2023irs}  & \multicolumn{1}{c|}{} & \multicolumn{1}{c|}{MU-SISO-DL} & \multicolumn{1}{c|}{Perfect} & Rician & \multicolumn{1}{c|}{Passive} & Single & \multicolumn{1}{c|}{-} & Single & SSR \\  \cline{1-1} \cline{3-10}

\cite{sun2023leveraging}  & \multicolumn{1}{c|}{} & \multicolumn{1}{c|}{SU-MISO-DL} & \multicolumn{1}{c|}{Perfect} & Rayleigh & \multicolumn{1}{c|}{Passive} & Single & \multicolumn{1}{c|}{-} & Single & ASR\\  \cline{1-1} \cline{3-10}

\cite{diao2023reflecting}  & \multicolumn{1}{c|}{} & \multicolumn{1}{c|}{SU-SISO-DL} & \multicolumn{1}{c|}{Perfect} & Nakagami-$m$ & \multicolumn{1}{c|}{Passive} & Single & \multicolumn{1}{c|}{-} & Multiple & \begin{tabular}[c]{@{}c@{}}Number of \\ RIS elements\end{tabular} \\  \cline{1-1} \cline{3-10}

\cite{guo2023secure}  & \multicolumn{1}{c|}{} & \multicolumn{1}{c|}{MU-SISO-UL} & \multicolumn{1}{c|}{Perfect} & Rician & \multicolumn{1}{c|}{Passive} & Single & \multicolumn{1}{c|}{-} & Multiple & Secrecy EE \\  \cline{1-1} \cline{3-10}

\cite{cheng2022irs}  & \multicolumn{1}{c|}{} & \multicolumn{1}{c|}{SU-MISO-DL} & \multicolumn{1}{c|}{Perfect} & Rician & \multicolumn{1}{c|}{Passive} & Single & \multicolumn{1}{c|}{BF} & Multiple & ASR \\  \cline{1-1} \cline{3-10}

\cite{kim2023energy}  & \multicolumn{1}{c|}{} & \multicolumn{1}{c|}{SU-SISO-UL} & \multicolumn{1}{c|}{Perfect} & Rayleigh & \multicolumn{1}{c|}{Passive} & Single & \multicolumn{1}{c|}{BF} & Multiple & \begin{tabular}[c]{@{}c@{}} Energy \\ consumption \end{tabular}\\  \cline{1-1} \cline{3-10}

\cite{liu2023deep}  & \multicolumn{1}{c|}{} & \multicolumn{1}{c|}{SU-SISO-DL} & \multicolumn{1}{c|}{Outdated} & \begin{tabular}[c]{@{}c@{}} Rayleigh/\\Rician \end{tabular} & \multicolumn{1}{c|}{Passive} & Single & \multicolumn{1}{c|}{BF} & Single & ASR\\  \cline{1-1} \cline{3-10}

\textcolor{black}{\cite{ye2023robust}}  & \multicolumn{1}{c|}{} & \multicolumn{1}{c|}{\textcolor{black}{SU-SISO-DL}} & \multicolumn{1}{c|}{\textcolor{black}{Imperfect}} & \begin{tabular}[c]{@{}c@{}} \textcolor{black}{Rayleigh/}\\\textcolor{black}{Rician} \end{tabular} & \multicolumn{1}{c|}{\textcolor{black}{Passive}} & \textcolor{black}{Single} & \multicolumn{1}{c|}{\textcolor{black}{Jamming}} & \textcolor{black}{Single} & \textcolor{black}{ASR} \\  \cline{1-1} \cline{3-10}

\textcolor{black}{\cite{jiang2023aerial}}  & \multicolumn{1}{c|}{} & \multicolumn{1}{c|}{\textcolor{black}{SU-SISO-DL}} & \multicolumn{1}{c|}{\textcolor{black}{Perfect}} & \textcolor{black}{Rayleigh} & \multicolumn{1}{c|}{\textcolor{black}{Passive}} & \textcolor{black}{Single} & \multicolumn{1}{c|}{\textcolor{black}{BF}} & \textcolor{black}{Single} & \textcolor{black}{ASR} \\  \cline{1-1} \cline{3-10}

\textcolor{black}{\cite{arzykulov2024aerial}}  & \multicolumn{1}{c|}{} & \multicolumn{1}{c|}{\textcolor{black}{SU-SISO-DL}} & \multicolumn{1}{c|}{\textcolor{black}{Perfect}} & \textcolor{black}{Rayleigh} & \multicolumn{1}{c|}{\textcolor{black}{Passive}} & \textcolor{black}{Single} & \multicolumn{1}{c|}{\textcolor{black}{AN}} & \textcolor{black}{Multiple} & \textcolor{black}{Ergodic SC} \\  \cline{1-1} \cline{3-10}

%\textcolor{black}{\cite{tariq2023reinforcement}}  & \multicolumn{1}{c|}{} & \multicolumn{1}{c|}{\textcolor{black}{SU-MISO-DL}} & \multicolumn{1}{c|}{\textcolor{black}{Perfect}} & \begin{tabular}[c]{@{}c@{}} \textcolor{black}{Rayleigh/}\\\textcolor{black}{Rician} \end{tabular} & \multicolumn{1}{c|}{\textcolor{black}{Passive}} & \textcolor{black}{Single} & \multicolumn{1}{c|}{} &  & \\  \cline{1-1} \cline{3-10} % No EAV - Jamming only paper - ignore

\textcolor{black}{\cite{guo2024ris}}  & \multicolumn{1}{c|}{} & \multicolumn{1}{c|}{\textcolor{black}{MU-MISO-DL}} & \multicolumn{1}{c|}{\textcolor{black}{Statistical}} & \begin{tabular}[c]{@{}c@{}}  \textcolor{black}{Rayleigh/}\\\textcolor{black}{Rician} \end{tabular} & \multicolumn{1}{c|}{\textcolor{black}{Passive}} & \textcolor{black}{Single} & \multicolumn{1}{c|}{\textcolor{black}{BF}} & \textcolor{black}{Multiple} & \textcolor{black}{Multicast rate}\\  \cline{1-1} \cline{3-10}

\textcolor{black}{\cite{diao2024secure}}  & \multicolumn{1}{c|}{} & \multicolumn{1}{c|}{\textcolor{black}{MU-SISO-UL}} & \multicolumn{1}{c|}{\textcolor{black}{Perfect}} & \textcolor{black}{Nakagami-$m$} & \multicolumn{1}{c|}{\textcolor{black}{Passive}} & \textcolor{black}{Multiple} & \multicolumn{1}{c|}{\textcolor{black}{BF}} & \textcolor{black}{Single} & \textcolor{black}{EC, ECP, ESC} \\  \cline{1-1} \cline{3-10}

\textcolor{black}{\cite{dong2024secure}}  & \multicolumn{1}{c|}{} & \multicolumn{1}{c|}{\textcolor{black}{SU-MISO-DL}} & \multicolumn{1}{c|}{\begin{tabular}[c]{@{}c@{}} \textcolor{black}{Perfect/}\\ \textcolor{black}{Statistical} \end{tabular}} & \textcolor{black}{Rayleigh} & \multicolumn{1}{c|}{\textcolor{black}{Passive}} & \textcolor{black}{Single} & \multicolumn{1}{c|}{\textcolor{black}{BF}} & \textcolor{black}{Single} & \textcolor{black}{ASR and SOP} \\  \cline{1-1} \cline{3-10}

\cite{ge2023active}  & \multicolumn{1}{c|}{} & \multicolumn{1}{c|}{MU-MISO-DL} & \multicolumn{1}{c|}{Imperfect} & Rician & \multicolumn{1}{c|}{Active} & Single & \multicolumn{1}{c|}{BF} & Single & Transmit power\\  \hline

%D2D
\cite{khalid2021ris} & \multicolumn{1}{c|}{\multirow{3}{*}{\begin{tabular}[c]{@{}c@{}}D2D \\ commun.\end{tabular}}} & \multicolumn{1}{c|}{SU-SIMO} & \multicolumn{1}{c|}{Perfect} & Rayleigh & \multicolumn{1}{c|}{Passive} & Multiple & \multicolumn{1}{c|}{AN/BF} & Single & Ergodic capacity \\ \cline{1-1} \cline{3-10} 
    
\cite{khoshafa2020reconfigurable} & \multicolumn{1}{c|}{} & \multicolumn{1}{c|}{SU-SISO} & \multicolumn{1}{c|}{Perfect} & Rayleigh & \multicolumn{1}{c|}{Passive} & Single & \multicolumn{1}{c|}{Jamming} & Single & SOP and PNSC \\ \cline{1-1} \cline{3-10} 
  
\cite{hu2023securing} & \multicolumn{1}{c|}{} & \multicolumn{1}{c|}{SU-SISO} & \multicolumn{1}{c|}{Perfect} & Rayleigh & \multicolumn{1}{c|}{Passive} & Single & \multicolumn{1}{c|}{BF} & Single & Data rate and SR \\ \hline

%Cognitive Radio   
\cite{khoshafa2023ris} & \multicolumn{1}{c|}{\multirow{6}{*}{\begin{tabular}[c]{@{}c@{}}CRN\end{tabular}}} & \multicolumn{1}{c|}{SU-SISO-DL} & \multicolumn{1}{c|}{Perfect} & Rayleigh & \multicolumn{1}{c|}{Passive} & Single & \multicolumn{1}{c|}{Jamming} & Single & SOP and PNSC \\ \cline{1-1} \cline{3-10} 

\cite{wu2022joint} & \multicolumn{1}{c|}{} & \multicolumn{1}{c|}{SU-MISO-DL} & \multicolumn{1}{c|}{Perfect} & Rayleigh & \multicolumn{1}{c|}{Passive} & Single & \multicolumn{1}{c|}{BF} & Multiple & SR  \\ \cline{1-1} \cline{3-10}

\cite{dong2021secure} & \multicolumn{1}{c|}{} & \multicolumn{1}{c|}{SU-MISO-DL} & \multicolumn{1}{c|}{Perfect} & Rayleigh & \multicolumn{1}{c|}{Passive} & Single & \multicolumn{1}{c|}{AN} & Single & SR \\ \cline{1-1} \cline{3-10} 

%Note: This paper is discussed in the Performance Analysis section, in text and table, where there is no PLS technique used, just performance analysis.
%\cite{sheng2023performance} \textcolor{red}{New1} & \multicolumn{1}{c|}{} & \multicolumn{1}{c|}{SU-SISO-DL} & \multicolumn{1}{c|}{Perfect} & Nakagami-$m$ & \multicolumn{1}{c|}{Passive} & Single & \multicolumn{1}{c|}{HARQ} & Single & ANT, COP and SOP \\ \cline{1-1} \cline{3-10} 

\cite{niu2022active} & \multicolumn{1}{c|}{} & \multicolumn{1}{c|}{SU-MISO-DL} & \multicolumn{1}{c|}{Perfect} & Rayleigh & \multicolumn{1}{c|}{Active} & Single & \multicolumn{1}{c|}{AN} & Single & SR \\ \cline{1-1} \cline{3-10} 

\cite{wu2021secure} & \multicolumn{1}{c|}{} & \multicolumn{1}{c|}{SU-MISO-DL} & \multicolumn{1}{c|}{Perfect} & Rayleigh & \multicolumn{1}{c|}{Passive} & Single & \multicolumn{1}{c|}{BF} & Multiple & Secrecy EE \\ \cline{1-1} \cline{3-10}

%\textcolor{black}{\cite{hema2024physical}} & \multicolumn{1}{c|}{} & \multicolumn{1}{c|}{\textcolor{black}{MU-SISO-DL}} & \multicolumn{1}{c|}{\textcolor{black}{Perfect}} & \textcolor{black}{Rayleigh} & \multicolumn{1}{c|}{\textcolor{black}{Passive}} &  & \multicolumn{1}{c|}{} &  & \begin{tabular}[c]{@{}c@{}}\textcolor{black}{SOP, secrecy EE,} \\ \textcolor{black}{secrecy throughput} \end{tabular} \\ \cline{1-1} \cline{3-10}  %ingore NO RIS in the paper

%\textcolor{black}{\cite{kumar2024secure}} & \multicolumn{1}{c|}{} & \multicolumn{1}{c|}{\textcolor{black}{SU-SISO-DL}} & \multicolumn{1}{c|}{\begin{tabular}[c]{@{}c@{}}\textcolor{black}{Perfect/} \\ \textcolor{black}{Imperfect} \end{tabular}}  & \textcolor{black}{Rayleigh} & \multicolumn{1}{c|}{\textcolor{black}{Passive}} &  & \multicolumn{1}{c|}{} &  &  \\ \cline{1-1} \cline{3-10} %ingore NO RIS in the paper

\cite{wu2022robust} & \multicolumn{1}{c|}{} & \multicolumn{1}{c|}{SU-MISO-DL} & \multicolumn{1}{c|}{Imperfect} & Quasi-static & \multicolumn{1}{c|}{Passive} & Single & \multicolumn{1}{c|}{BF} & Multiple & Worst-case SR \\ \hline

%SWIPT
\cite{zhai2022improving} & \multicolumn{1}{c|}{\multirow{27}{*}{\begin{tabular}[c]{@{}c@{}}WPC/\\SWIPT\end{tabular}}} & \multicolumn{1}{c|}{MU-SISO-DL} & \multicolumn{1}{c|}{Perfect} & Rayleigh & \multicolumn{1}{c|}{Passive} & single & \multicolumn{1}{c|}{BF} & Multiple &  \begin{tabular}[c]{@{}c@{}} Consumed \\energy \end{tabular} \\ \cline{1-1} \cline{3-10} 

\cite{thien2022secure} & \multicolumn{1}{c|}{} & \multicolumn{1}{c|}{SU-SISO-DL} & \multicolumn{1}{c|}{Perfect} & \begin{tabular}[c]{@{}c@{}} Rayleigh/\\Rician \end{tabular} & \multicolumn{1}{c|}{Passive} & Single & \multicolumn{1}{c|}{-} & Single & ASR \\ \cline{1-1} \cline{3-10} 

\cite{shi2022secrecy} & \multicolumn{1}{c|}{} & \multicolumn{1}{c|}{MU-MISO-DL/UL} & \multicolumn{1}{c|}{Perfect} & Rayleigh & \multicolumn{1}{c|}{Passive} & Single & \multicolumn{1}{c|}{BF} & Single & \begin{tabular}[c]{@{}c@{}} Secrecy \\ throughput \end{tabular} \\ \cline{1-1} \cline{3-10}

\cite{cao2023physical} & \multicolumn{1}{c|}{} & \multicolumn{1}{c|}{SU-SISO-DL} & \multicolumn{1}{c|}{Perfect} & \begin{tabular}[c]{@{}c@{}} Rayleigh/\\Rician \end{tabular} & \multicolumn{1}{c|}{Passive} & Single & \multicolumn{1}{c|}{BF} & Single &  SOP, and EST \\ \cline{1-1} \cline{3-10} 

\cite{cao2022ergodic} & \multicolumn{1}{c|}{} & \multicolumn{1}{c|}{SU-SISO-DL} & \multicolumn{1}{c|}{Perfect} & \begin{tabular}[c]{@{}c@{}} Rayleigh/\\Rician \end{tabular} & \multicolumn{1}{c|}{Passive} & Single & \multicolumn{1}{c|}{BF} & Single & ESC \\ \cline{1-1} \cline{3-10} 

\cite{zhao2022secrecy} & \multicolumn{1}{c|}{} & \multicolumn{1}{c|}{SU-MISO-DL} & \multicolumn{1}{c|}{Statistical} & Rician & \multicolumn{1}{c|}{Passive} & Single & \multicolumn{1}{c|}{BF} & Multiple & ASR \\ \cline{1-1} \cline{3-10} 

\cite{jin2022secure} & \multicolumn{1}{c|}{} & \multicolumn{1}{c|}{SU-MISO-DL} & \multicolumn{1}{c|}{Perfect} & Rician & \multicolumn{1}{c|}{Passive} & Single & \multicolumn{1}{c|}{BF} & Single & SR \\ \cline{1-1} \cline{3-10}

\cite{jiang2023robust} & \multicolumn{1}{c|}{} & \multicolumn{1}{c|}{MU-MISO-DL} & \multicolumn{1}{c|}{Statistical} & Rician & \multicolumn{1}{c|}{Passive} & Multiple & \multicolumn{1}{c|}{AN} & Multiple & \begin{tabular}[c]{@{}c@{}} Power \\ consumption \end{tabular} \\ \cline{1-1} \cline{3-10}

\cite{wang2022robust} & \multicolumn{1}{c|}{} & \multicolumn{1}{c|}{MU-MISO-DL} & \multicolumn{1}{c|}{Perfect} & \begin{tabular}[c]{@{}c@{}} Rayleigh/\\Rician \end{tabular} & \multicolumn{1}{c|}{Passive} & Single & \multicolumn{1}{c|}{Jamming} & Multiple & Transmit power \\ \cline{1-1} \cline{3-10}

\cite{niu2021robust} & \multicolumn{1}{c|}{} & \multicolumn{1}{c|}{MU-MISO-DL} & \multicolumn{1}{c|}{Imperfect} & \begin{tabular}[c]{@{}c@{}} Rayleigh/\\Rician \end{tabular} & \multicolumn{1}{c|}{Passive} & Single & \multicolumn{1}{c|}{AN} & Multiple & Outage rate \\ \cline{1-1} \cline{3-10}

\cite{qiao2022irs} & \multicolumn{1}{c|}{} & \multicolumn{1}{c|}{SU-SISO-UL} & \multicolumn{1}{c|}{Perfect} & \begin{tabular}[c]{@{}c@{}} Rayleigh/\\Rician \end{tabular} & \multicolumn{1}{c|}{Passive} & Single & \multicolumn{1}{c|}{Jamming} & Single & SR \\ \cline{1-1} \cline{3-10}

%\textcolor{black}{\cite{gao2023exploiting}} & \multicolumn{1}{c|}{} & \multicolumn{1}{c|}{} & \multicolumn{1}{c|}{} & \begin{tabular}[c]{@{}c@{}}  \end{tabular} & \multicolumn{1}{c|}{} &  & \multicolumn{1}{c|}{} &  & \\ \cline{1-1} \cline{3-10} % Ignore: No PLS in this paper

\textcolor{black}{\cite{shen2023outage}} & \multicolumn{1}{c|}{} & \multicolumn{1}{c|}{\textcolor{black}{MU-MISO-DL}} & \multicolumn{1}{c|}{\textcolor{black}{Statistical}} & \textcolor{black}{Rayleigh} & \multicolumn{1}{c|}{\textcolor{black}{Passive}} & \textcolor{black}{Single} & \multicolumn{1}{c|}{\textcolor{black}{AN}} & \textcolor{black}{Multiple} & \textcolor{black}{SR} \\ \cline{1-1} \cline{3-10}

\textcolor{black}{\cite{kong2024secrecy}} & \multicolumn{1}{c|}{} & \multicolumn{1}{c|}{\textcolor{black}{SU-MISO-DL/UL}} & \multicolumn{1}{c|}{\textcolor{black}{Perfect}} & \begin{tabular}[c]{@{}c@{}} \textcolor{black}{Rayleigh/}\\ \textcolor{black}{Rician} \end{tabular} & \multicolumn{1}{c|}{\textcolor{black}{Active}} & \textcolor{black}{Single} & \multicolumn{1}{c|}{\textcolor{black}{BF}} & \textcolor{black}{Multiple} & \textcolor{black}{SR} \\ \cline{1-1} \cline{3-10}

\textcolor{black}{\cite{ma2023secure}} & \multicolumn{1}{c|}{} & \multicolumn{1}{c|}{\textcolor{black}{SU-MISO-DL/UL}} & \multicolumn{1}{c|}{\textcolor{black}{Imperfect}} & \begin{tabular}[c]{@{}c@{}} \textcolor{black}{Rayleigh/}\\ \textcolor{black}{Rician} \end{tabular} & \multicolumn{1}{c|}{\textcolor{black}{Passive}} & \textcolor{black}{Single} & \multicolumn{1}{c|}{\textcolor{black}{BF}} & \textcolor{black}{Single} & \textcolor{black}{SR} \\ \cline{1-1} \cline{3-10}

\cite{hehao2020intelligent} & \multicolumn{1}{c|}{} & \multicolumn{1}{c|}{SU-MIMO-DL} & \multicolumn{1}{c|}{Perfect} & \begin{tabular}[c]{@{}c@{}} Rayleigh/\\Rician \end{tabular} & \multicolumn{1}{c|}{Passive} & Single & \multicolumn{1}{c|}{AN} & Single & SR \\ \hline

\end{tabular}%
}

\end{table*} 

\begin{table*}[!t]
\centering
\vspace{-2em} 
\caption{Summary of \gls{RIS}-Assisted \gls{PLS} adopted system models in \gls{RF} Communication Systems (continued)} %\textcolor{red}{Please check the text highlighted in red}
\vspace{-0.7em} 
\label{tab:RF-table-3}
\resizebox{0.95\textwidth}{!}{%
\begin{tabular}{|c|cccc|cc|cc|c|}
\hline
\multirow{2}{*}{\textbf{{[}\#{]}}} & \multicolumn{4}{c|}{\textbf{System related content}} & \multicolumn{2}{c|}{\textbf{RIS related content}} & \multicolumn{2}{c|}{\textbf{Security related content}} & \multirow{2}{*}{\textbf{\begin{tabular}[c]{@{}c@{}}Performance\\ Metric(s)\end{tabular}}} \\ \cline{2-9}

& \multicolumn{1}{c|}{\textbf{Technology}} & \multicolumn{1}{c|}{\textbf{System Model}} & \multicolumn{1}{c|}{\textbf{\begin{tabular}[c]{@{}c@{}}CSI\\ Condition\end{tabular}}} & \textbf{\begin{tabular}[c]{@{}c@{}}Channel\\ Model\end{tabular}} & \multicolumn{1}{c|}{\textbf{\begin{tabular}[c]{@{}c@{}}RIS\\ Type\end{tabular}}} & \textbf{RIS(s)} & \multicolumn{1}{c|}{\textbf{\begin{tabular}[c]{@{}c@{}}PLS\\ Technique\end{tabular}}} & \textbf{Eve(s)} &  \\ \hline

%STAR-RIS
%\cite{han2022artificial} & \multicolumn{1}{c|}{\multirow{7}{*}{\textcolor{black}{STAR-RIS}}} & \multicolumn{1}{c|}{} & \multicolumn{1}{c|}{} &  & \multicolumn{1}{c|}{} &  & \multicolumn{1}{c|}{} &  &  \\ \cline{1-1} \cline{3-10} 

%\cite{han2023secure} & \multicolumn{1}{c|}{} & \multicolumn{1}{c|}{} & \multicolumn{1}{c|}{} &  & \multicolumn{1}{c|}{} &  & \multicolumn{1}{c|}{} &  &  \\ \cline{1-1} \cline{3-10} 

%\cite{zhang2022secrecy} & \multicolumn{1}{c|}{} & \multicolumn{1}{c|}{} & \multicolumn{1}{c|}{} &  & \multicolumn{1}{c|}{} &  & \multicolumn{1}{c|}{} &  &  \\ \cline{1-1} \cline{3-10} 

%\cite{fang2022intelligent} & \multicolumn{1}{c|}{} & \multicolumn{1}{c|}{} & \multicolumn{1}{c|}{} &  & \multicolumn{1}{c|}{} &  & \multicolumn{1}{c|}{} &  &  \\ \cline{1-1} \cline{3-10} 

%\cite{zhang2023security} & \multicolumn{1}{c|}{} & \multicolumn{1}{c|}{} & \multicolumn{1}{c|}{} &  & \multicolumn{1}{c|}{} &  & \multicolumn{1}{c|}{} &  &  \\ \cline{1-1} \cline{3-10} 

%\cite{lv2023security} & \multicolumn{1}{c|}{} & \multicolumn{1}{c|}{} & \multicolumn{1}{c|}{} &  & \multicolumn{1}{c|}{} &  & \multicolumn{1}{c|}{} &  &  \\ \cline{1-1} \cline{3-10} 

%\cite{niu2021simultaneous} & \multicolumn{1}{c|}{} & \multicolumn{1}{c|}{} & \multicolumn{1}{c|}{} &  & \multicolumn{1}{c|}{} &  & \multicolumn{1}{c|}{} &  &  \\ \hline

%MEC
\cite{li2021intelligent} & \multicolumn{1}{c|}{\multirow{9}{*}{MEC}} & \multicolumn{1}{c|}{MU-FD-UL/DL} & \multicolumn{1}{c|}{Perfect} & Rayleigh & \multicolumn{1}{c|}{Passive} & Single & \multicolumn{1}{c|}{AN} & Multiple & \begin{tabular}[c]{@{}c@{}}Secure energy \\ consumption\end{tabular} \\ \cline{1-1} \cline{3-10} 

\cite{xu2023deep}  & \multicolumn{1}{c|}{ } & \multicolumn{1}{c|}{MU-SISO-UL} & \multicolumn{1}{c|}{Perfect} & Rayleigh & \multicolumn{1}{c|}{Passive} & Single & \multicolumn{1}{c|}{-} & Multiple & \begin{tabular}[c]{@{}c@{}} Sum secrecy\\ computation\\ efficiency \end{tabular}\\ \cline{1-1} \cline{3-10} 

\cite{zhang2023deep}  & \multicolumn{1}{c|}{ } & \multicolumn{1}{c|}{MU-MISO-UL} & \multicolumn{1}{c|}{Perfect} & \begin{tabular}[c]{@{}c@{}} Rayleigh/\\Rician \end{tabular} & \multicolumn{1}{c|}{Passive} & Single & \multicolumn{1}{c|}{BF} & single &  SR, secure EE \\ \cline{1-1} \cline{3-10} 

\textcolor{black}{\cite{zhou2023secure}}  & \multicolumn{1}{c|}{ } & \multicolumn{1}{c|}{\textcolor{black}{MU-SISO-UL}} & \multicolumn{1}{c|}{\textcolor{black}{Perfect}} & \textcolor{black}{Rician} & \multicolumn{1}{c|}{\textcolor{black}{Passive}} & \textcolor{black}{Single} & \multicolumn{1}{c|}{\textcolor{black}{Jamming}} & \textcolor{black}{Single} & \begin{tabular}[c]{@{}c@{}}\textcolor{black}{Secure}\\ \textcolor{black}{computational}\\ \textcolor{black}{tasks}\end{tabular} \\ \cline{1-1} \cline{3-10} 

\cite{liao2023intelligent} & \multicolumn{1}{c|}{ } & \multicolumn{1}{c|}{MU-FD-UL/DL} & \multicolumn{1}{c|}{Perfect} & Rayleigh & \multicolumn{1}{c|}{Passive} & Single & \multicolumn{1}{c|}{BF} & Single & \begin{tabular}[c]{@{}c@{}}Secure\\ computational\\ bits\end{tabular}\\ \hline

%Satellite Communications
\cite{ngo2023physical} & \multicolumn{1}{c|}{\multirow{10}{*}{\begin{tabular}[c]{@{}c@{}}Satellite \\ commun.\end{tabular}}} & \multicolumn{1}{c|}{SU-SISO-DL} & \multicolumn{1}{c|}{Statistical} & \begin{tabular}[c]{@{}c@{}} Rayleigh/\\Rician \end{tabular} & \multicolumn{1}{c|}{Passive} & Single & \multicolumn{1}{c|}{-} & Single & \begin{tabular}[c]{@{}c@{}} Secure\\ transmission\\ probability\end{tabular} \\ \cline{1-1} \cline{3-10}

\cite{ge2022active} & \multicolumn{1}{c|}{ } & \multicolumn{1}{c|}{SU-MISO-DL} & \multicolumn{1}{c|}{Imperfect} & \begin{tabular}[c]{@{}c@{}} Rayleigh/\\Rician \end{tabular} & \multicolumn{1}{c|}{Active} & Single & \multicolumn{1}{c|}{BF} & Single & Transmit power \\ \cline{1-1} \cline{3-10}

\cite{hoang2023secrecy} & \multicolumn{1}{c|}{ } & \multicolumn{1}{c|}{SU-MISO-DL} & \multicolumn{1}{c|}{Imperfect} & Rician & \multicolumn{1}{c|}{Passive} & Single & \multicolumn{1}{c|}{BF} & Multiple & SR\\ \cline{1-1} \cline{3-10}

\cite{zhou2022secrecy} & \multicolumn{1}{c|}{ } & \multicolumn{1}{c|}{SU-SISO-DL} & \multicolumn{1}{c|}{Perfect} & \begin{tabular}[c]{@{}c@{}} Rayleigh/\\Rician \end{tabular} & \multicolumn{1}{c|}{Passive} & Single & \multicolumn{1}{c|}{-} & Multiple & SOP\\ \cline{1-1} \cline{3-10}

\cite{zhao2022irs} & \multicolumn{1}{c|}{ } & \multicolumn{1}{c|}{MU-MISO-DL} & \multicolumn{1}{c|}{Imperfect} & Rayleigh & \multicolumn{1}{c|}{Passive} & Single & \multicolumn{1}{c|}{\begin{tabular}[c]{@{}c@{}} Unintended\\interference \end{tabular}} & Multiple & Sum rate\\ \cline{1-1} \cline{3-10}

%\textcolor{black}{\cite{dong2023optimization}} & \multicolumn{1}{c|}{ } & \multicolumn{1}{c|}{} & \multicolumn{1}{c|}{} &  & \multicolumn{1}{c|}{} &  & \multicolumn{1}{c|}{\begin{tabular}[c]{@{}c@{}} \end{tabular}} &  & \\ \cline{1-1} \cline{3-10} % Ignore No Eve

\cite{xu2021intelligent} & \multicolumn{1}{c|}{ } & \multicolumn{1}{c|}{SU-MISO-DL} & \multicolumn{1}{c|}{Perfect} & \begin{tabular}[c]{@{}c@{}} Rayleigh/\\Rician \end{tabular} & \multicolumn{1}{c|}{Passive} & Single & \multicolumn{1}{c|}{Jamming} & Single & Eve SINR\\ \hline

%Multicast Communications
%\cite{han2021reconfigurable} & \multicolumn{1}{c|}{\multirow{3}{*}{\begin{tabular}[c]{@{}c@{}}Multicast \\ Commun.\end{tabular}}} & \multicolumn{1}{c|}{} & \multicolumn{1}{c|}{} &  & \multicolumn{1}{c|}{} &  & \multicolumn{1}{c|}{} &  & \\ \cline{1-1} \cline{3-10} %Note (Omar): No PLS (ignored)

\cite{du2020reconfigurable} & \multicolumn{1}{c|}{\multirow{7}{*}{\begin{tabular}[c]{@{}c@{}} Multicast \\ commun.\end{tabular}}} & \multicolumn{1}{c|}{MU-MISO-DL} & \multicolumn{1}{c|}{Perfect} &  Rician & \multicolumn{1}{c|}{Passive} & Single & \multicolumn{1}{c|}{-} & Single & SC\\ \cline{1-1} \cline{3-10} 

\cite{lin2023secure}  & \multicolumn{1}{c|}{ } & \multicolumn{1}{c|}{MU-MISO-DL} & \multicolumn{1}{c|}{Imperfect} & Rayleigh & \multicolumn{1}{c|}{Passive} & Single  & \multicolumn{1}{c|}{BF} & Single & SR \\ \cline{1-1} \cline{3-10} 

\cite{lyu2023robust}  & \multicolumn{1}{c|}{ } & \multicolumn{1}{c|}{SU-MISO-DL} &  \multicolumn{1}{c|}{\begin{tabular}[c]{@{}c@{}} Bounded/ \\ Statistical \end{tabular}} & \begin{tabular}[c]{@{}c@{}} Rayleigh/\\Rician \end{tabular} & \multicolumn{1}{c|}{Active} & Single & \multicolumn{1}{c|}{BF} & Multiple & \begin{tabular}[c]{@{}c@{}} Power \\consumption \end{tabular}\\ \cline{1-1} \cline{3-10} 

\cite{li2022secure} & \multicolumn{1}{c|}{ } & \multicolumn{1}{c|}{MU-MISO-DL} & \multicolumn{1}{c|}{Imperfect} & Rayleigh & \multicolumn{1}{c|}{Passive} & Single & \multicolumn{1}{c|}{Jamming} & Multiple & Secure EE \\ \cline{1-1} \cline{3-10} 

\cite{wang2021distributionally} & \multicolumn{1}{c|}{ } & \multicolumn{1}{c|}{MU-MISO-DL} & \multicolumn{1}{c|}{Imperfect} & \begin{tabular}[c]{@{}c@{}} Rayleigh/\\Rician \end{tabular} & \multicolumn{1}{c|}{Passive} & Single & \multicolumn{1}{c|}{BF} & Multiple & \begin{tabular}[c]{@{}c@{}} worst \\ Bob SNR \end{tabular} \\ \hline

%Cell-free Networks
\cite{hao2022securing} & \multicolumn{1}{c|}{\multirow{3}{*}{\begin{tabular}[c]{@{}c@{}}Cell-free \\ networks\end{tabular}}} & \multicolumn{1}{c|}{MU-MISO-DL} & \multicolumn{1}{c|}{\begin{tabular}[c]{@{}c@{}} Perfect/\\Imperfect \end{tabular}} & \begin{tabular}[c]{@{}c@{}} Rayleigh/\\Rician \end{tabular} & \multicolumn{1}{c|}{Passive} & Multiple & \multicolumn{1}{c|}{BF} & Multiple  & Weighted SSR \\ \cline{1-1} \cline{3-10} 

\textcolor{black}{\cite{li2024exploiting}} & \multicolumn{1}{c|}{ } & \multicolumn{1}{c|}{\textcolor{black}{MU-MISO-DL}} & \multicolumn{1}{c|}{\textcolor{black}{Perfect}} & \textcolor{black}{Rician} & \multicolumn{1}{c|}{\textcolor{black}{Passive}} & \textcolor{black}{Multiple} & \multicolumn{1}{c|}{\textcolor{black}{-}} & \textcolor{black}{Single} & \textcolor{black}{Secure EE}\\ \cline{1-1} \cline{3-10} 

\cite{elhoushy2021exploiting} & \multicolumn{1}{c|}{ } & \multicolumn{1}{c|}{MU-MIMO-DL} & \multicolumn{1}{c|}{Perfect} & Rician & \multicolumn{1}{c|}{Passive} & Multiple & \multicolumn{1}{c|}{BF} & Single & ASR\\ \hline

%Relay Networks
\cite{gong2023joint} & \multicolumn{1}{c|}{\multirow{8}{*}{\begin{tabular}[c]{@{}c@{}}Relay \\ networks\end{tabular}}} & \multicolumn{1}{c|}{SU-SISO-DL} & \multicolumn{1}{c|}{Perfect} & Rayleigh & \multicolumn{1}{c|}{Passive} & Single & \multicolumn{1}{c|}{Jamming} & Single & SR \\ \cline{1-1} \cline{3-10} 

\cite{huang2021multi} & \multicolumn{1}{c|}{} & \multicolumn{1}{c|}{SU-SISO-DL} & \multicolumn{1}{c|}{Perfect} & \begin{tabular}[c]{@{}c@{}} Rayleigh/\\Rician \end{tabular} & \multicolumn{1}{c|}{Passive} & Single & \multicolumn{1}{c|}{BF} & Single & \begin{tabular}[c]{@{}c@{}} ASR and \\ throughput \end{tabular} \\ \cline{1-1} \cline{3-10} 

\cite{liu2023irs} & \multicolumn{1}{c|}{} & \multicolumn{1}{c|}{SU-MISO-DL} & \multicolumn{1}{c|}{Perfect} & Rayleigh & \multicolumn{1}{c|}{Passive} & Single & \multicolumn{1}{c|}{BF} & Single & ASR \\ \cline{1-1} \cline{3-10}

\textcolor{black}{\cite{zhang2024enhancing}} & \multicolumn{1}{c|}{} & \multicolumn{1}{c|}{\textcolor{black}{MU-MISO-DL/UL}} & \multicolumn{1}{c|}{\begin{tabular}[c]{@{}c@{}} \textcolor{black}{Perfect/}\\ \textcolor{black}{Imperfect} \end{tabular}} & \textcolor{black}{Rayleigh} & \multicolumn{1}{c|}{\textcolor{black}{Active}} & \textcolor{black}{Single} & \multicolumn{1}{c|}{\textcolor{black}{BF}} & \textcolor{black}{Single} & \textcolor{black}{SSR} \\ \cline{1-1} \cline{3-10}

\textcolor{black}{\cite{yuan2023security}} & \multicolumn{1}{c|}{} & \multicolumn{1}{c|}{\textcolor{black}{SU-SISO-DL}} & \multicolumn{1}{c|}{\textcolor{black}{Statistical}} & \textcolor{black}{Rayleigh} & \multicolumn{1}{c|}{\textcolor{black}{Active}} & \textcolor{black}{Single} & \multicolumn{1}{c|}{\textcolor{black}{-}} & \textcolor{black}{Single} & \textcolor{black}{SOP and IP} \\ \cline{1-1} \cline{3-10}

\cite{zhang2023intelligent} & \multicolumn{1}{c|}{} & \multicolumn{1}{c|}{SU-MISO-DL} & \multicolumn{1}{c|}{Perfect} & Rayleigh & \multicolumn{1}{c|}{Semi-active} & Single & \multicolumn{1}{c|}{-} & Single & Secrecy EE \\ \hline

%Vehicular Commun.
\cite{ai2021secure} & \multicolumn{1}{c|}{\multirow{5}{*}{\begin{tabular}[c]{@{}c@{}}Vehicular \\ commun.\end{tabular}}} & \multicolumn{1}{c|}{SU-SISO-DL} & \multicolumn{1}{c|}{Perfect} & Rayleigh & \multicolumn{1}{c|}{Passive} & Single & \multicolumn{1}{c|}{-} & Single & SOP\\ \cline{1-1} \cline{3-10} 

\cite{liu2022ris}   & \multicolumn{1}{c|}{} & \multicolumn{1}{c|}{SU-SISO-DL} & \multicolumn{1}{c|}{Imperfect} & Rayleigh  & \multicolumn{1}{c|}{Passive} & Single & \multicolumn{1}{c|}{BF} & Single & ASR\\ \cline{1-1} \cline{3-10}

\textcolor{black}{\cite{shang2024secure}}   & \multicolumn{1}{c|}{} & \multicolumn{1}{c|}{\textcolor{black}{SU-MISO-DL}} & \multicolumn{1}{c|}{\textcolor{black}{Unavailable}} & \textcolor{black}{Rayleigh}  & \multicolumn{1}{c|}{\textcolor{black}{Passive}} & \textcolor{black}{Single} & \multicolumn{1}{c|}{\textcolor{black}{AN}} & \textcolor{black}{Single} & \textcolor{black}{SR} \\ \cline{1-1} \cline{3-10} 

\textcolor{black}{\cite{li2024secure}}   & \multicolumn{1}{c|}{} & \multicolumn{1}{c|}{\textcolor{black}{MU-SISO-DL}} & \multicolumn{1}{c|}{\textcolor{black}{Imperfect}} & \textcolor{black}{Rayleigh} & \multicolumn{1}{c|}{\textcolor{black}{Passive}} & \textcolor{black}{Single} & \multicolumn{1}{c|}{\textcolor{black}{-}} & \textcolor{black}{Single} & \textcolor{black}{OP and IP} \\ \cline{1-1} \cline{3-10} 

\cite{chen2024physical}   & \multicolumn{1}{c|}{} & \multicolumn{1}{c|}{SU-SISO-DL} & \multicolumn{1}{c|}{Perfect} & Rayleigh  & \multicolumn{1}{c|}{Passive} & Single & \multicolumn{1}{c|}{BF} & Multiple & SOP and SC\\ \hline

%WBAN
\cite{xiao2022irs} & \multicolumn{1}{c|}{WBAN} & \multicolumn{1}{c|}{SU-SISO-D2D} & \multicolumn{1}{c|}{Estimated} & Rayleigh & \multicolumn{1}{c|}{\begin{tabular}[c]{@{}c@{}} Semi-\\passive\end{tabular}} & Single & \multicolumn{1}{c|}{\begin{tabular}[c]{@{}c@{}} Anti-\\jamming\end{tabular}} &  Single & \begin{tabular}[c]{@{}c@{}} Eve rate, energy,\\ latency, SC \end{tabular}\\ \hline
 
%Ad-hoc
\cite{hoang2021ris} & \multicolumn{1}{c|}{\begin{tabular}[c]{@{}c@{}} Ad-hoc\\ Networks\end{tabular}} & \multicolumn{1}{c|}{SU-MISO-DL} & \multicolumn{1}{c|}{Perfect} & Rician & \multicolumn{1}{c|}{Passive} & Single & \multicolumn{1}{c|}{BF} & Single & ASR\\ \hline

%Integrated communication and sensing  (ISAC)
\cite{salem2022active} & \multicolumn{1}{c|}{\multirow{8}{*}{ISAC}} & \multicolumn{1}{c|}{MU-MISO-DL} & \multicolumn{1}{c|}{Perfect} & Rician & \multicolumn{1}{c|}{Active} & Single & \multicolumn{1}{c|}{BF} & Single & ASR \\ \cline{1-1} \cline{3-10} 

\cite{chu2023joint}  & \multicolumn{1}{c|}{} & \multicolumn{1}{c|}{MU-MISO-DL} & \multicolumn{1}{c|}{Perfect} & \begin{tabular}[c]{@{}c@{}} Rayleigh/\\Rician \end{tabular} & \multicolumn{1}{c|}{Passive} & Single & \multicolumn{1}{c|}{BF} & Single & SNR \\ \cline{1-1} \cline{3-10} 

\cite{xing2023reconfigurable}  & \multicolumn{1}{c|}{} & \multicolumn{1}{c|}{SU-MISO-DL} & \multicolumn{1}{c|}{Perfect} & Rayleigh & \multicolumn{1}{c|}{Passive} & Single & \multicolumn{1}{c|}{BF} & Single & Ergodic SR  \\ \cline{1-1} \cline{3-10}

\cite{liu2023drl} & \multicolumn{1}{c|}{} & \multicolumn{1}{c|}{MU-MISO-DL} & \multicolumn{1}{c|}{Perfect} & Rayleigh & \multicolumn{1}{c|}{Passive} & Single & \multicolumn{1}{c|}{BF} & Single & SR \\ \cline{1-1} \cline{3-10}

%\textcolor{black}{\cite{jia2023physical}} & \multicolumn{1}{c|}{} & \multicolumn{1}{c|}{} & \multicolumn{1}{c|}{} &  & \multicolumn{1}{c|}{} &  & \multicolumn{1}{c|}{} &  &  \\ \cline{1-1} \cline{3-10} % ingore NO RIS

\textcolor{black}{\cite{sun2024secure}} & \multicolumn{1}{c|}{} & \multicolumn{1}{c|}{\textcolor{black}{SU-MISO-DL}} & \multicolumn{1}{c|}{\textcolor{black}{Perfect}} & \textcolor{black}{Rayleigh} & \multicolumn{1}{c|}{\textcolor{black}{Active}} & \textcolor{black}{Single} & \multicolumn{1}{c|}{\textcolor{black}{BF}} & \textcolor{black}{Single} & \textcolor{black}{SR} \\ \cline{1-1} \cline{3-10}

\textcolor{black}{\cite{wei2024star}} & \multicolumn{1}{c|}{} & \multicolumn{1}{c|}{\textcolor{black}{MU-MISO-DL}} & \multicolumn{1}{c|}{\textcolor{black}{Perfect}} & \textcolor{black}{Rician} & \multicolumn{1}{c|}{\textcolor{black}{Passive}} & \textcolor{black}{Single} & \multicolumn{1}{c|}{\textcolor{black}{Jamming}} & \textcolor{black}{Single} & \textcolor{black}{SR} \\ \cline{1-1} \cline{3-10}

\cite{jiang2023secure} & \multicolumn{1}{c|}{} & \multicolumn{1}{c|}{MU-MISO-DL} & \multicolumn{1}{c|}{Perfect} & Rayleigh & \multicolumn{1}{c|}{Passive} & Single & \multicolumn{1}{c|}{BF} & Single & SSR \\ \hline

%Radar 
\cite{wang2023star}  & \multicolumn{1}{c|}{\multirow{4}{*}{\begin{tabular}[c]{@{}c@{}} Radar \\ commun. \end{tabular}}} & \multicolumn{1}{c|}{MU-MISO-DL} & \multicolumn{1}{c|}{Perfect} & \begin{tabular}[c]{@{}c@{}} Rayleigh/\\Rician \end{tabular} & \multicolumn{1}{c|}{Passive} & Single &\multicolumn{1}{c|}{BF} & Multiple & \begin{tabular}[c]{@{}c@{}} Radar sensing\\ power \end{tabular} \\ \cline{1-1} \cline{3-10}

\textcolor{black}{\cite{liu2024intelligent}} & \multicolumn{1}{c|}{} & \multicolumn{1}{c|}{\textcolor{black}{MU-MISO-DL}} & \multicolumn{1}{c|}{\textcolor{black}{Statistical}} & \begin{tabular}[c]{@{}c@{}} \textcolor{black}{Rayleigh/}\\\textcolor{black}{Rician} \end{tabular} & \multicolumn{1}{c|}{\textcolor{black}{Passive}} & \textcolor{black}{Single} & \multicolumn{1}{c|}{\textcolor{black}{BF}} & \textcolor{black}{Single} & \textcolor{black}{Radar SINR} \\ \cline{1-1} \cline{3-10}

\cite{zhang2023irs}   & \multicolumn{1}{c|}{} & \multicolumn{1}{c|}{MU-MIMO-DL} & \multicolumn{1}{c|}{Perfect} & \begin{tabular}[c]{@{}c@{}} Rayleigh/\\Rician \end{tabular} & \multicolumn{1}{c|}{Passive} & Single & \multicolumn{1}{c|}{AN} & Single & SR \\ \hline

\end{tabular}%
}

\end{table*}

\begin{table*}[!t]
\centering
\vspace{-2em} 
\caption{Summary of \gls{RIS}-Assisted \gls{PLS} adopted system models in \gls{RF} Communication Systems (continued)} %\textcolor{red}{Please check the text highlighted in red}
\vspace{-0.7em} 
\label{tab:RF-table-4}
\resizebox{0.95\textwidth}{!}{%
\begin{tabular}{|c|cccc|cc|cc|c|}
\hline
\multirow{2}{*}{\textbf{{[}\#{]}}} & \multicolumn{4}{c|}{\textbf{System related content}} & \multicolumn{2}{c|}{\textbf{RIS related content}} & \multicolumn{2}{c|}{\textbf{Security related content}} & \multirow{2}{*}{\textbf{\begin{tabular}[c]{@{}c@{}}Performance\\ Metric(s)\end{tabular}}} \\ \cline{2-9}

& \multicolumn{1}{c|}{\textbf{Technology}} & \multicolumn{1}{c|}{\textbf{System Model}} & \multicolumn{1}{c|}{\textbf{\begin{tabular}[c]{@{}c@{}}CSI\\ Condition\end{tabular}}} & \textbf{\begin{tabular}[c]{@{}c@{}}Channel\\ Model\end{tabular}} & \multicolumn{1}{c|}{\textbf{\begin{tabular}[c]{@{}c@{}}RIS\\ Type\end{tabular}}} & \textbf{RIS(s)} & \multicolumn{1}{c|}{\textbf{\begin{tabular}[c]{@{}c@{}}PLS\\ Technique\end{tabular}}} & \textbf{Eve(s)} &  \\ \hline

%IoT
% Ignore it in the table [not a well-known conference paper]
%\cite{long2021physical} & \multicolumn{1}{c|}{\multirow{5}{*}{\begin{tabular}[c]{@{}c@{}} IoT \\ Networks \end{tabular}}} & \multicolumn{1}{c|}{MU-SISO-DL} & \multicolumn{1}{c|}{Perfect} & Nakagami-$m$ & \multicolumn{1}{c|}{Passive} & Single & \multicolumn{1}{c|}{\colorbox{red}{xxx}} & Single & SOP \\ \cline{1-1} \cline{3-10} 

\cite{mao2022reconfigurable} & \multicolumn{1}{c|}{\multirow{6.5}{*}{\begin{tabular}[c]{@{}c@{}} IoT \\ Networks \end{tabular}}} & \multicolumn{1}{c|}{MU-SISO-UL} & \multicolumn{1}{c|}{Perfect} & \begin{tabular}[c]{@{}c@{}} Rayleigh/\\Rician \end{tabular} & \multicolumn{1}{c|}{Passive} & Single & \multicolumn{1}{c|}{BF} & Multiple & Computation EE \\ \cline{1-1} \cline{3-10} 

\cite{benaya2023physical} & \multicolumn{1}{c|}{} & \multicolumn{1}{c|}{SU-SISO-DL} & \multicolumn{1}{c|}{Perfect} & Rayleigh & \multicolumn{1}{c|}{Passive} & Single & \multicolumn{1}{c|}{Jamming} & Single & ASR \\ \cline{1-1} \cline{3-10}

\cite{saleem2022deep}  & \multicolumn{1}{c|}{ } & \multicolumn{1}{c|}{MU-MISO-DL} & \multicolumn{1}{c|}{Perfect} & Rayleigh & \multicolumn{1}{c|}{Passive} & Single & \multicolumn{1}{c|}{BF} & Multiple & SSR \\ \cline{1-1} \cline{3-10}

\cite{khoshafa2023securing} & \multicolumn{1}{c|}{ } & \multicolumn{1}{c|}{SU-SISO-DL} & \multicolumn{1}{c|}{Perfect} & Nakagami-$m$ & \multicolumn{1}{c|}{Passive} & Single  & \multicolumn{1}{c|}{-} & Single & \begin{tabular}[c]{@{}c@{}} SOP, PNSC,\\ and ASR \end{tabular} \\ \cline{1-1} \cline{3-10}

%\textcolor{black}{\cite{lin2023physical}} & \multicolumn{1}{c|}{ } & \multicolumn{1}{c|}{\textcolor{black}{MU-SISO-DL}} & \multicolumn{1}{c|}{\textcolor{black}{Perfect}} & \textcolor{black}{Rayleigh} & \multicolumn{1}{c|}{} &   & \multicolumn{1}{c|}{} &  & \begin{tabular}[c]{@{}c@{}}  \end{tabular} \\ \cline{1-1} \cline{3-10} %Ignore NO RIS

%\textcolor{black}{\cite{li2024privacy}} & \multicolumn{1}{c|}{ } & \multicolumn{1}{c|}{} & \multicolumn{1}{c|}{} &  & \multicolumn{1}{c|}{} &   & \multicolumn{1}{c|}{} &  & \begin{tabular}[c]{@{}c@{}}  \end{tabular} \\ \cline{1-1} \cline{3-10} % Ignore NO eve

\cite{niu2022joint} & \multicolumn{1}{c|}{ } & \multicolumn{1}{c|}{MU-MISO-DL} & \multicolumn{1}{c|}{Perfect} & Rician & \multicolumn{1}{c|}{\begin{tabular}[c]{@{}c@{}} Passive/\\ Active\end{tabular}} & Multiple  & \multicolumn{1}{c|}{BF} & Multiple & \begin{tabular}[c]{@{}c@{}} SSR and\\ Secrecy EE\end{tabular} \\ \hline

% Backscatter communications
\cite{han2023broadcast}  & \multicolumn{1}{c|}{\multirow{11}{*}{\begin{tabular}[c]{@{}c@{}} Backscatter \\ Commun. \end{tabular}}} & \multicolumn{1}{c|}{MU-MISO-DL} & \multicolumn{1}{c|}{Perfect} & Rician & \multicolumn{1}{c|}{Active} & Single & \multicolumn{1}{c|}{BF} & Multiple & ASR \\ \cline{1-1} \cline{3-10} 

\cite{wang2023intelligent} & \multicolumn{1}{c|}{} & \multicolumn{1}{c|}{SU-MISO-DL} & \multicolumn{1}{c|}{Unavailable} & Nakagami-$m$ & \multicolumn{1}{c|}{Passive} & Single & \multicolumn{1}{c|}{AN} & Single & SR \\ \cline{1-1} \cline{3-10} 

\cite{wang2022multicast}  & \multicolumn{1}{c|}{ } & \multicolumn{1}{c|}{MU-MISO-DL} & \multicolumn{1}{c|}{Perfect} & Rayleigh & \multicolumn{1}{c|}{Active} & Single & \multicolumn{1}{c|}{BF} & Multiple & SR \\ \cline{1-1} \cline{3-10}  

\cite{wu2022irs}  & \multicolumn{1}{c|}{} & \multicolumn{1}{c|}{SU-MISO-DL} & \multicolumn{1}{c|}{Perfect} & Rician & \multicolumn{1}{c|}{Passive} & Single & \multicolumn{1}{c|}{BF} & Multiple & \begin{tabular}[c]{@{}c@{}} Transmit \\ power \end{tabular} \\ \cline{1-1} \cline{3-10}

\textcolor{black}{\cite{xia2024joint}} & \multicolumn{1}{c|}{ } & \multicolumn{1}{c|}{\textcolor{black}{MU-SIMO-UL}} & \multicolumn{1}{c|}{\textcolor{black}{Estimated}} & \textcolor{black}{Rayleigh} & \multicolumn{1}{c|}{\textcolor{black}{Passive}} & \textcolor{black}{Multiple} & \multicolumn{1}{c|}{\textcolor{black}{BF}} & \textcolor{black}{Single} & \textcolor{black}{SSR} \\ \cline{1-1} \cline{3-10} 

\textcolor{black}{\cite{pei2024secrecy}}  & \multicolumn{1}{c|}{} & \multicolumn{1}{c|}{\textcolor{black}{SU-SISO-DL}} & \multicolumn{1}{c|}{\begin{tabular}[c]{@{}c@{}} \textcolor{black}{Perfect/} \\ \textcolor{black}{Imperfect} \end{tabular}} & \textcolor{black}{Rayleigh} & \multicolumn{1}{c|}{\textcolor{black}{Passive}} & \textcolor{black}{Single} & \multicolumn{1}{c|}{\textcolor{black}{-}} & \textcolor{black}{Single} &  \textcolor{black}{SOP} \\ \cline{1-1} \cline{3-10}

\cite{cao2022proactive}  & \multicolumn{1}{c|}{} & \multicolumn{1}{c|}{SU-SIMO-UL} & \multicolumn{1}{c|}{Perfect} & Rician & \multicolumn{1}{c|}{\textcolor{black}{Passive}} & Single & \multicolumn{1}{c|}{BF} & Single & \begin{tabular}[c]{@{}c@{}} Destination \\ SINR \end{tabular} \\ \hline

%Note: I went through the paper, and the quality is not good.
% SDN 
%\cite{chiti2023secure} \textcolor{red}{New1} & \multicolumn{1}{c|}{SDN} & \multicolumn{1}{c|}{SU-SIMO-UL} & \multicolumn{1}{c|}{Perfect} & Rayleigh & \multicolumn{1}{c|}{Passive} & Multiple & \multicolumn{1}{c|}{BF} & Multiple & SSR \\ \hline

% WSNs
\cite{ahmed2022joint} & \multicolumn{1}{c|}{\multirow{2}{*}{WSNs}} & \multicolumn{1}{c|}{MU-SISO-UL} & \multicolumn{1}{c|}{Perfect} & Rayleigh & \multicolumn{1}{c|}{Passive} & Single & \multicolumn{1}{c|}{BF} & Single & SNR \\ \cline{1-1} \cline{3-10}

\textcolor{black}{\cite{zhou2024securing}} & \multicolumn{1}{c|}{} & \multicolumn{1}{c|}{\textcolor{black}{MU-SIMO-UL}} & \multicolumn{1}{c|}{\textcolor{black}{Perfect}} & \textcolor{black}{Rayleigh} & \multicolumn{1}{c|}{\textcolor{black}{Active}} & \textcolor{black}{Single} & \multicolumn{1}{c|}{\textcolor{black}{-}} & \textcolor{black}{Single} & \textcolor{black}{SR} \\ \hline

% HAPs
\cite{sun2022energy} & \multicolumn{1}{c|}{\multirow{4}{*}{HAPs}} & \multicolumn{1}{c|}{MU-MIMO-DL} & \multicolumn{1}{c|}{Perfect} & Rayleigh & \multicolumn{1}{c|}{Passive} & Single & \multicolumn{1}{c|}{BF} & Multiple & EE \\ \cline{1-1} \cline{3-10} 

\cite{wang2023uplink} & \multicolumn{1}{c|}{} & \multicolumn{1}{c|}{MU-SISO-UL} & \multicolumn{1}{c|}{Perfect} & Nakagami-$m$ & \multicolumn{1}{c|}{Passive} & Single & \multicolumn{1}{c|}{-} & Single & SOP and PPSC \\ \cline{1-1} \cline{3-10} 

\cite{yuan2022secure} & \multicolumn{1}{c|}{} & \multicolumn{1}{c|}{SU-SISO-DL} & \multicolumn{1}{c|}{\begin{tabular}[c]{@{}c@{}}Perfect/ \\ Statistical\end{tabular}} & \begin{tabular}[c]{@{}c@{}} Gamma-\\Gamma \end{tabular} & \multicolumn{1}{c|}{Passive} & Single & \multicolumn{1}{c|}{-} & Single &  Ergodic SR\\ \hline

\end{tabular}%
}

\end{table*}

\end{subtables}

\subsection{Integrating RIS with Enabling Technologies for Secure Communications}
Current research contributions have demonstrated that \gls{RIS} plays a pivotal role in increasing the levels of security and confidentiality within wireless networks. This notable secrecy performance enhancement holds considerable potential for application across diverse wireless communication networks. This subsection dives into the integration of \gls{RIS} with emerging and cutting-edge technologies, which encompass \gls{mmWave}, \gls{THz}, \gls{UAV}s, \gls{NOMA}, \gls{CRN}, \gls{D2D} communications, and other pertinent technologies. In this subsection, we explore the efficacy of integrating the \gls{RIS} with emerging technologies to heighten wireless networks' security and confidentiality, offering tangible benefits for various wireless communication networks. In the following, we delve into integrating \gls{RIS} with current technologies, clarifying these integrations' contributions to wireless communication security. The integration of RIS in PLS-aided systems with RF-enabling technologies for secure communications is summarized in Fig.~\ref{Fig:0_RF_Chart}.

%\subsubsection{RIS-Assisted Secure multi-antenna communications}

\subsubsection{RIS-Assisted Secure mmWave Communications}
\textcolor{black}{Integrating \gls{RIS} into \gls{mmWave} communications represents a significant advancement in enhancing security. \Gls{RIS}'s adaptability enables dynamic adjustments to the propagation environment, establishing secure communication channels in \gls{mmWave} contexts. Its role in optimizing \textcolor{black}{\gls{BF}} and channel characteristics addresses the directional communication requirements of \gls{mmWave}, minimizing the risk of unauthorized interception.} %The \gls{RIS} further contributes to security by employing dynamic signal jamming and controlled reflections to deter potential eavesdropping. Additionally, its collaboration with \gls{PLS} techniques ensures secure communication between legitimate users. Incorporating \gls{RIS} and \gls{mmWave} technologies addresses security concerns and holds promise for developing efficient and adaptive communication networks in future wireless systems.
The authors of~\cite{xiu2021secrecy} investigated the secrecy rate of a \gls{mmWave} system improved by an \gls{RIS} with low-resolution digital-to-analog converters, focusing on mitigating hardware losses and improving secrecy rates by integrating with \gls{RIS}. The obtained results demonstrated the effectiveness of \gls{RIS} in reducing hardware losses. Secure \textcolor{black}{\gls{BF}} in the \gls{mmWave} \gls{MISO} system, aided by an \gls{RIS}, was explored in~\cite{lu2020robust}. The presence of multiple single-antenna eavesdroppers near the legitimate receiver was addressed, taking into account imperfect knowledge of cascaded wiretap \gls{CSI} at the transmitter and the colluding and non-colluding eavesdropping scenarios. Simulation results showed the superior performance of the proposed scheme in terms of \gls{ASR}.

\subsubsection{RIS-Assisted Secure THz Communications}
\textcolor{black}{In the \gls{THz} communications domain, incorporating \gls{RIS} represents a noteworthy advancement, elevating security within this frequency band. Leveraging the adaptability of \gls{RIS}, the propagation environment undergoes dynamic modulation to optimize \textcolor{black}{\gls{BF}} and channel characteristics, particularly salient in the distinctive context of \gls{THz} communications.} The authors in~\cite{xu2023sum} investigated a secure \gls{THz}-empowered network, which involved transmitting confidential information from a low earth orbit satellite to a \gls{UAV} via \gls{RIS}-mounted \gls{HAPS} in the presence of an untrustworthy \gls{UAV}. The \gls{RIS} phase shifts were optimized to enhance the secrecy performance. In~\cite{yuan2022secure}, a \gls{THz} \gls{MIMO}-\gls{NOMA} system with the assistance of a \gls{RIS} was investigated, considering the presence of an eavesdropper. Results demonstrated that deploying \gls{RIS} led to substantial \textcolor{black}{\gls{BF}} gains, effectively mitigating eavesdropping threats.

\subsubsection{RIS-Assisted Secure UAV Communications}
\textcolor{black}{Integrating \gls{RIS} with \gls{UAV}s offers significant potential for increasing security measures within wireless communication networks. This convergence between \gls{RIS} and \gls{UAV} technologies represents a pivotal advancement, as it harnesses the unique capabilities of both systems to address security challenges across various operational scenarios.}
%With its adaptive \textcolor{black}{\gls{EM}} surface ability, \gls{RIS} can effectively manipulate \textcolor{black}{\gls{EM}} wave propagation, facilitating secure communication by altering signal characteristics. 
%The integrated \gls{RIS}-\gls{UAV} system becomes proficient in delivering secure communication links and surveillance capabilities in dynamic and demanding settings when combined with \gls{UAV}s, which provide mobility and adaptability in deployment.} %This integration introduces a multifaceted approach to security, encompassing surveillance, communication, and the ability to establish secure connections in scenarios where traditional infrastructure may be deficient or compromised. In this context, we explore the complexities, potential application scenarios, and security implications of \gls{RIS}-\gls{UAV} integration, shedding light on its significance in contemporary models of wireless communication security.
Utilizing a swarm of \glspl{UAV} with \gls{ARIS} presents an effective avenue 
for increasing ground users' \gls{PLS} through control of phase shifts, spatial placement, and strategic movement. Additionally, a traceable \gls{PLS} mechanism conducive to secure \gls{V2X} communications was introduced in~\cite{shang2021uav}. Utilizing a \gls{RIS} in the presence of \gls{UAV}s acting as potential eavesdroppers was analyzed in~\cite{wei2022secrecy}, where enhanced security was achieved through \gls{RIS} deployment. The authors of~\cite{tang2023robust} proposed a strategy for enhancing wireless security by employing aerial reflection and jamming techniques to address channel uncertainties effectively. ARIS mounted on a UAV was introduced to improve the wireless secrecy of fixed-deployed \glspl{RIS} and optimize the ARIS trajectory to maximize average secrecy rates~\cite{tang2023secure}. The authors of~\cite{cheng2023irs} investigated a secure multi-user UAV communication system supported by a \gls{RIS} and considered hardware impairments in the transceiver and the \gls{RIS}. In~\cite{diao2023reflecting}, the minimum number of reflecting elements required for secure and energy-efficient \gls{RIS}-assisted \gls{UAV} systems was investigated, considering the phase estimation errors. Robust optimization methods were utilized, and aerial deployment was accomplished through \gls{DRL}. Additionally, the superiority of flexible deployment of \gls{ARIS} over fixed \gls{RIS} was highlighted, and significant security improvements were achieved through the collaboration between fixed \gls{RIS} and \gls{ARIS}.

\subsubsection{RIS-Assisted Secure D2D Communications}
\textcolor{black}{\gls{RIS} presents a compelling avenue for strengthening the security infrastructure of \gls{D2D} communications. The capacity of \gls{RIS} to dynamically modulate surface properties affords the selective manipulation of wireless signals, establishing secure communication channels by redirecting signals away from potential eavesdroppers or unauthorized users.} Through careful manipulation of the propagation environment, \gls{RIS} engenders the dynamic reconfiguration of wireless communication channels, confusing interception and interference by unauthorized devices, thereby adding an additional level of security. Furthermore, the precision of \gls{RIS} in producing highly directional communication links heightens privacy and security by constricting signal exposure to unintended recipients, thereby attenuating the susceptibility to signal interception. These diverse capabilities, encompassing the mitigation of jamming, interference, and dynamic signal \textcolor{black}{\gls{BF}}, collectively form a robust security framework for \gls{D2D} communications. \textcolor{black}{The authors of~\cite{khoshafa2020reconfigurable} explored the RIS technology's application to enhance the reliability and robustness of D2D communication and improve the security level of cellular networks. New analytical expressions were derived for the cellular SOP, PNSC, and the D2D outage probability. In~\cite{khalid2021ris}, a RIS was introduced to enhance system robustness and security against eavesdropping threats, employing an FD jamming receiver. In~\cite{hu2023securing}, the RIS-aided secure uplink communication scheme was investigated by maximizing the ASR for the cellular network while improving the achievable rate of D2D communication.}

\subsubsection{RIS-Assisted Secure Cognitive Radio Networks}
\textcolor{black}{Regarding the \gls{CRN}, the utilization of \gls{RIS} aims to enhance the \gls{PLS} of the network. Leveraging the numerous reflecting elements in the \gls{RIS} facilitates establishing a highly directional and manipulable wireless transmission environment. 
%This capability enables selective enhancement or nullation of the signal strength of legitimate or malicious signals, thereby improving the network's overall security. 
Furthermore, integrating \gls{RIS} allows for generating multiple virtual channels, which can effectively separate legitimate and malicious signals, providing an additional layer of protection against eavesdropping and jamming attacks.} The authors in~\cite{khoshafa2023ris} proposed a novel system model 
that utilized the \gls{RIS} technology to enhance simultaneous wireless communication and security in \gls{CRN} environments. 
It aimed to improve the transmission of the secondary network 
and enhance the secrecy performance of the primary network concurrently. In~\cite{dong2021secure}, utilizing an \gls{RIS} 
in \gls{CRN} was explored, focusing on improving secrecy rates 
under various scenarios, including perfect/imperfect Eve's \gls{CSI}. Additionally, an \gls{AN}-aided approach was proposed for scenarios without Eve's \gls{CSI} to enhance the secrecy rate. The authors in~\cite{niu2022active}  focused on employing an active \gls{RIS} 
in a secure cognitive satellite-terrestrial network to optimize the secrecy rate. This involved concurrently optimizing the design of the \gls{BF}, \gls{AN}, and reflection coefficients. The provided results revealed the superior performance of the active \gls{RIS} scheme in enhancing secrecy within the network.
%%%%%%%%%%%%%%%%%%%%%%%%%%%%%%%%%%%%%%%%%%%%%%%%%%%%%%%%%%%%%%%%
\subsubsection{RIS-Assisted Secure WPT and SWIPT}
\textcolor{black}{\gls{RIS}'s utility extends further to secure \gls{WPT} and \gls{SWIPT} communications, optimizing energy transfer efficiency while ensuring secure communication channels. By manipulating \textcolor{black}{\gls{EM}} waves, RIS improves signal strength, reduces interference, and establishes secure communication links. This integration ensures safe and efficient power and data transmission, protecting against eavesdropping and unauthorized access in modern wireless technologies.} Integrating \gls{RIS} into a secure WPCN multicast setup was proposed in~\cite{zhai2022improving} to enhance energy transfer efficiency and ensure secure communication. Specifically,  the energy was initially harvested from a power station and subsequently utilized to transmit data to multiple \gls{IoT} devices in the presence of multiple eavesdroppers. The \textcolor{black}{\gls{BF}} optimization challenge within a \gls{RIS}-enhanced \gls{SWIPT} framework was investigated in~\cite{zhao2022secrecy}, with energy users accounted for as potential eavesdroppers. The purpose was to maximize the average worst-case \gls{SR} while adhering to power and energy harvesting constraints. The authors of~\cite{cao2023physical, cao2022ergodic} explored the \gls{PLS} of a \gls{WPC} system with enhancements provided by an \gls{RIS} in the presence of a passive eavesdropper. Three secure modes for \gls{RIS}-\gls{WPC} systems were introduced and examined. In~\cite{thien2022secure}, \gls{RIS} was utilized to maximize secure \gls{SWIPT} systems with a power splitting scheme. The objective was to optimize the system's \gls{SR} by choosing the optimal transmitter power and \gls{RIS} phase shifts while ensuring user energy-harvesting requirements and adhering to transmit power constraints at the transmitter.
%%%%%%%%%%%%%%%%%%%%%%%%%%%%%%%%%%%%%%%%%%%%%%%%%%%%%%%
\subsubsection{RIS-Assisted Secure MEC}
\textcolor{black}{The integration of \gls{RIS} into \gls{MEC} enhances the security of edge networks. Through the dynamic configuration of the wireless environment, \gls{RIS} contributes to securing data transmission and computation processes in \gls{MEC} systems. This integration enhances data processing and storage security at the network edge, which is crucial for protecting sensitive information and preventing unauthorized access in MEC environments.} The authors of~\cite{mao2022reconfigurable}  presented an approach for ensuring secure task offloading and efficient wireless resource management in \gls{MEC} networks equipped with \gls{RIS}, considering \gls{IoT} devices' secure computation rate constraints. The secure computation performance in an \gls{RIS}-assisted \gls{WPT} and \gls{MEC} system with a passive eavesdropper was investigated in~\cite{liao2023intelligent}. A harvest-then-offload protocol was focused on, where users were charged by the \gls{AP} in the first slot, and harvested energy was used to offload computation tasks in the second slot, assuming concurrent local computation during energy harvesting. Strategies to enhance the security of \gls{MEC} systems by leveraging \gls{RIS} technology and \gls{AN} in the \gls{IoT} were explored in~\cite{li2021intelligent}, aiming to strengthen users' signals while simultaneously attenuating eavesdroppers' signals by manipulating \gls{RIS} phase configurations.

%%%%%%%%%%%%%%%%%%%%%%%%%%%%%%%%%%%%%%%%%%%%%%%%%%%%%%%
\subsubsection{RIS-Assisted Secure Satellite Networks}
\textcolor{black}{The sophisticated integration of \gls{RIS} technology into satellite communication systems provides insights into applications for enhancing communication security in space-based systems. 
%This integration leverages \gls{RIS}'s capability to manipulate \textcolor{black}{\gls{EM}} waves, enhancing signal strength, minimizing interference, and establishing secure communication channels. 
In doing so, security protocols are strengthened, ensuring secure and efficient data transmission while mitigating risks such as unauthorized access and eavesdropping. This strategy is essential for protecting sensitive data and upholding the reliability of satellite communication systems.} The authors of~\cite{ngo2023physical} introduced a two-hop content delivery strategy within a cache-enabled satellite-terrestrial network supported by an \gls{RIS}. The system incorporated probabilistic caching policies at both the satellite and ground station. The analysis evaluated the system's connection and secrecy probability, utilizing asymptotic and closed-form expressions. In~\cite{ge2022active}, a secure \textcolor{black}{\gls{BF}} design for cognitive satellite-terrestrial networks with active \gls{RIS} was introduced, considering \gls{CSI}. The objective was to minimize transmission power at the \gls{BS} while ensuring an acceptable \gls{SR} for primary users and an acceptable rate for secondary users. Two configurations of \gls{RIS}-aided space-ground networks, double-\gls{RIS} and single-\gls{RIS} schemes, were explored in~\cite{hoang2023secrecy} to maximize the \gls{SR}. The superiority of the double-\gls{RIS} scheme in terms of security performance compared to the single-\gls{RIS} method was demonstrated.

%%%%%%%%%%%%%%%%%%%%%%%%%%%%%%%%%%%%%%%%%%%%%%%%%%%%%%%
\subsubsection{RIS-Assisted Secure ISAC}

\textcolor{black}{Integrating an \gls{RIS} with \glspl{ISAC} systems enhances secrecy performance by dynamically controlling signal propagation. The \gls{RIS} enables precise communication through dynamic \textcolor{black}{\gls{BF}}, directing signals to intended recipients and reducing interception threats. By incorporating RIS into ISAC systems, security measures can be significantly enhanced, ensuring safe and efficient data transmission while preventing unauthorized access or eavesdropping. } The combined optimization of transmit and reflection \textcolor{black}{\gls{BF}} for secure \gls{RIS}-\gls{ISAC} systems was analyzed in~\cite{chu2023joint}. The results validated that the proposed design enhances radar \gls{SNR} performance, confirming the benefits of incorporating \gls{RIS} in secure \gls{ISAC} systems. The authors of~\cite{xing2023reconfigurable} introduced a \textcolor{black}{\gls{BF}} design assisted by \gls{RIS} in an \gls{ISAC} system to enhance \gls{PLS}. The objective was to maximize secrecy performance while meeting the minimum requirements for communication and sensing performance. \gls{RIS}-assisted secure \gls{ISAC} system was examined in~\cite{liu2023drl} wherein the objective was to maximize the \gls{SR} by concurrently optimizing the transmit \textcolor{black}{\gls{BF}}, \gls{AN} matrix, and \gls{RIS} phase-shift matrix. The results confirmed the significance of the proposed \gls{DRL} algorithm compared to benchmark techniques. \textcolor{black}{The secure transmission in an active RIS-assisted ISAC in THz was discussed in~\cite{sun2024secure}, where the target was considered a possible eavesdropper. The objective was to maximize the secrecy rate while ensuring the lowest illumination power for the target. In~\cite{wei2024star}, a secure transmission within a STAR-RIS-assisted ISAC system was investigated, where the system was partitioned into distinct sensing and communication regions. The 
STAR-RIS was strategically positioned 
to establish LoS links for confidential 
signal transmission to users and for 
target sensing and confidential signal transmission to users using NOMA. The 
PLS of RIS-assisted ISAC systems was investigated in~\cite{jiang2023secure}, where AN was utilized. The main objective was maximizing the multi-user sum secrecy rate by optimizing joint active and passive BF.}
%%%%%%%%%%%%%%%%%%%%%%%%%%%%%%%%%%%%%%%%
\subsubsection{RIS-Assisted Secure IoT}
\textcolor{black}{Incorporating \gls{RIS} into \gls{IoT} networks optimizes security through dynamic signal reflection control, interference mitigation, secure zone establishment, adaptability to changes, jamming prevention, and enhanced privacy measures~\cite{zappone2022surface}. This collaborative integration strengthens the overall security resilience of \gls{IoT} networks, providing robust protection against a spectrum of potential threats. An innovative approach was presented in~\cite{khoshafa2023securing} to enhance confidentiality in \gls{LPWAN}s by combining \gls{RIS} technology with UAV,  improving secure data transmission between \gls{IoT} sensors and gateways within LPWAN environments. Analytical expressions were derived for the \gls{SOP}, \gls{PNSC}, and \gls{ASR}, considering Nakagami-$m$ fading channels. The impact of eavesdropper locations was also explored. An IRS-based model was introduced in~\cite{saleem2022deep} to tackle security challenges in \gls{IoT} environments with diverse trusted and untrusted devices, where untrusted devices could pose eavesdropping risks. The secrecy rate of trusted devices was optimized for the model while ensuring \gls{QoS} for all legitimate devices. The low-latency computing demands of IoT devices with limited resources have been acknowledged and addressed by MEC. However, security risks could be created for these devices due to the inherent broadcast nature of wireless \textcolor{black}{\gls{EM}} communications. In response, a secure MEC network framework was proposed in~\cite{mao2022reconfigurable}, leveraging RIS to enhance security in task offloading.}
%~\cite{mao2022reconfigurable, benaya2023physical, saleem2022deep, zappone2022surface,khoshafa2023securing}. 
%%%%%%%%%%%%%%%%%%%%%%
\subsubsection{RIS-Assisted Secure Other Cutting-Edge Technologies}
The exploration extends to \gls{RIS}-assisted secure multicast communications, strategically deploying \gls{RIS} to reinforce security in the face of the unique challenges posed by one-to-many communication paradigms~\cite{du2020reconfigurable, li2022secure, lin2023secure, lyu2023robust}. \Gls{RIS}-assisted secure cell-free networks utilize \gls{RIS} to optimize signal coverage and mitigate interference 
in distributed antenna systems~\cite{hao2022securing, elhoushy2021exploiting}. Transitioning to relay networks, \gls{RIS} dynamically configures relay paths 
and signal reflections, enhancing the security and reliability of relayed communications~\cite{gong2023joint, huang2021multi}. In vehicular networks, 
a critical domain for secure communication, the strategic deployment of \gls{RIS} to optimize communication links in vehicular environments, addressing security concerns and improving overall network performance was proposed in~\cite{ai2021secure}. \glspl{WBAN} find \gls{RIS} enhancing communication security and reliability in healthcare and wearable technologies, especially 
in wearable health monitoring devices~\cite{xiao2022irs}.

Integrating \gls{RIS} with a radar system enhances security by dynamically controlling \textcolor{black}{\gls{EM}} wave propagation. This integration improves signal directionality, supports stealth and camouflage, facilitates adaptive radar sensing, and mitigates interference~\cite{wang2023star, zhang2023irs}. Moreover, backscatter communications can be integrated with \gls{RIS} to improve security performance by dynamically optimizing signal reflection, establishing secure signal directionality, adapting security configurations, and creating secure zones~\cite{han2023broadcast, wang2023intelligent, wang2022multicast}. Furthermore, the benefits of using RIS to improve the secrecy performance of \gls{WSNs} were investigated in~\cite{ahmed2022joint}. As summarized in  Tables~\ref{tab:RF-table-1} to~\ref{tab:RF-table-4}, the attractive advantages of RIS-assisted PLS have generated enormous studies investigating its applicability to 5G/6G cutting-edge wireless technologies such as multi-antenna communications, mmWave communications, THz communications, \gls{UAV} communications, \gls{D2D} communications, \glspl{CRN}, \gls{WPC}/\gls{SWIPT}, \gls{MEC}, satellite-enabled networks, multicast communications, cell-free networks, relay-aided networks, vehicular communications, \gls{WBAN}, ad-hoc networks, \gls{ISAC}, radar communications, IoT networks, backscatter communications, wireless sensor networks, and \gls{HAPS}.
{\color{black}{
\subsection{Lessons Learnt}
This section explores how RIS can enhance PLS in wireless communication systems. Here are the key takeaways:
\begin{itemize}
    \item \glspl{RIS} can be combined with various techniques to enhance \gls{PLS}. It can control the direction of \glspl{AN} to confuse eavesdroppers, manipulate signal reflection paths for \textcolor{black}{\gls{BF}}, and optimize jamming strategies in cooperative jamming approaches. Understanding how these techniques address attacks like eavesdropping and jamming provides valuable insights. Moreover, understanding how cooperative strategies, possibly involving multiple RIS elements, impact the overall security of the communication system can provide valuable insights into the potential for collaborative security tools against potential threats.
    \item RISs empower a versatile PLS approach applicable to various communication systems. By manipulating high-frequency propagation such as mmWave and THz, RIS mitigates eavesdropping across diverse scenarios. Integrating RIS with UAV communication enables flexible and secure links, while in D2D communication, RIS controls the environment for directional beams and enhanced security. The CRNs and WPT systems benefit from improved secrecy and optimized energy transfer, respectively. Similarly,  the MEC leverages RIS for secure data transmission and computation. RIS's dynamic control extends security to satellite networks, ISAC, and the IoT, maintaining overall network resilience. Furthermore, RIS-assisted PLS presents promising applications in emerging technologies like secure multicast communications, cell-free networks, and various network types.
    \item The real-world potential of \gls{RIS}-assisted \gls{PLS} hinges on addressing implementation challenges, scalability, and adaptability. Research explores how well \gls{RIS} scales with complex networks and adapts to dynamic communication scenarios. Additionally, active \gls{RIS} is a potential solution for the dual-fading issue in \gls{RIS}-supported links, although it comes with increased complexity and power consumption. Active \gls{RIS} elements amplify reflected signals, offering improved \gls{PLS} compared to mostly passive \gls{RIS} modules.
    \item The success of reliable beamforming and optimal reflection in RIS-assisted wireless communication systems is intrinsically linked to the precise acquisition of CSI. Despite the well-established conventional channel estimation methods in traditional wireless systems, applying these techniques to RIS-assisted secured wireless communications presents significant challenges. As observed in Table II, much of the existing research assumes perfect CSI for RIS-assisted secured wireless communications. Additionally, the Rayleigh fading channel is predominantly utilized, whereas more complex channel fading models are seldom employed, as indicated in Table II. Furthermore, although various types of RIS exist, passive RIS is the most prevalent in the literature due to its relative simplicity in analysis.
\end{itemize}}}
\section{RIS-Assisted PLS in OWC Systems}
\label{Section: RIS-Assisted PLS in OWC Systems}

\begin{figure*}[!th]
\centering
\includegraphics[width=6.8in]{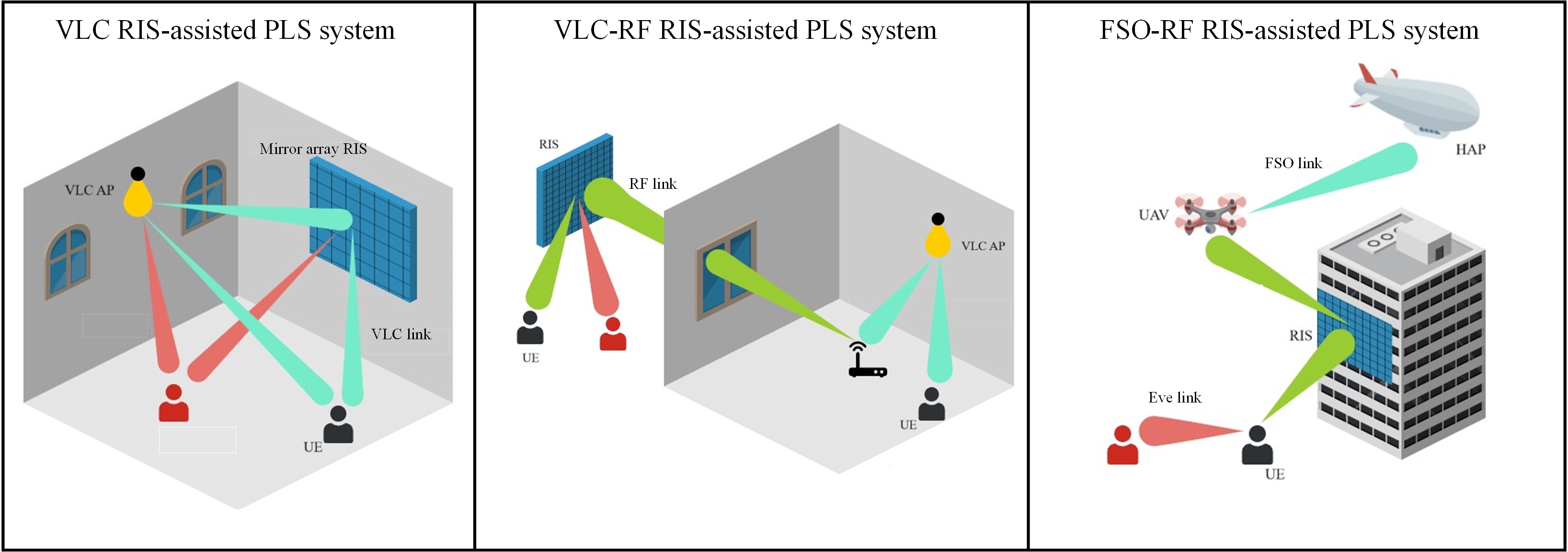}
\caption{{The integration of \gls{RIS} in \gls{PLS}-aided systems with \gls{OWC} enabling technologies for secure communications.}}
\label{Fig:0_VLC_Chart_1}
\end{figure*}

\textcolor{black}{This section discusses the potential of the adoption of the \gls{RIS} technology to secure \gls{OWC} networks. Specifically, the integration with \gls{VLC}, \gls{VLC}-\gls{RF}, and \gls{FSO}-\gls{RF}, which are summarized in both Fig.~\ref{Fig:0_VLC_Chart_1} and Table~\ref{tab: OWC-table}.}

%%%%%%%%%%%%%%%%%%%%%%%%%%%%
%\begin{figure*}[!t]\centering\includegraphics[width=1.0\textwidth]{Figures/Omar/0_VLC_Chart.pdf}\caption{The integration of \gls{RIS} in \gls{PLS}-aided systems with \gls{OWC} enabling technologies for secure communications..}\label{Fig:0_VLC_Chart}\end{figure*}
%%%%%%%%%%%%%%%%%%%%%%%%%%%%

\begin{table*}[!t]
\centering
%\vspace{-2.5em} 
\caption{Summary of \gls{RIS}-Assisted \gls{PLS} adopted system models in \gls{OWC} Systems}
\vspace{-0.7em} 
\label{tab: OWC-table}
\resizebox{0.98\textwidth}{!}{%
\begin{tabular}{|c|ccc|ccc|cc|c|}
\hline
\multirow{2}{*}{\textbf{{[}\#{]}}} & \multicolumn{3}{c|}{\textbf{System related content}} & \multicolumn{3}{c|}{\textbf{RIS related content}} & \multicolumn{2}{c|}{\textbf{Security related content}} & \multirow{2}{*}{\textbf{\begin{tabular}[c]{@{}c@{}}Performance\\ Metric(s)\end{tabular}}} \\ \cline{2-9}

 & \multicolumn{1}{c|}{\textbf{Technology}} & \multicolumn{1}{c|}{\textbf{\begin{tabular}[c]{@{}c@{}}System\\  Model\end{tabular}}} & \multicolumn{1}{c|}{\textbf{\begin{tabular}[c]{@{}c@{}}CSI\\ Condition\end{tabular}}} & \multicolumn{1}{c|}{\textbf{\begin{tabular}[c]{@{}c@{}}RIS\\ Type\end{tabular}}} & \multicolumn{1}{c|}{\textbf{RIS(s)}} & \textbf{\begin{tabular}[c]{@{}c@{}}RIS\\ Arch.\end{tabular}} & \multicolumn{1}{c|}{\textbf{\begin{tabular}[c]{@{}c@{}}PLS\\ Technique\end{tabular}}} & \textbf{Eve(s)} &  \\ \hline

%VLC
\cite{abumarshoud2022intelligent} & \multicolumn{1}{c|}{\multirow{9}{*}{VLC}} & \multicolumn{1}{c|}{SU-SISO-DL} & \multicolumn{1}{c|}{Perfect}  & \multicolumn{1}{c|}{Passive} & \multicolumn{1}{c|}{Single} & \begin{tabular}[c]{@{}c@{}} Mirror \\ array \end{tabular} & \multicolumn{1}{c|}{\begin{tabular}[c]{@{}c@{}} Channel\\ gain\\ manipulation\end{tabular}}  & Single & SC \\ \cline{1-1} \cline{3-10} 

\cite{sun2022optimization} & \multicolumn{1}{c|}{} & \multicolumn{1}{c|}{MU-SISO-DL} & \multicolumn{1}{c|}{Perfect} & \multicolumn{1}{c|}{Passive} & \multicolumn{1}{c|}{Single} & \begin{tabular}[c]{@{}c@{}} Mirror \\ array \end{tabular} & \multicolumn{1}{c|}{\begin{tabular}[c]{@{}c@{}} Unintended\\ interference \end{tabular}} & Single & SR \\ \cline{1-1} \cline{3-10} 

\cite{saifaldeen2022dr} & \multicolumn{1}{c|}{} & \multicolumn{1}{c|}{SU-SISO-DL} & \multicolumn{1}{c|}{Perfect}  & \multicolumn{1}{c|}{Passive} & \multicolumn{1}{c|}{Single} & \begin{tabular}[c]{@{}c@{}} Mirror \\ array \end{tabular} & \multicolumn{1}{c|}{BF}  & Single & SC \\ \cline{1-1} \cline{3-10} 

\cite{qian2021secure} & \multicolumn{1}{c|}{} & \multicolumn{1}{c|}{SU-SISO-DL} & \multicolumn{1}{c|}{Perfect}  & \multicolumn{1}{c|}{Passive} & \multicolumn{1}{c|}{Single} & \begin{tabular}[c]{@{}c@{}} Mirror \\ array \end{tabular} & \multicolumn{1}{c|}{\begin{tabular}[c]{@{}c@{}} Channel\\ gain\\ manipulation\end{tabular}}  & Single & SR  \\ \cline{1-1} \cline{3-10}

\cite{soderi2022vlc}  & \multicolumn{1}{c|}{} & \multicolumn{1}{c|}{SU-SISO-DL} & \multicolumn{1}{c|}{Perfect}  & \multicolumn{1}{c|}{Passive} & \multicolumn{1}{c|}{Single} & \begin{tabular}[c]{@{}c@{}} Mirror \\ array \end{tabular} & \multicolumn{1}{c|}{Jamming} & Single & SC \\ \hline

%VLC-RF
% The system model is not realistic
%\cite{zhao2022secure} & \multicolumn{1}{c|}{\multirow{5}{*}{VLC-RF}} & \multicolumn{1}{c|}{MU-SISO-DL} & \multicolumn{1}{c|}{Perfect} & \multicolumn{1}{c|}{Passive} & \multicolumn{1}{c|}{Single} & \begin{tabular}[c]{@{}c@{}} Meta-\\ surface \end{tabular} & \multicolumn{1}{c|}{-} & Single & SOP \\ \cline{1-1} \cline{3-10} 

\cite{zhang2022physical} & \multicolumn{1}{c|}{\multirow{6}{*}{VLC-RF}} & \multicolumn{1}{c|}{SU-SISO-DL} & \multicolumn{1}{c|}{Perfect} & \multicolumn{1}{c|}{Passive} & \multicolumn{1}{c|}{Single} & \begin{tabular}[c]{@{}c@{}} Meta-\\ surface \end{tabular} & \multicolumn{1}{c|}{-} & Single & SOP and SPSC \\ \cline{1-1} \cline{3-10}

%\textcolor{black}{\cite{rakib2024ris}} & \multicolumn{1}{c|}{} & \multicolumn{1}{c|}{} & \multicolumn{1}{c|}{} & \multicolumn{1}{c|}{} & \multicolumn{1}{c|}{} & \begin{tabular}[c]{@{}c@{}}  \end{tabular} & \multicolumn{1}{c|}{} &  & \\ \cline{1-1} \cline{3-10} % Ignore no EVE 

\textcolor{black}{\cite{saifaldeen2024drl}} & \multicolumn{1}{c|}{} & \multicolumn{1}{c|}{\textcolor{black}{SU-MISO-DL}} & \multicolumn{1}{c|}{\textcolor{black}{Perfect}} & \multicolumn{1}{c|}{\textcolor{black}{Passive}} & \multicolumn{1}{c|}{\textcolor{black}{Multiple}} & \begin{tabular}[c]{@{}c@{}} \textcolor{black}{Mirror} \\ \textcolor{black}{array} \\ \textcolor{black}{and} \\ \textcolor{black}{Meta-}\\ \textcolor{black}{surface} \end{tabular} & \multicolumn{1}{c|}{\textcolor{black}{BF}} & \textcolor{black}{Single} & \textcolor{black}{SC} \\ \cline{1-1} \cline{3-10} 

\cite{zhang2023security-Phys} & \multicolumn{1}{c|}{} & \multicolumn{1}{c|}{SU-SISO-DL} & \multicolumn{1}{c|}{Perfect} & \multicolumn{1}{c|}{Passive} & \multicolumn{1}{c|}{Single} & \begin{tabular}[c]{@{}c@{}} Meta-\\ surface \end{tabular} & \multicolumn{1}{c|}{-} & Single &  SOP and SPSC \\ \hline

%FSO-RF
\cite{rahman2023ris} & \multicolumn{1}{c|}{\multirow{5}{*}{FSO-RF}} & \multicolumn{1}{c|}{SU-SISO-DL} & \multicolumn{1}{c|}{Perfect} & \multicolumn{1}{c|}{Passive} & \multicolumn{1}{c|}{Multiple} & \begin{tabular}[c]{@{}c@{}} Meta-\\ surface \end{tabular} & \multicolumn{1}{c|}{-}  & Multiple & \begin{tabular}[c]{@{}c@{}} SOP, SPSC, IP,\\ ASC, and EST \end{tabular}  \\ \cline{1-1} \cline{3-10}

\cite{wang2023uplink} & \multicolumn{1}{c|}{} & \multicolumn{1}{c|}{MU-SISO-UL} & \multicolumn{1}{c|}{Perfect} & \multicolumn{1}{c|}{Passive} & \multicolumn{1}{c|}{Single} & \begin{tabular}[c]{@{}c@{}} Meta-\\ surface \end{tabular} & \multicolumn{1}{c|}{-}  & Single & SOP and PPSC \\ \cline{1-1} \cline{3-10}

\cite{ahmed2023enhancing} & \multicolumn{1}{c|}{} & \multicolumn{1}{c|}{SU-SISO-DL} & \multicolumn{1}{c|}{Perfect} & \multicolumn{1}{c|}{Passive} & \multicolumn{1}{c|}{Single} & \begin{tabular}[c]{@{}c@{}} Meta-\\ surface \end{tabular} & \multicolumn{1}{c|}{-}  & Single & \begin{tabular}[c]{@{}c@{}} SOP, ASC, SPSC,\\ and EST \end{tabular} \\ \hline
 
%\cite{hossain2022physical} & \multicolumn{1}{c|}{\multirow{2}{*}{UOWC}} & \multicolumn{1}{c|}{SU-SISO-DL} & \multicolumn{1}{c|}{Perfect} & \begin{tabular}[c]{@{}c@{}} Nakagami \\ and \\ Mixture \\ exponential \\ generalized \\ Gamma \\ (mEGG) \end{tabular} & \multicolumn{1}{c|}{Passive} & \multicolumn{1}{c|}{\begin{tabular}[c]{@{}c@{}} Single/ \\  Two \end{tabular}} & \begin{tabular}[c]{@{}c@{}}Meta-\\surface \end{tabular} & \multicolumn{1}{c|}{} & Single & \begin{tabular}[c]{@{}c@{}} SC, SOP, \\ and secrecy throughput \end{tabular} \\ \cline{1-1} \cline{3-11} 
 
\end{tabular}%
}
\end{table*}

\subsection{Integrating RIS with OWC Enabling Technologies for Secure Communications}

\subsubsection{RIS-Assisted Secure VLC}
when it comes to the integration of \gls{RIS}-aided \gls{PLS} in \gls{VLC} systems, the research started with single user scenarios for a static user~\cite{soderi2022vlc,qian2021secure, abumarshoud2022intelligent} or a mobile user~\cite{saifaldeen2022dr}, then evolved to multi-user scenarios~\cite{sun2022optimization}. These works employed different \gls{PLS} techniques to improve the secrecy rate performance. Specifically, in~\cite{qian2021secure, abumarshoud2022intelligent}, the mirror-array-based \gls{RIS} elements are optimized to increase the channel gain difference between the trusted user and the un-trusted user. In~\cite{sun2022optimization}, the closest \gls{LED} to the eavesdropper is employed to send an unintended interference to the eavesdropper in an attempt to degrade its \gls{SINR}. In~\cite{saifaldeen2022dr}, the transmit \gls{BF} of the \gls{VLC} \gls{AP} is optimized to maximize the system's \textcolor{black}{\gls{SC}}. Lastly, in~\cite{soderi2022vlc}, an \gls{RIS} is deployed close to a jamming receiver, which transmits a jamming interference toward the eavesdropper.  

\subsubsection{RIS-Assisted Secure VLC-RF}
The high data rates of \gls{VLC} and the extended coverage of \gls{RF} communications have motivated the research of \gls{VLC}-\gls{RF} hybrid systems. In~\cite{zhang2022physical,zhang2023security-Phys}, a two-hop scenario is considered, where in the first hop, \gls{VLC} is used for data transmission in an \textcolor{black}{\gls{EM}}-sensitive environment, and in the second hop, a relay that can perform an optical-to-electrical conversion is deployed next to an \gls{RIS} to extend the communication coverage. Additionally, there exists an eavesdropper in the second hop trying to wiretap the \gls{RF} link. The aforementioned works evaluated both the \gls{SOP} and the \gls{SPSC} of the proposed systems without employing a specific \gls{PLS} technique. Rather, the introduction of the \gls{RIS} is shown to increase the \gls{SOP} to a certain extent.  

\subsubsection{RIS-Assisted Secure FSO-RF}
The motivation for \gls{RF}/\gls{FSO} networks arises from the need for high-speed, reliable, and flexible communication solutions in various scenarios, ranging from urban environments with high data demands to remote areas with limited infrastructure. The security of the \gls{FSO} links is inherently assured by the laser beam's narrow and imperceptible nature, while the inherent broadcast nature of wireless \gls{RF} communication implies the likelihood of information leakage to eavesdroppers~\cite{wang2023uplink}. However, the propagation characteristics of the \gls{FSO} link need to be taken into account, such as the pointing errors, the path loss due to random atmospheric radio medium, and the atmospheric turbulence. In~\cite{rahman2023ris}, a secure mixed \gls{RF}/\gls{FSO} \gls{RIS}-aided system is proposed, where a secure message is conveyed from an \gls{RF} transmitter to an \gls{FSO} receiver with the assistance of a decode-and-forward intermediary relay. This work deploys one \gls{RIS} between the \gls{RF} transmitter and the relay and another one between the relay and the \gls{FSO} receiver. Such deployment is suitable in urban environments. In~\cite{wang2023uplink}, a secure mixed \gls{RF}/\gls{FSO} \gls{RIS}-aided \gls{HAPS}-\gls{UAV} collaborative system is proposed, which is suitable to serve remote areas users. A wiretap \gls{RF} link is assumed from the users to an eavesdropper.

\subsection{Lessons Learnt}
The lessons learnt from this section can be summarized as follows. %\textcolor{red}{Please rewrite the lessons learnt for this section based on the Reviewer's comment and remove the references for consistency}
\begin{itemize}
    \item \textcolor{black}{The literature on \gls{RIS}-assisted \gls{PLS} \gls{VLC} agrees that increasing the number of \gls{RIS} elements improves the overall security performance of \gls{VLC} systems. Specifically, additional \gls{RIS} elements enable spatial \gls{BF} and nulling, allowing precise directional control of transmitted light and the formation of constructive interference patterns directed toward intended receivers, while creating destructive interference for potential eavesdroppers. The greater flexibility provided by a larger number of \gls{RIS} elements supports continuous optimization to adapt to changing environmental conditions, thereby enhancing security against eavesdropping attempts. This increased spatial diversity and improved \gls{SNR} at the legitimate receiver enhances the reliability of communication links, thereby raising the difficulty for unauthorized entities attempting to intercept communications.}
    %\item There is consensus in the provided \gls{RIS}-aided \gls{PLS} \gls{VLC} literature that increasing the number of \gls{RIS} elements results in enhancing the overall secrecy performance of \gls{VLC} systems. Specifically, the additional \gls{RIS} elements enable spatial \textcolor{black}{\gls{BF}} and nulling, allowing for precise directionality of transmitted light and constructive interference patterns towards intended receivers while creating interference for potential eavesdroppers. The dynamic adaptability of a larger number of \gls{RIS} elements facilitates continuous optimization in response to changing environmental conditions, contributing to improved security against eavesdropping attempts. This increased spatial diversity and improved \gls{SNR} at the legitimate receiver enhances the reliability of communication links, making it more challenging for unauthorized entities to intercept the communication.
    \item \textcolor{black}{\gls{VLC} offers inherent properties that make it suitable for advanced \gls{PLS} techniques. These properties, combined with the potential of using \gls{RGB} \glspl{LED} as transmitters, open doors for improved security in \gls{VLC} systems. For instance, advancements like the Watermark Blind \gls{PLS} algorithm, which uses a combination of spread-spectrum watermarking and a jamming receiver to achieve confidentiality, have been implemented in \gls{VLC}. Recent research has explored leveraging \gls{RIS} technology alongside such advanced PLS techniques, demonstrating further enhancements in \gls{VLC} security. Overall, the capabilities of \gls{RIS} technology pave the way for the effective adoption of novel \gls{PLS} techniques in \gls{VLC}, significantly improving its security.}
    %\item The core characteristics of \gls{VLC} and the potential of employing \gls{RGB} \glspl{LED} as \gls{VLC} transmitters facilitate the adoption of advanced \gls{PLS} techniques in \gls{VLC} systems. For example, the Watermark Blind \gls{PLS}  algorithm~\cite{soderi2017physical} that utilizes a combination of spread-spectrum watermarking and a jamming receiver to provide confidentiality and protect the transmitted information from eavesdroppers has been adopted in \gls{VLC} in~\cite{soderi20216g}. In~\cite{soderi2022vlc}, leveraging the \gls{RIS} technology with such advanced \gls{PLS} technique attain further improvement in the security properties of \gls{VLC} systems. Overall, utilizing the capabilities of the \gls{RIS} technology facilitates the effective adoption of new \gls{PLS} techniques in \gls{VLC} systems.
    \item \textcolor{black}{The research of \gls{RIS}-assisted \gls{PLS} in \gls{OWC} is still in its infancy compared to the more developed investigations in \gls{RF} systems. This is evident in the amount of presented works in Table III as compared to Table II. Hence, most works presented in Table III considered simple scenarios (e.g., the assumption of perfect \gls{CSI}, passive eavesdropper, etc.) to illustrate the feasibility of such integration. Exploring more practical scenarios (e.g., multi-cell scenarios, imperfect \gls{CSI}, active eavesdropper, and \gls{LoS} blockage) is needed to better understand the potential and complexity of the integration of \gls{RIS}-assisted \gls{PLS} in \gls{OWC} systems.}
\end{itemize}
%%%%%%%%%%%%%%%%%%%%%%%%%%%%%%%%%%%%%%%%%%%%%%%%%%%%%%%%%%%%%%%%%%%%%%%%%%%%
%\begin{figure*}     \centering     \begin{subfigure}[t]{0.30\textwidth}         \centering         \includegraphics[width=\textwidth]{Figures/Majid/NVLC02.png}         \caption{VLC RIS-assisted PLS system.}         \label{VLC-RIS}     \end{subfigure}     \hfill     \begin{subfigure}[t]{0.35\textwidth}         \centering        \includegraphics[width=\textwidth]{Figures/Majid/NVLC01.png}         \caption{VLC-RF RIS-assisted PLS system.}         \label{VLC-RF}     \end{subfigure}     \hfill     \begin{subfigure}[t]{0.28\textwidth}         \centering        \includegraphics[width=\textwidth]{Figures/Majid/NVLC033.png}         \caption{FSO-RF RIS-assisted PLS system.}         \label{FSO-RIS}     \end{subfigure}\caption{The integration of \gls{RIS} in \gls{PLS}-aided systems with \gls{OWC} enabling technologies for secure communications.}        \label{Fig:0_VLC_Chart_1}\end{figure*}
\section{Optimization Techniques for RIS-Assisted PLS}
\label{Section: Optimization Techniques for RIS-Assisted PLS}

\begin{table*}[!t]
\centering
\vspace*{-6mm}
\caption{\textcolor{black}{Summary of Secrecy Rate Maximization related literature for \gls{RIS}-Assisted \gls{PLS} systems}}
\vspace*{-3mm}
\label{Table:opt_tech1}
\resizebox{\textwidth}{!}{%
\begin{tabular}{|l|l|l|l|l|}
\hline
\textbf{[\#]} &
\textbf{\begin{tabular}[c]{@{}l@{}}Phase-shift\\ resolution\end{tabular}} &
  \textbf{Optimization variables} &
  \textbf{Optimization techniques} &
  \textbf{Constraints} 
\\ \hline

\cite{li2021reconfigurable}  & Continuous &  \begin{tabular}[c]{@{}l@{}} RIS phase shifts, drones' trajectory, \\transmit power  \end{tabular}& \begin{tabular}[c]{@{}l@{}} BCD, AO,  SCA, One-dimension search \end{tabular} & \begin{tabular}[c]{@{}l@{}}Drone mobility, Power, and RIS phase \\constraints \end{tabular}\\ \hline

\cite{guo2022joint} & Continuous   &\begin{tabular}[c]{@{}l@{}} RIS location, RIS phase shifts\end{tabular}& \begin{tabular}[c]{@{}l@{}} Heuristic search, Charnes-Cooper transformation, \\ Sequential rank-one constraint relaxation \end{tabular} & \begin{tabular}[c]{@{}l@{}}RIS location and RIS phase constraints \end{tabular}\\ \hline

\cite{hao2022securing} &Continuous   &\begin{tabular}[c]{@{}l@{}} BS beamforming vector, RIS phase shifts\end{tabular}& \begin{tabular}[c]{@{}l@{}} AO, SDR, SCA, Linear conic relaxation \end{tabular} & \begin{tabular}[c]{@{}l@{}}Min. SNR and RIS phase constraints \end{tabular}\\ \hline

\cite{tang2021securing} &Continuous   &\begin{tabular}[c]{@{}l@{}} BS beamforming vector, RIS phase shifts\end{tabular}& \begin{tabular}[c]{@{}l@{}} AO, SDR, BCD \end{tabular} & \begin{tabular}[c]{@{}l@{}}Power and RIS phase constraints \end{tabular}\\ \hline

\cite{dong2021secure} &Continuous   &\begin{tabular}[c]{@{}l@{}} BS beamforming vector, RIS phase shifts\end{tabular}& \begin{tabular}[c]{@{}l@{}} AO, Dinkelbach Method, SCA, CCP  \end{tabular} & \begin{tabular}[c]{@{}l@{}} Interference threshold, power,\\ and RIS phase constraints \end{tabular}\\ \hline

\cite{jiang2020intelligent} &\begin{tabular}[c]{@{}l@{}}Continuous/\\Discrete \end{tabular}  &\begin{tabular}[c]{@{}l@{}} Transmit covariance matrix, RIS phase shifts\end{tabular}& \begin{tabular}[c]{@{}l@{}} AO, SCA, Dinkelbach Method  \end{tabular} & \begin{tabular}[c]{@{}l@{}} Power, transmit covariance matrix, and\\ RIS phase constraints\end{tabular}\\ \hline

\cite{shi2022secrecy} &\begin{tabular}[c]{@{}l@{}}Continuous\end{tabular}  &\begin{tabular}[c]{@{}l@{}}Time allocation, energy transmit covariance\\ matrix, information transmit beamforming \\ matrix, and RIS phase shifts\end{tabular}& \begin{tabular}[c]{@{}l@{}} Mean-square error method, AO, \\dual subgradient method, MM, SCA, \\ Second-order cone programming  \end{tabular} & \begin{tabular}[c]{@{}l@{}} Energy and information transmit power \\constraints, RIS phase constraints\end{tabular}\\ \hline

\cite{sun2023joint} &\begin{tabular}[c]{@{}l@{}}Continuous\end{tabular}  &\begin{tabular}[c]{@{}l@{}}BS beamforming vector, phase shifts for transmitting \\ and reflecting RISs, users' digital decorder\end{tabular}& \begin{tabular}[c]{@{}l@{}} Akaike information criterion, BCD, AO,\\ Double Deterministic Transformation, \\Lagrange multiplier method, Penalty method  \end{tabular} & \begin{tabular}[c]{@{}l@{}} SR, power, and RIS phase shifts\\ constraints\end{tabular}\\ \hline

\cite{alexandropoulos2023counteracting} &\begin{tabular}[c]{@{}l@{}}Continuous\end{tabular}  &\begin{tabular}[c]{@{}l@{}}BS precoding matrix, number of data streams, AN \\covariance matrix, combining matrix, RIS phase shifts  \end{tabular}& \begin{tabular}[c]{@{}l@{}} AO, MM, Manifold optimization  \end{tabular} & \begin{tabular}[c]{@{}l@{}} Power, thermal noise, and RIS phase\\ shifts constraints\end{tabular}\\ \hline

\cite{shen2019secrecy} &\begin{tabular}[c]{@{}l@{}}Continuous\end{tabular}  &\begin{tabular}[c]{@{}l@{}}BS covariance matrix, RIS phase shifts  \end{tabular}& \begin{tabular}[c]{@{}l@{}} AO, Bisection search \end{tabular} & \begin{tabular}[c]{@{}l@{}} Power and RIS phase shifts constraints\end{tabular}\\ \hline

\cite{cui2019secure} &\begin{tabular}[c]{@{}l@{}}Continuous\end{tabular}  &\begin{tabular}[c]{@{}l@{}}BS beamforming matrix, RIS phase shifts  \end{tabular}& \begin{tabular}[c]{@{}l@{}} AO, SDR \end{tabular} & \begin{tabular}[c]{@{}l@{}} Power and RIS phase shifts constraints\end{tabular}\\ \hline

\cite{keming2021physical} &\begin{tabular}[c]{@{}l@{}}Continuous\end{tabular}  &\begin{tabular}[c]{@{}l@{}}BS beamforming vector, RIS phase shifts  \end{tabular}& \begin{tabular}[c]{@{}l@{}} AO, Manifold optimization, \\Fractional programming \end{tabular} & \begin{tabular}[c]{@{}l@{}} Power and RIS phase shifts constraints\end{tabular}\\ \hline

\cite{zhou2021secure} &\begin{tabular}[c]{@{}l@{}}Continuous\end{tabular}  &\begin{tabular}[c]{@{}l@{}}BS beamforming vector, RIS phase shifts  \end{tabular}& \begin{tabular}[c]{@{}l@{}} AO, SCA, SDR \end{tabular} & \begin{tabular}[c]{@{}l@{}} Power and RIS phase shifts constraints\end{tabular}\\ \hline

\cite{du2020reconfigurable} &\begin{tabular}[c]{@{}l@{}}Continuous\end{tabular}  &\begin{tabular}[c]{@{}l@{}}Transmit covariance matrix, RIS phase shifts  \end{tabular}& \begin{tabular}[c]{@{}l@{}} Logarithmic barrier method \end{tabular} & \begin{tabular}[c]{@{}l@{}} Power and RIS phase shifts constraints\end{tabular}\\ \hline

\cite{dong2023robust} &\begin{tabular}[c]{@{}l@{}}Continuous\end{tabular}  &\begin{tabular}[c]{@{}l@{}}BS beamforming vector, RIS amplification and \\ phase shifts matrix \end{tabular}& \begin{tabular}[c]{@{}l@{}} AO, Dual-SCA, SDR\end{tabular} & \begin{tabular}[c]{@{}l@{}}  BS and RIS reflecting power budget, \\and SOP constraints\end{tabular}\\ \hline

\cite{han2023secure} &\begin{tabular}[c]{@{}l@{}}Continuous\end{tabular}  &\begin{tabular}[c]{@{}l@{}}BS beamforming vector, RIS reflective and transmissive \\ coefficients \end{tabular}& \begin{tabular}[c]{@{}l@{}} AO, SOCP, SCA\end{tabular} & \begin{tabular}[c]{@{}l@{}}  BS power budget, Min. SNR requirement,\\ transmissive and reflective phase shifts and \\ SIC order constraint\end{tabular}\\ \hline

\cite{zhang2023robust} &\begin{tabular}[c]{@{}l@{}}Continuous\end{tabular}  &\begin{tabular}[c]{@{}l@{}}BS beamforming vector, RIS phase shifts \end{tabular}& \begin{tabular}[c]{@{}l@{}} AO, SCA, SDR, multi-dimensional quadratic\\ transform\end{tabular} & \begin{tabular}[c]{@{}l@{}}  BS power budget, Min. SINR requirement, \\RIS phase shift, and SIC order constraint\end{tabular}\\ \hline

\cite{shu2022beamforming} &\begin{tabular}[c]{@{}l@{}}Continuous\end{tabular}  &\begin{tabular}[c]{@{}l@{}}BS beamforming vector, RIS phase shifts  \end{tabular}& \begin{tabular}[c]{@{}l@{}} AO, SDR, SCA, Dual ascent, Quadratic \\transform method \end{tabular} & \begin{tabular}[c]{@{}l@{}} Power and RIS phase shifts constraints\end{tabular}\\ \hline

\cite{sun2022optimization} &\begin{tabular}[c]{@{}l@{}}N/A\end{tabular}  &\begin{tabular}[c]{@{}l@{}}RIS-LED association matrix  \end{tabular}& \begin{tabular}[c]{@{}l@{}} Bipartite graph, Kuhn-Munkres algorithm \end{tabular} & \begin{tabular}[c]{@{}l@{}} RIS-LED association and fairness constraints\end{tabular}\\ \hline

\cite{wang2023uav} &\begin{tabular}[c]{@{}l@{}}Continuous\end{tabular}  &\begin{tabular}[c]{@{}l@{}}BS beamforming vector, RIS phase shifts, UAV-mounted\\ RIS location\end{tabular}& \begin{tabular}[c]{@{}l@{}} \gls{SCA}, Sequential rank-one\\ constraint relaxation, Penalty method \end{tabular} & \begin{tabular}[c]{@{}l@{}} UAV and BS power budgets, RIS phase shifts \\ constraint\end{tabular}\\ \hline

\cite{pang2021irs} &\begin{tabular}[c]{@{}l@{}}Continuous\end{tabular}  &\begin{tabular}[c]{@{}l@{}}UAV trajectory, BS beamforming vector, RIS phase shifts\end{tabular}& \begin{tabular}[c]{@{}l@{}} Dinkelbach method, SCA \end{tabular} & \begin{tabular}[c]{@{}l@{}} RIS phase shifts constraints, UAV power \\budget, UAV mobility constraints\end{tabular}\\ \hline

\cite{salem2022active} & Continuous  &\begin{tabular}[c]{@{}l@{}} Radar receive beamforming, RIS phase shifts, BS transmit\\ beamforming\end{tabular} & \begin{tabular}[c]{@{}l@{}}Quadratic transformation, MM \end{tabular} & \begin{tabular}[c]{@{}l@{}} Radar detection QoS,  power budgets of \\BS and RIS\end{tabular}\\ \hline

\cite{cheng2023irs} & Continuous  & \begin{tabular}[c]{@{}l@{}} BS transmit precoding matrix, RIS phase shifts, UAV trajectory \end{tabular}& \begin{tabular}[c]{@{}l@{}}AO, SDR, SCA, Riemannian
Manifold gradient   \end{tabular}& \begin{tabular}[c]{@{}l@{}} UAV trajectory constraints, UAV power\\ budget, RIS phase shifts constraint   \end{tabular}\\ \hline

\cite{benaya2023physical}  & Continuous  &\begin{tabular}[c]{@{}l@{}} AP and UAV transmit power, UAV trajectory, RIS transmit\\ and reflective phase shifts, power splitting factor  \end{tabular}& SCA, SDR   & \begin{tabular}[c]{@{}l@{}} UAV trajectory constraints, jamming and \\transmit power budget, power splitting ratio\\ constraint, harvested energy and RIS phase shifts\end{tabular}\\ \hline

\cite{lin2023secure} & Continuous  & \begin{tabular}[c]{@{}l@{}} BS beamforming vector, RIS phase shifts  \end{tabular}& AO, SCA& \begin{tabular}[c]{@{}l@{}}BS transmit power budget, RIS phase shift constraint   \end{tabular}\\ \hline

\cite{xing2023reconfigurable} & Continuous  & \begin{tabular}[c]{@{}l@{}} BS transmit and receive beamforming vectors, AN covariance\\ matrix, RIS phase shifts  \end{tabular}& AO, SCA, SDR & \begin{tabular}[c]{@{}l@{}}BS transmit power budget, RIS phase shift constraint,\\ Min. data rate and sensing SNR requirements \end{tabular}\\ \hline

\cite{han2023broadcast} & Continuous & \begin{tabular}[c]{@{}l@{}} UAV BS beamforming, UAV's trajectory,\\ RIS phase shifts  \end{tabular}& \begin{tabular}[c]{@{}l@{}} BCD, SDP, SCA, Reinforcement learning \end{tabular}& \begin{tabular}[c]{@{}l@{}}BS transmit power budget, RIS phase shift constraint,\\ UAV trajectory constraints    \end{tabular}\\ \hline

\cite{wang2022multicast} & Continuous  & \begin{tabular}[c]{@{}l@{}} BS beamforming vector, RIS phase shifts  \end{tabular}& \begin{tabular}[c]{@{}l@{}} AO, SDP  \end{tabular}& \begin{tabular}[c]{@{}l@{}}BS transmit power budget, RIS phase shifts constraint \end{tabular}\\ \hline

\cite{abumarshoud2022intelligent} & Continuous  & \begin{tabular}[c]{@{}l@{}} BS transmit power, RIS allocation matrix  \end{tabular}& \begin{tabular}[c]{@{}l@{}} Heuristics, Adaptive restart genetic algorithm  \end{tabular}& \begin{tabular}[c]{@{}l@{}}BS transmit power budget, minimum rate requirement, \\RIS allocation constraints    \end{tabular}\\ \hline

\cite{hu2023securing} & Continuous  & \begin{tabular}[c]{@{}l@{}} User transmit power, RIS phase shifts  \end{tabular}& \begin{tabular}[c]{@{}l@{}} AO, MM, Difference of concave programming \end{tabular}& \begin{tabular}[c]{@{}l@{}}User transmit power budget, RIS phase shifts constraints    \end{tabular}\\ \hline

\cite{xu2023sum} & Continuous  & \begin{tabular}[c]{@{}l@{}} BS transmit power, RIS phase shifts  \end{tabular}& \begin{tabular}[c]{@{}l@{}} AO, SDR \end{tabular}& \begin{tabular}[c]{@{}l@{}}BS transmit power budget, RIS phase shifts constraints,\\ Min. data rate requirement    \end{tabular}\\ \hline

\cite{yuan2022secure} & Continuous  & \begin{tabular}[c]{@{}l@{}}  RIS phase shifts  \end{tabular}& \begin{tabular}[c]{@{}l@{}} SDR \end{tabular}& \begin{tabular}[c]{@{}l@{}} RIS phase shifts constraints \end{tabular}\\ \hline

\cite{niu2022active} & Continuous  & \begin{tabular}[c]{@{}l@{}}BS beamforming vector, BS AN,  RIS phase shifts  \end{tabular}& \begin{tabular}[c]{@{}l@{}} AO, QCQP \end{tabular}& \begin{tabular}[c]{@{}l@{}} BS transmit power budget, interference threshold,\\ RIS phase shifts constraints \end{tabular}\\ \hline

\cite{hoang2023secrecy}  & Continuous  & \begin{tabular}[c]{@{}l@{}}Ground BS beamforming vector, ground and satellite \\ RIS phase shifts  \end{tabular}& \begin{tabular}[c]{@{}l@{}} Dinkelbach's method, AO \end{tabular}& \begin{tabular}[c]{@{}l@{}} BS transmit power budget, RIS phase shifts constraints \end{tabular}\\ \hline

\cite{gong2023joint}  & Continuous  & \begin{tabular}[c]{@{}l@{}}BS transmit power, RIS phase shifts  \end{tabular}& \begin{tabular}[c]{@{}l@{}} AO, SDR  \end{tabular}& \begin{tabular}[c]{@{}l@{}} BS transmit power budget, RIS phase shifts constraints \end{tabular}\\ \hline

\cite{feng2020physical} & Continuous  & \begin{tabular}[c]{@{}l@{}}BS beamforming vector, RIS phase shifts  \end{tabular}& \begin{tabular}[c]{@{}l@{}} Fractional programming Manifold optimization  \end{tabular}& \begin{tabular}[c]{@{}l@{}} BS transmit power budget, RIS phase shifts constraints \end{tabular}\\ \hline

\cite{qian2021secure} & Continuous  & \begin{tabular}[c]{@{}l@{}}RIS orientation angles  \end{tabular}& \begin{tabular}[c]{@{}l@{}} Particle swarm optimization\end{tabular}& \begin{tabular}[c]{@{}l@{}} RIS orientation angle constraints \end{tabular}\\ \hline

\cite{sun2021intelligent} &\begin{tabular}[c]{@{}l@{}}Continuous\end{tabular}  &\begin{tabular}[c]{@{}l@{}}UAV BS beamforming vector, RIS phase shifts, UAV BS \\and RIS positions  \end{tabular}&\begin{tabular}[c]{@{}l@{}} AO, Exhaustive search, SDR \end{tabular} & \begin{tabular}[c]{@{}l@{}} Power, minimum rate, RIS phase shifts, and\\ UAV position constraints\end{tabular}\\\hline

% Sylvester

%\textcolor{black}{\cite{huang2023smart}} &   &\begin{tabular}[c]{@{}l@{}}  \end{tabular}&\begin{tabular}[c]{@{}l@{}}  \end{tabular} & \begin{tabular}[c]{@{}l@{}} \end{tabular}\\\hline

%\textcolor{black}{\cite{ma2023physical}} &  &\begin{tabular}[c]{@{}l@{}}  \end{tabular}&\begin{tabular}[c]{@{}l@{}}  \end{tabular} & \begin{tabular}[c]{@{}l@{}} \end{tabular}\\\hline

\textcolor{black}{\cite{li2023sum}} & \textcolor{black}{Continuous} &\begin{tabular}[c]{@{}l@{}} \textcolor{black}{BS receive beamforming vector,}\\ \textcolor{black}{Active RIS reflection coefficients, Users power budget}  \end{tabular}&\begin{tabular}[c]{@{}l@{}}  \textcolor{black}{BCD, SCA}\end{tabular} & \begin{tabular}[c]{@{}l@{}} \textcolor{black}{BS amplification power budget}\\ \textcolor{black}{Users' transmit power budget}\end{tabular}\\\hline

\textcolor{black}{\cite{tu2024physical}} & \begin{tabular}[c]{@{}l@{}} \textcolor{black}{Continuous}\\ \textcolor{black}{Discrete}\end{tabular}&\begin{tabular}[c]{@{}l@{}}  \textcolor{black}{BS transmit beamforming vector, RIS phase shifts}\end{tabular}&\begin{tabular}[c]{@{}l@{}}  \textcolor{black}{Fractional programming, SCA}\end{tabular} & \begin{tabular}[c]{@{}l@{}} \textcolor{black}{BS power budget, RIS unit-modulus}\end{tabular}\\\hline

\textcolor{black}{\cite{jiang2023aerial}} & \textcolor{black}{Continuous} &\begin{tabular}[c]{@{}l@{}}  \textcolor{black}{ARIS 3D deployment, ARIS phase shifts}\end{tabular}&\begin{tabular}[c]{@{}l@{}}   \textcolor{black}{Iterative algorithm, SCA,}\\ \textcolor{black}{Phase alignment methods}\end{tabular} & \begin{tabular}[c]{@{}l@{}} \textcolor{black}{UAV's mobility,} \textcolor{black}{RIS phase shifts} \end{tabular}\\\hline

% Omar
\textcolor{black}{\cite{arzykulov2024aerial}} & \textcolor{black}{Continuous} &\begin{tabular}[c]{@{}l@{}} \textcolor{black}{ARIS 3D deployment, ARIS user portions} \end{tabular} &\begin{tabular}[c]{@{}l@{}} \textcolor{black}{Iterative algorithm} \end{tabular} & \begin{tabular}[c]{@{}l@{}} \textcolor{black}{QoS constraints for user and Eves,} \\ \textcolor{black}{total number of active RIS elements}\end{tabular} \\\hline

\textcolor{black}{\cite{kong2024secrecy}} & \textcolor{black}{Continuous} &\begin{tabular}[c]{@{}l@{}} \textcolor{black}{AP energy transmit beamforming vector,} \\ \textcolor{black}{AP receive beamforming vector, } \\ \textcolor{black}{ARIS time allocation, ARIS phase shifts} \end{tabular} & \textcolor{black}{AO, SCA} & \begin{tabular}[c]{@{}l@{}} \textcolor{black}{AP transmit power budget,} \\ \textcolor{black}{ARIS amplification power budget,} \\ \textcolor{black}{RIS amplitude constraint}\end{tabular}\\\hline

\textcolor{black}{\cite{ma2023secure}} &  \textcolor{black}{Continuous} & \begin{tabular}[c]{@{}l@{}} \textcolor{black}{AP transmit beamforming vector, time allocation,} \\ \textcolor{black}{RIS energy  and information reflection coefficients} \end{tabular} &\begin{tabular}[c]{@{}l@{}} \textcolor{black}{AO, SDR, Gaussian randomization procedure} \end{tabular} & \begin{tabular}[c]{@{}l@{}} \textcolor{black}{Transmission time allocation constraints, AP transmit} \\ \textcolor{black}{power budget, RIS modulus constraints of the energy} \\ \textcolor{black}{and information refection coefficients} \end{tabular}\\\hline

\textcolor{black}{\cite{zhang2024enhancing}} & \textcolor{black}{Discrete} & \begin{tabular}[c]{@{}l@{}} \textcolor{black}{BS beamforming matrix, RIS phase shifts} \end{tabular}&\begin{tabular}[c]{@{}l@{}} \textcolor{black}{AO, SCA, augmented Lagrange method,} \\ \textcolor{black}{and the quasi-Newton method} \end{tabular} & \begin{tabular}[c]{@{}l@{}} \textcolor{black}{RIS unit-modulus, BS power budget} \end{tabular}\\\hline

\textcolor{black}{\cite{sun2024secure}} & \textcolor{black}{Continuous} &\begin{tabular}[c]{@{}l@{}} \textcolor{black}{BS transmit beamforming, RIS phase shifts} \end{tabular}&\begin{tabular}[c]{@{}l@{}}  \textcolor{black}{AO, majorization-minimization, SDR}\end{tabular} & \begin{tabular}[c]{@{}l@{}} \textcolor{black}{Target minimum illumination power, BS power budget} \end{tabular}\\\hline

\textcolor{black}{\cite{wei2024star}} & \textcolor{black}{Continuous} &\begin{tabular}[c]{@{}l@{}} \textcolor{black}{BS transmit beamforming, BS artificial jamming,} \\ \textcolor{black}{STAR-RIS passive transmission and reflection beamformings}  \end{tabular} &\begin{tabular}[c]{@{}l@{}} \textcolor{black}{AO, SCA} \end{tabular} & \begin{tabular}[c]{@{}l@{}} \textcolor{black}{Target minimum beampattern gain} \end{tabular}\\\hline

\end{tabular}
}
\vspace*{-3mm}
\end{table*}

{\textcolor{black}{A general mathematical optimization problem typically has the form 
\begin{equation}\label{opt_form}
\begin{array}{l}
\mathop {\min }\limits_{{\mathbf{x}}} f\left({\mathbf {x}}\right) \\
{\rm{s}}{\rm{.t}}{\rm{.}}\\
 g_i\left({\mathbf{x}}\right)\le b_i,\,\,i=1,2,\ldots, m,\\
 h_i\left({\mathbf{x}}\right) = c_j,\,\,i=1,2,\ldots, n,\\
\end{array}
\end{equation}
where the function $f\left({\mathbf {x}}\right)$ is the objective function that needs to be minimized, ${\mathbf{x}}$ is the set of optimization variables, $g_i\left({\mathbf{x}}\right)\le b_i$ and $h_i\left({\mathbf{x}}\right) = c_j,$ are the inequality and equality constraints that restrict the feasible region, respectively, $m$ denotes the number of inequality constraints, and $n$ is the number of equality constraints. The optimization problem in~\eqref{opt_form} is a constrained optimization problem and can be a convex or non-convex problem. It is  convex if $f\left({\mathbf {x}}\right)$ and $g_i\left({\mathbf{x}}\right)$ are convex functions and $h_i\left({\mathbf{x}}\right) - c_j$ is affine~\cite{boyd2004convex}. In the context of optimizing \gls{RIS}-assisted \gls{PLS}, typical objective functions include \gls{SR}, \gls{SC}, \gls{SOP}, \gls{PNSC}, \gls{IP}, secure \gls{EE}, and transmit power. The decision variables include resource such as power and bandwidth, allocation, \gls{RIS} phase shifts and placement,  beamformer/precoder design, and antenna selection. A brief introduction of the various optimization techniques in this section can be found in~\cite{zhou2023survey,aboagye2024multi}. This section reviews state-of-the-art optimization techniques for \gls{RIS}-assisted \gls{PLS} and summarizes existing optimization schemes for \gls{SR} maximization and other secrecy metric optimization in Tables IV and V, respectively.}}

%How to classify these works?

\subsection{Secrecy Rate Optimization}
In~\cite{li2021reconfigurable}, the authors investigated \gls{SR} maximization by optimizing the phase shifts of the \gls{RIS}, the trajectory of a drone BS, and its transmit power. In this paper, the authors first used  \gls{BCD} approach, \textcolor{black}{an iterative optimization method}, to decompose the joint problem into three subproblems. A closed-form expression of the optimal phase shifts given the drone's trajectory and transmit power was proposed. Then, an \gls{SCA} method was developed to obtain an approximate solution for the drone trajectory subproblem. Finally, a one-dimension search method was designed for the transmit power allocation subproblem. Under the criteria of maximizing the \gls{SR}, the authors in~\cite{cheng2023ris} proposed a low-complexity technique based on the Riemannian conjugate gradient approach to optimize the \gls{BS} \textcolor{black}{\gls{BF}} vector and the \gls{RIS} phase shift under transmit power budget and \gls{RIS} phase constraint. In~\cite{guo2022joint}, the \gls{RIS} phase shift and location were optimized to maximize the \gls{SR} by exploiting the Charnes-Cooper transformation and sequential rank-one constraint relaxation methods while considering constraints on the \gls{RIS} phase shifts and the \gls{RIS} location. In~\cite{sun2022ris}, the authors examined a worst-case sum rate maximization problem under minimum achievable rate, maximum wiretap rate, and maximum transmit power constraints. An algorithm to design the receive decoder at the users, both the digital precoder and the \gls{AN} at the \gls{BS}, and the analog precoder at the \gls{RIS} was developed by leveraging on \gls{AO}, {\textcolor{black}{another iterative optimization method}}, the \gls{MM} method, \gls{QCQP}, and the Riemannian manifold optimization technique. In~\cite{zhao2022secrecy}, the authors investigated a stochastic \gls{SCA}-based algorithm to optimize the \textcolor{black}{\gls{BF}} vectors at the \gls{BS} and the phase shifts of an \gls{RIS} to maximize the average worst-case \gls{SR} subject to both the power constraint at the BS and energy harvesting constraints at the energy harvesting users.

A \gls{SR} maximization problem to optimize the \gls{BS} \textcolor{black}{\gls{BF}} vector and \gls{RIS} phase shifts was explored and solved using \gls{AO}, \gls{SDR}, \gls{SCA}, and the linear conic relaxation method in~\cite{hao2022securing}. The authors in~\cite{tang2021securing} developed an algorithm based on \gls{SDR} and the \gls{BCD} framework to optimize BS \textcolor{black}{\gls{BF}} vector and \gls{RIS} phase shifts while considering constraints on the BS transmit power and \gls{RIS} phase shifts. In~\cite{dong2021secure}, the authors proposed an algorithm to optimize the \textcolor{black}{\gls{BF}} vector of the BS and the \gls{RIS} phase shift subject to total power constraint at the secondary transmitter, interference power constraint at the primary receiver as well as unit modulus constraint at \gls{RIS}. Under the assumption of both continuous and discrete \gls{RIS} phase shift resolution, the joint optimization of transmit covariance matrix at the \gls{BS} and \gls{RIS} phase shifts to maximize the \gls{SR} under power budget, the positive semi-definite constraint on transmit covariance matrix, and \gls{RIS} phase constraints was investigated in~\cite{jiang2020intelligent}. In that paper, the authors proposed an \gls{AO}-based algorithm in which, given the reflecting coefficients at the IRS and by leveraging the \gls{SCA}, a convex approach was used to solve the transmit covariance matrix optimization at the \gls{AP}, while given the transmit covariance matrix at the \gls{AP}, \gls{AO} was explored to find the \gls{RIS} phase shits. In~\cite{shi2022secrecy}, the authors maximized sum \gls{SR} by optimizing the downlink/uplink time allocation, the energy transmit covariance matrix of a \gls{BS}, the information transmit \textcolor{black}{\gls{BF}} matrix of users and the \gls{RIS} phase shit matrix subject to constraints on energy/information transmit power at the hybrid BS/users and the \gls{RIS} phase shifts. The proposed \gls{AO} algorithm involved the \gls{MSE} method to reformulate the original problem, an \gls{AO} algorithm and the dual subgradient method to optimize the energy covariance and information \textcolor{black}{\gls{BF}} matrices, a one-dimensional search to obtain the optimal time allocation, and \gls{SOCP}, \gls{SCA}, and the \gls{MM} algorithm to find the \gls{RIS} phase shifts.

A worst-case rate maximization problem to optimize the joint transmissive and reflective phase shifts, cascaded angular uncertainties, and the unknown jamming and \glspl{BS} \textcolor{black}{\gls{BF}} vectors, under power budgets, secure QoS requirements, and \gls{RIS}s' phase shift constraints was explored in~\cite{sun2023joint}. The authors proposed several optimization techniques for the individual subproblems. First, an Akaike information criterion-based diagonalization method was designed to estimate the unknown jamming covariance matrix. Second, a double deterministic transformation method was proposed to tackle the cascaded angular uncertainties. Third, a two-layer iterative Lagrange multiplier algorithm was developed to obtain the globally optimal solution of the digital precoder. Finally, a polyblock-based multiple penalty method was designed to obtain the solutions to \gls{RIS}s' phase shifts. Focusing on \gls{SR} maximization, the authors in~\cite{alexandropoulos2023counteracting} proposed an algorithm to optimize the BS precoding matrix, the number of data streams, the AN covariance matrix, the receive combining matrices for the legitimate user and eavesdropper, and the \gls{RIS} phase shifts for the legitimate and malicious \gls{RIS}s under constraints on undesired amplification of thermal noise, power budget, and \gls{RIS} phase shifts constraint. The designed algorithm was based on \gls{AO}, \gls{MM}, and manifold optimization. In~\cite{lu2020robust}, the authors proposed a scheme based on \gls{AO} and \gls{SDR} techniques to optimize the BS \textcolor{black}{\gls{BF}} matrix and the \gls{RIS} phase shifts under power and unit-modulus constraints to maximize worst-case \gls{SR}.   

In~\cite{chen2019intelligent}, the authors explored a minimum \gls{SR} maximization problem to optimize BS \textcolor{black}{\gls{BF}} vector and \gls{RIS} phase shifts under power budget and \gls{RIS} phase shifts constraint. The proposed solution involved \gls{AO} to deal with the coupled optimization variables and the path-following algorithm to handle the non-convexity of the objective function. A \gls{SR} maximization algorithm that optimizes the covariance matrix of the BS and the \gls{RIS} phase shifts subject to transmit power and \gls{RIS} phase shifts constraint was explored in~\cite{shen2019secrecy}. By exploiting \gls{AO} and the bisection search method, the authors developed closed-form and semi-closed-form solutions for the \gls{BS} transmit covariance and the \gls{RIS} phase shift matrix, respectively.  An algorithm based on \gls{AO} and \gls{SDR} techniques to maximize \gls{SR} by optimizing the BS \textcolor{black}{\gls{BF}} vector and \gls{RIS} phase shifts under the \gls{RIS} unit modulus constraints and BS power budget was proposed in~\cite{cui2019secure}. In~\cite{keming2021physical}, the authors developed a \gls{SR} maximization algorithm based on fractional programming and manifold optimization to optimize the \gls{BS} \textcolor{black}{\gls{BF}} vector and \gls{RIS} phase shifts under BS power budget and \gls{RIS} phase shifts constraint. In~\cite{zhou2021secure}, the authors investigated \gls{SR} maximization while ensuring the transmit power constraint on the BS \textcolor{black}{\gls{BF}} vector and the unit modulus constraint on the \gls{RIS} phase shifts. An \gls{AO}-based minimum \gls{SR} maximization algorithm that exploits the \gls{SCA} and \gls{SDR} techniques to optimize BS \textcolor{black}{\gls{BF}} vector and \gls{RIS} phase shifts under power budget and \gls{RIS} phase shifts constraint was proposed in~\cite{liu2022secure}.

In~\cite{li2021secure}, the authors investigated a max-min problem regarding \gls{SR} by optimizing the BS \textcolor{black}{\gls{BF}} matrix, \gls{RIS} phase shifts, \gls{AN} matrix, and the \gls{RIS}-user association matrix under constraints on BS power budget, \gls{RIS} phase shifts, and that each user should be served by one \gls{RIS}. A \gls{SR} maximization algorithm based on the difference of convex programming, \gls{SCA}, and manifold optimization to optimize \gls{BS} \textcolor{black}{\gls{BF}} vector and \gls{RIS} phase shifts under power and \gls{RIS} phase shifts constraint was explored in~\cite{xiu2020secure}. Aiming to maximize \gls{SR} in a \gls{UAV} and \gls{RIS} assisted \gls{mmWave} wireless network, the optimization of  \gls{UAV} \gls{BS} and \gls{RIS} positions, the \gls{UAV} \gls{BS} \textcolor{black}{\gls{BF}} vector, and the \gls{RIS} phase shift
was investigated in~\cite{sun2021intelligent}. To tackle this non-convex problem, the \gls{AO} approach, \gls{SDR} technique, and the exhaustive search method were exploited to propose a solution method. A logarithmic barrier method-based \gls{SR} maximization algorithm was proposed to optimize the transmit covariance matrix and \gls{RIS} phase shift under power budget and \gls{RIS} phase shifts constraints in~\cite{du2020reconfigurable}. Focusing on maximizing the worst-case \gls{SR} and weighted sum \gls{SR} by optimizing BS \textcolor{black}{\gls{BF}} vector, \gls{RIS} phase shifts and amplification coefficients under \gls{BS} and \gls{RIS} power budgets, maximum \gls{RIS} amplification coefficient, and \gls{SOP} constraints, two \gls{AO}-based algorithms were proposed in~\cite{dong2023robust}. In~\cite{zhang2021securing}, the authors maximized minimum \gls{SR} by designing BS \textcolor{black}{\gls{BF}} vector and \gls{RIS} phase shifts subject to transmit power constraint at the BS, phase shifts constraints of the \gls{RIS}, the \gls{SIC} decoding constraints and the \gls{SOP} constraints for single-antenna and multi-antenna BS scenarios. For the single-antenna \gls{BS} scenario, the authors derived the exact \gls{SOP} in closed-form expressions and proposed a ring-penalty-based \gls{SCA} algorithm. In addition, a Bernstein-type inequality approximation based \gls{AO} algorithm that uses the Dinkelbach method and the ring-penalty based \gls{SCA} algorithm was proposed for the multi-antenna BS case. The study in~\cite{han2023secure} considered a maximum \gls{SR} problem and proposed an \gls{SCA}-based \gls{AO} algorithm to design the \gls{BS} \textcolor{black}{\gls{BF}} vector and the reflective and transmissive phase shift vectors of \gls{STAR}-\gls{RIS} under constraints on the minimum \gls{SINR} of legitimate users, \gls{BS} maximum transmit power, \gls{SIC} decoding order, and the reflective and transmissive coefficients of the \gls{RIS}.

In~\cite{zhang2023robust}, the authors investigated sum \gls{SR} maximization by optimizing the BS transmit \textcolor{black}{\gls{BF}} vector and the \gls{RIS} phase shifts under constraints on minimum \gls{SINR} requirements for legitimate users, maximum transmit power, \gls{SIC} decoding condition, and the \gls{RIS} phase shifts. An \gls{AO}-based solution was proposed that uses the multi-dimensional quadratic transform method and the \gls{SDR} technique to transform the non-convex objective functions into convex forms and the \gls{SCA} algorithm to deal with the non-convex constraints. Optimization techniques to maximize the average \gls{SR} by adjusting the \gls{BS} \textcolor{black}{\gls{BF}} vector and the \gls{RIS} phase shift under transmit power budget and \gls{RIS} phase shifts constraints were studied in~\cite{shu2022beamforming}. An iterative Khun-Munkres algorithm was proposed in~\cite{sun2022optimization} to maximize the sum secrecy rate in an \gls{RIS}-aided \gls{VLC} system by optimizing the association of \glspl{LED} to \gls{RIS} elements. In~\cite{wei2023secure}, the authors optimized the hovering position of UAV, the transmit \textcolor{black}{\gls{BF}} of AP, and the phase shift matrix of \gls{RIS} to maximize the worst-case sum secrecy rate. An \gls{AO}-based algorithm whereby the non-convex hovering position optimization sub-problem is transformed into a convex problem by the \gls{SCA} and the \textcolor{black}{\gls{BF}} and \gls{RIS} phase shift optimization sub-problems are converted into rank-constrained problems by the \gls{SCA} and tackled by \gls{SDR} is developed. A secrecy rate maximization problem was formulated in a UAV-mounted multi-functional-\gls{RIS} assisted secure communication system by optimizing BS \textcolor{black}{\gls{BF}}, \gls{RIS} phase shifts, and the 3D location of the UAV-enabled \gls{RIS} in~\cite{wang2023uav}. Under the consideration of BS and UAV power budgets and \gls{RIS} phase shift constraints, an \gls{AO} algorithm that exploits \gls{SCA} and the sequential rank-one constraint relaxation method was proposed. In~\cite{tang2023robust}, the authors proposed a two-layer worst-case secrecy rate maximization algorithm for a UAV-enabled \gls{RIS} and fixed \gls{RIS} and secured communication system. The inner layer optimizes \gls{RIS} phase shifts and jamming by the aerial \gls{RIS} via the \gls{BCD} framework while the outer layer solves the \gls{UAV} deployment problem using \gls{DRL} technique. An \gls{AO} algorithm based on fractional programming and \gls{SCA} was developed in~\cite{pang2021irs} to maximize secrecy rate via the optimization of \gls{BS} \textcolor{black}{\gls{BF}} vector, trajectory of \gls{UAV}, and \gls{RIS} phase shifts. In~\cite{rafieifar2023secure}, the authors proposed an algorithm that exploits continuous relaxation, \gls{BCD} approach, \gls{SCA}, and penalizing methods to maximize the minimum secrecy rate by optimizing BS \textcolor{black}{\gls{BF}} vector and RIS phase shifts matrix under BS power budget and RIS phase constraints. The optimization problem to maximize the worst-case secrecy rate by designing \gls{UAV}'s trajectory, \gls{RIS} phase shifts, and \gls{UAV} and \gls{BS} transmit power under constraints on \gls{UAV} trajectory, \gls{RIS} phase shifts and transmit power budgets was considered and a solution based on \gls{SCA} and \gls{SDR} was proposed in~\cite{li2021robust}. To achieve the maximum worst-case secrecy rate via optimizing power allocation, \gls{RIS} phase shifts, and \gls{UAV} trajectory, an \gls{AO}-based algorithm that leverages \gls{SDR} and \gls{SCA} was investigated in~\cite{liu2022ris}.

A particle swarm optimization-based algorithm was proposed in~\cite{qian2021secure} to optimize the orientation of \gls{RIS} mirrors and maximize the secrecy rate. An algorithm that exploits fractional programming and manifold optimization to optimize \gls{BS} \textcolor{black}{\gls{BF}} and \gls{RIS} phase shifts for secrecy rate maximization was proposed in~\cite{feng2020physical}. An iterative algorithm based on \gls{SDR} for optimizing RIS phase shift and \gls{BS} power allocation to maximize secrecy rate under the constraints of unit modulus and total power is developed in~\cite{gong2023joint}. In~\cite{hoang2023secrecy}, the authors proposed two \gls{AO}-based algorithms to maximize secrecy rate by optimizing the \textcolor{black}{\gls{BF}} vector of a ground \gls{BS} and phase shifts of a pair of \gls{RIS} under power budget and RIS phase constraints. In~\cite{niu2022active}, the authors focused on maximizing the secrecy rate subject to the transmit power constraint and interference threshold by proposing an \gls{AO} scheme to optimize the beamformer and \gls{AN} at the BS and the \gls{RIS} phase shifts. In~\cite{yuan2022secure}, the authors developed an algorithm to optimize \gls{RIS} phase shifts by using the \gls{SDR} method. A secrecy rate maximization problem via the optimization of radar receive beamformers, \gls{RIS} phase shifts, and radar transmit beamformers under minimum radar detection \gls{SNR} and total system power budget constraints was studied in~\cite{salem2022active}. The authors tackled this problem by exploiting \gls{MM} and fractional programming techniques to achieve a solution. In~\cite{cheng2023irs}, the authors focused on maximizing the secrecy rate by optimizing \gls{BS} transmit precoding matrices, \gls{RIS} phase shifts matrices, and \gls{UAV} trajectory under constraints on \gls{UAV} flying distance, maximum transmit power, and \gls{RIS} phase shifts. For this design problem, an \gls{AO}-based procedure where \gls{SDP} is used to solve the transmit precoding and RIS phase shift matrices while SCA is used to obtain the UAV trajectory was proposed.

An \gls{SCA}-based procedure to maximize secrecy rate by optimizing BS and UAV transmit powers, UAV trajectory, transmit and reflect phase shifts of RIS, and power splitting factor at the legitimate receiver under UAV trajectory constraints, power budgets, power splitting and energy harvesting constraints, and RIS phase constraints was examined in~\cite{benaya2023physical}. The study in~\cite{yuan2022secure} focused on maximizing the secrecy rate by optimizing BS \textcolor{black}{\gls{BF}} vector and RIS phase shifts under transmit power budget and RIS phase shift constraint. An optimization problem to maximize ergodic secrecy rate by optimizing BS transmit and receive beamformer, the covariance matrix of the \gls{AN}, and RIS phase shift matrix was considered and solved by exploiting \gls{SCA} and Rayleigh-quotient techniques in~\cite{xing2023reconfigurable}. In~\cite{han2023broadcast}, the authors investigated secrecy rate maximization by optimizing \gls{UAV} \gls{BS} \textcolor{black}{\gls{BF}} vector, \gls{UAV} trajectory, and the \gls{RIS} phase shifts with constraints on power budgets, \gls{UAV} trajectory, and \gls{RIS} phase shifts. The proposed \gls{AO}-based algorithm involved \gls{SDP} and \gls{SCA} techniques for beamformer and RIS phase optimization and a \gls{DRL} scheme for the trajectory optimization. An alternating algorithm to optimize \gls{RIS} phase shifts and \textcolor{black}{\gls{BF}} vector at the \gls{BS} under constraints on total transmit power and \gls{RIS} phase shifts was developed in~\cite{wang2022multicast}.  In~\cite{abumarshoud2022intelligent}, the authors proposed an algorithm to maximize secrecy rate of the trusted user under minimum rate requirements for the untrusted user and transmit power budget by optimizing \gls{NOMA} power allocation and \gls{RIS} allocation matrix. In~\cite{hu2023securing}, the authors studied the optimization of \gls{D2D} transmit power and \gls{RIS} phase shifts to maximize secrecy rate while guaranteeing transmit power budgets and \gls{RIS} phase constraints. An optimization problem to maximize secrecy rate by optimizing \gls{BS} power allocation and \gls{RIS} phase shifts subject to the total transmit power budget, minimal achievable rate requirements, and \gls{RIS} phase shifts was examined and solved in~\cite{xu2023sum}.

%Table:opt_tech2

\begin{table*}[!t]
\centering
\vspace*{-2mm}
\caption{\textcolor{black}{Summary of secrecy metrics optimization (other than secrecy rate maximization) related literature for \gls{RIS}-Assisted \gls{PLS} systems}}
\vspace*{-3mm}
\label{Table:opt_tech3}
\resizebox{\textwidth}{!}{%
\begin{tabular}{|l|l|l|c|l|l|}
\hline
\textbf{[\#]} &
\textbf{\begin{tabular}[c]{@{}l@{}}Phase-shift\\ resolution\end{tabular}} &
  \textbf{Optimization variables} &
  \textbf{Objective} &
  \textbf{Optimization techniques} &
  \textbf{Constraints} 
\\ \hline

\cite{zhao2022secrecy} &Continuous   &\begin{tabular}[c]{@{}l@{}} BS beamforming vector, RIS phase shifts\end{tabular}& \multirow{27}{*}{Maxmin SR} & Stochastic SCA  & \begin{tabular}[c]{@{}l@{}}Energy harvesting and Power constraints \end{tabular}\\ \cline{1-3} \cline{5-6}

\cite{chen2019intelligent} &\begin{tabular}[c]{@{}l@{}}Continuous/\\Discrete\end{tabular}  &\begin{tabular}[c]{@{}l@{}}BS beamforming vector, RIS phase shifts  \end{tabular}&  &\begin{tabular}[c]{@{}l@{}} AO, Path-following algorithm \end{tabular} & \begin{tabular}[c]{@{}l@{}} Power and RIS phase shifts constraints\end{tabular}\\ \cline{1-3} \cline{5-6}

\cite{lu2020robust} & Continuous  &\begin{tabular}[c]{@{}l@{}}BS beamforming vector, RIS phase shifts  \end{tabular}&  &\begin{tabular}[c]{@{}l@{}} AO, SDR  \end{tabular} & \begin{tabular}[c]{@{}l@{}} Power and RIS phase shifts constraints\end{tabular}\\ \cline{1-3} \cline{5-6}

\cite{li2021secure} & Continuous  &\begin{tabular}[c]{@{}l@{}}BS beamforming matrix, RIS phase shifts,\\ AN matrix, RIS assignment matrix  \end{tabular}&  &\begin{tabular}[c]{@{}l@{}} AO, SDR, SCA \end{tabular} & \begin{tabular}[c]{@{}l@{}} RIS-user single connectivity, power,\\ and RIS phase shifts constraints\end{tabular}\\ \cline{1-3} \cline{5-6}

\cite{liu2022secure} & Continuous  &\begin{tabular}[c]{@{}l@{}}BS beamforming vector, RIS phase shifts  \end{tabular}&  &\begin{tabular}[c]{@{}l@{}} AO, SCA, SDR \end{tabular} & \begin{tabular}[c]{@{}l@{}} Power and RIS phase shifts constraints\end{tabular}\\ \cline{1-3} \cline{5-6}

\cite{dong2023robust} & Continuous  &\begin{tabular}[c]{@{}l@{}}BS beamforming vector, RIS amplification\\ and phase shifts matrix \end{tabular}&  &\begin{tabular}[c]{@{}l@{}} AO, Dual-SCA, SDR, Dinkelbach \\ method, Bisection search \end{tabular} & \begin{tabular}[c]{@{}l@{}} BS and RIS reflecting power budget, maximum \\amplification coefficient and SOP constraints\end{tabular}\\\cline{1-3} \cline{5-6}

\cite{wei2023secure} & Continuous  &\begin{tabular}[c]{@{}l@{}}UAV position, AP transmit beamforming \\ phase shifts matrix \end{tabular}&  &\begin{tabular}[c]{@{}l@{}} AO, SCA, SDR \end{tabular} & \begin{tabular}[c]{@{}l@{}}  Min. rate requirement of legitimate users, LoS\\ connection requirement, power budget, RIS phase \\constraints, UAV position constraints\end{tabular}\\\cline{1-3} \cline{5-6}

\cite{tang2023robust} &\begin{tabular}[c]{@{}l@{}}Continuous\end{tabular}  &\begin{tabular}[c]{@{}l@{}}UAV position, covariance of jamming signal\\ by UAV RIS, fixed RIS and UAV RIS phase\\ shifts matrix \end{tabular}&  &\begin{tabular}[c]{@{}l@{}} AO, BCD, SDP, SCA, DRL \end{tabular} & \begin{tabular}[c]{@{}l@{}}Jamming Power budget, RIS phase constraints, \\UAV position constraints\end{tabular}\\\cline{1-3} \cline{5-6}

\cite{liu2022ris} & \begin{tabular}[c]{@{}l@{}}Continuous\end{tabular}  &\begin{tabular}[c]{@{}l@{}}BS Power allocation, RIS phase shifts, \\RIS-UAV trajectory \end{tabular}&  &\begin{tabular}[c]{@{}l@{}} AO, SCA, SDR  \end{tabular} & \begin{tabular}[c]{@{}l@{}}BS Power budget, RIS phase constraints, \\ UAV position constraints\end{tabular}\\\cline{1-3} \cline{5-6}

\cite{li2021robust} & Continuous  &\begin{tabular}[c]{@{}l@{}}UAV BS transmit power, RIS phase shifts, \\ UAV trajectory \end{tabular}&  &\begin{tabular}[c]{@{}l@{}} AO, SCA, SDR \end{tabular} & \begin{tabular}[c]{@{}l@{}}BS and legitimate user power budget, RIS phase \\constraints, UAV position constraints\end{tabular}\\\cline{1-3} \cline{5-6}

\cite{rafieifar2023secure} & Continuous  &\begin{tabular}[c]{@{}l@{}} BS beamforming vector, RIS phase shifts \end{tabular}&  &\begin{tabular}[c]{@{}l@{}} BCD, SCA, SDP \end{tabular} & \begin{tabular}[c]{@{}l@{}}BS  power budget, RIS phase constraints\end{tabular}\\\cline{1-3} \cline{5-6}

\textcolor{black}{\cite{shen2023outage}} & \textcolor{black}{Continuous} & \begin{tabular}[c]{@{}l@{}} \textcolor{black}{AP transmit beamforming, AP AN} \\ \textcolor{black}{covariance matrix, RIS phase shifts,} \\ \textcolor{black}{User power allocation} \end{tabular} & & \begin{tabular}[c]{@{}l@{}} \textcolor{black}{AO, SDR, SCA} \end{tabular} & \begin{tabular}[c]{@{}l@{}} \textcolor{black}{AP power budget, user minimum energy} \\ \textcolor{black}{harvesting requirement, SOP constraint,} \\ \textcolor{black}{non-negativity of AN convariance matrix,} \\ \textcolor{black}{RIS phase shift}\end{tabular}\\\cline{1-3} \cline{5-6}

\textcolor{black}{\cite{zhou2024securing}} & \textcolor{black}{Continuous} &\begin{tabular}[c]{@{}l@{}} \textcolor{black}{Transmit power allocation, receiver filter,} \\ \textcolor{black}{RIS amplification matrix, RIS phase shifts} \end{tabular}&  & \begin{tabular}[c]{@{}l@{}} \textcolor{black}{AO algorithm} \end{tabular} & \begin{tabular}[c]{@{}l@{}} \textcolor{black}{User and RIS transmit power thresholds} \end{tabular}\\\cline{1-3} \cline{5-6}

\cite{zhang2021securing} & Continuous  &\begin{tabular}[c]{@{}l@{}}BS beamforming vector, RIS phase shifts \end{tabular}&  &\begin{tabular}[c]{@{}l@{}} AO, Ring-penalty based SCA, \\ Dinkelbach method \end{tabular} & \begin{tabular}[c]{@{}l@{}}  BS power budget, RIS phase shifts, \\users' SIC decoding,  and SOP constraints\end{tabular}\\
\hline

\cite{sun2022ris} & Continuous   &\begin{tabular}[c]{@{}l@{}} Receive decoder, digital precoder, analog \\ precoder, artificial noise \end{tabular}& Maxmin rate &\begin{tabular}[c]{@{}l@{}} AO, MM, QCQP, Riemannian \\ manifold optimization \end{tabular} & \begin{tabular}[c]{@{}l@{}}Minimum  user and eavesdropper rates,\\ transmit power budget and RIS phase constraint \end{tabular}\\ \hline

\cite{cai2022symbol} & Continuous  & \begin{tabular}[c]{@{}l@{}}RIS phase shifts, BS precoding vector  \end{tabular}  & Max. SINR  & \begin{tabular}[c]{@{}l@{}}  BCD, Brute force, Penalty method \end{tabular}  &\begin{tabular}[c]{@{}l@{}}Interference constraints, Power and \\RIS phase constraints \end{tabular} \\ \hline

\cite{elhoushy2021exploiting} &Continuous   &\begin{tabular}[c]{@{}l@{}} BS transmit power, RIS phase shifts\end{tabular}& Min. SINR &\begin{tabular}[c]{@{}l@{}} AO, SDR, Dinkelbach method \end{tabular} & \begin{tabular}[c]{@{}l@{}}Minimum SINR, power, and RIS phase constraints \end{tabular}\\ \hline

\cite{wang2023zero}  & Continuous &  \begin{tabular}[c]{@{}l@{}} BS beamforming vector, RIS phase shifts\end{tabular}& \multirow{3}{*}{Max. SNR}  & \begin{tabular}[c]{@{}l@{}} AO, CCP, SDR \end{tabular} & \begin{tabular}[c]{@{}l@{}}Power, SNR,  and RIS phase constraints \end{tabular}\\ \cline{1-3} \cline{5-6}

\cite{ahmed2022joint} & Discrete &  \begin{tabular}[c]{@{}l@{}} BS precoding vector, RIS phase shifts\end{tabular} &  & AO, SDR & \begin{tabular}[c]{@{}l@{}}Power budget, eavesdropper SNR constraint \end{tabular}\\ \hline

\cite{chu2023joint} & Continuous  &  \begin{tabular}[c]{@{}l@{}}BS beamforming vector, RIS phase shifts, \\ radar receive filter    \end{tabular}& \begin{tabular}[c]{@{}l@{}} Max. radar  SNR \end{tabular} & \begin{tabular}[c]{@{}l@{}} BCD,SDR, MM, Quadratic\\ transformation   \end{tabular}& \begin{tabular}[c]{@{}l@{}}Communication QoS, secure transmission rate, \\ power budget and RIS phase shifts constraint  \end{tabular}\\ \hline

\cite{peng2023robust}  & Continuous &  \begin{tabular}[c]{@{}l@{}} BS beamforming vector, RIS phase shifts\end{tabular}& \begin{tabular}[c]{@{}l@{}} Maxmin approximate\\ ergodic SR \end{tabular}   & \begin{tabular}[c]{@{}l@{}} AO, CCP, SDR, MM, \\Quadratic transform, BCD, SOCP \end{tabular} & \begin{tabular}[c]{@{}l@{}}Power and RIS phase constraints \end{tabular}\\ \hline

\cite{wei2023adversarial}  & Continuous &  \begin{tabular}[c]{@{}l@{}} RIS phase shifts\end{tabular}& \begin{tabular}[c]{@{}l@{}} Max. variance of \\Eve-RIS channel \end{tabular}   &  SDR & \begin{tabular}[c]{@{}l@{}}RIS phase shifts constraints \end{tabular}\\ \hline

\cite{lu2023joint}  & Continuous &  \begin{tabular}[c]{@{}l@{}} BS precoding matrix, RIS phase shifts\end{tabular}& \begin{tabular}[c]{@{}l@{}} Max. secret key rate \end{tabular}   & \begin{tabular}[c]{@{}l@{}} Water-filling algorithm, Bisection\\ search,  Grassmann manifold\\ optimization\end{tabular} & \begin{tabular}[c]{@{}l@{}}RIS unit-module constraint, channel measurement \\constraints, power budget \end{tabular}\\ \hline

\cite{chu2019intelligent} &\begin{tabular}[c]{@{}l@{}}Continuous\end{tabular}  &\begin{tabular}[c]{@{}l@{}}BS BF vector, RIS phase shifts  \end{tabular}& \multirow{4}{*}{Min. power}  &\begin{tabular}[c]{@{}l@{}} AO, SDR \end{tabular} & \begin{tabular}[c]{@{}l@{}} Minimum rate requirement, and RIS phase shifts  \end{tabular}\\ \cline{1-3} \cline{5-6}

\textcolor{black}{\cite{hong2023outage}} &Continuous   &\begin{tabular}[c]{@{}l@{}} \textcolor{black}{BS transmit beamforming, AN spatial distribution at the BS,}\\ \textcolor{black}{RIS phase shifts}\end{tabular}&  &\begin{tabular}[c]{@{}l@{}}\textcolor{black}{AO, SDR, penalty CCP method} \end{tabular} & \begin{tabular}[c]{@{}l@{}} \textcolor{black}{Minimum rate requirements,}\\ \textcolor{black} {Outage probability limitation}  \end{tabular}\\ \cline{1-3} \cline{5-6}

%\textcolor{black}{\cite{hong2023outage}} &  &\begin{tabular}[c]{@{}l@{}}  \end{tabular}&\begin{tabular}[c]{@{}l@{}}  \end{tabular} & \begin{tabular}[c]{@{}l@{}} \end{tabular}\\\hline

\cite{wu2022joint} &Continuous   &\begin{tabular}[c]{@{}l@{}} BS BF vector, RIS phase shifts\end{tabular}&  &\begin{tabular}[c]{@{}l@{}} SOCP \end{tabular} & \begin{tabular}[c]{@{}l@{}} Minimum rate requirements, Interference \\threshold, and RIS phase  \end{tabular}\\ \cline{1-3} \cline{5-6}

\cite{lv2022ris} &Continuous   &\begin{tabular}[c]{@{}l@{}} BS BF vector, RIS phase shifts\end{tabular}&  &\begin{tabular}[c]{@{}l@{}} AO, SCA  \end{tabular} & \begin{tabular}[c]{@{}l@{}} SR and Power \end{tabular}\\ \hline

\cite{cheng2023ris} &Continuous   &\begin{tabular}[c]{@{}l@{}} BS BF vector, RIS phase shifts\end{tabular}& \begin{tabular}[c]{@{}l@{}} Min. power, \\ Max. SR \end{tabular} &\begin{tabular}[c]{@{}l@{}} Riemannian gradient-\\based method \end{tabular} & \begin{tabular}[c]{@{}l@{}}Power budget, Minimum SR and RIS phase  \end{tabular}\\ \hline

\cite{niu2022efficient} &\begin{tabular}[c]{@{}l@{}}Continuous\end{tabular}  &\begin{tabular}[c]{@{}l@{}}BS BF vector, RIS phase shifts  \end{tabular}& Max. power &\begin{tabular}[c]{@{}l@{}} Alternating direction algorithm \end{tabular} & \begin{tabular}[c]{@{}l@{}} Power, minimum SNR, and RIS phase shifts \end{tabular}\\ \hline

\cite{xiu2020secure} &\begin{tabular}[c]{@{}l@{}}Continuous\end{tabular}  &\begin{tabular}[c]{@{}l@{}}\gls{BS} BF vector, RIS phase shifts \end{tabular}& \begin{tabular}[c]{@{}l@{}} Max. AN power,\\ Max. SR \end{tabular}&\begin{tabular}[c]{@{}l@{}}  difference of convex programming,\\ SCA Manifold optimization \end{tabular} & \begin{tabular}[c]{@{}l@{}} Power, RIS phase shifts and minimum rate \end{tabular}\\ \hline

\cite{mao2022reconfigurable} &Continuous   &\begin{tabular}[c]{@{}l@{}} IoT transmit power, RIS phase shifts,\\ computing frequencies, time assignment\end{tabular}& \begin{tabular}[c]{@{}l@{}} Maxmin  computation\\ efficiency \end{tabular} &\begin{tabular}[c]{@{}l@{}} AO, SCA, BCD,\\ Dinkelbach method  \end{tabular} & \begin{tabular}[c]{@{}l@{}} SR, Minimum computation requirements, Power,\\ Time, Computing frequencies, and RIS phase  \end{tabular}\\ \hline

\cite{li2021intelligent} &Continuous   &\begin{tabular}[c]{@{}l@{}} BS receive BF vector, AN covariance matrix,\\ RIS phase shifts, users' offloading time \\transmit power, local computation tasks\end{tabular}& \begin{tabular}[c]{@{}l@{}} Min. secure  energy \end{tabular} & \begin{tabular}[c]{@{}l@{}} AO, SDR, \\ Dinkelbach Method \end{tabular}  & \begin{tabular}[c]{@{}l@{}} Task offloading, BS and users' Power, Delay,\\ Computation, BF, and RIS phase\end{tabular}\\ \hline

\cite{wang2020energy} &\begin{tabular}[c]{@{}l@{}}Continuous\end{tabular}  &\begin{tabular}[c]{@{}l@{}}BS BF vector, Jamming BF vector,\\  RIS phase shifts  \end{tabular}& \multirow{4}{*}{\begin{tabular}[c]{@{}l@{}} Max. secure EE  \end{tabular}} &\begin{tabular}[c]{@{}l@{}} AO, SDR,  Dinkelbach method  \end{tabular} & \begin{tabular}[c]{@{}l@{}} Minimum rate requirement, power and RIS phase \\shifts\end{tabular}\\ \cline{1-3} \cline{5-6}

\textcolor{black}{\cite{bao2023secrecy}} & \textcolor{black}{Continuous} &\begin{tabular}[c]{@{}l@{}} \textcolor{black}{Active RIS reflecting matrix,}\\ \textcolor{black} {BS beamforming vector} \end{tabular}&  &\begin{tabular}[c]{@{}l@{}} \textcolor{black}{SCA, Lagrange dual method} \end{tabular} & \begin{tabular}[c]{@{}l@{}}  \textcolor{black}{Min. SR threshold, BS power budget,}\\ \textcolor{black} {RIS amplitude constrain} \end{tabular}\\ \cline{1-3} \cline{5-6}

\cite{li2023enhancing}  & continuous &\begin{tabular}[c]{@{}l@{}} Transmit covariance matrix, RIS phase shifts, \\and redundancy rate \end{tabular}&  &\begin{tabular}[c]{@{}l@{}} AO, concave-convex procedure \end{tabular} & \begin{tabular}[c]{@{}l@{}}  SOP, Power and RIS phase shifts \end{tabular}\\ \cline{1-3} \cline{5-6}

\cite{li2022secure} &\begin{tabular}[c]{@{}l@{}}Continuous/\\Discrete\end{tabular}  &\begin{tabular}[c]{@{}l@{}}BS BF vector, RIS phase shifts, \\ and transmission rate  \end{tabular}&  &\begin{tabular}[c]{@{}l@{}} AO, Path-following procedure,\\ Quadratic transformation  \end{tabular} & \begin{tabular}[c]{@{}l@{}} SOP, Power and RIS phase shifts \end{tabular}\\ \hline

\cite{wang2023star}  & Continuous &\begin{tabular}[c]{@{}l@{}} Transmission waveform signal, BS BF vector,\\ RIS transmit and reflective phase shifts   \end{tabular} & \begin{tabular}[c]{@{}l@{}} Max. received radar\\ sensing power \end{tabular} &\begin{tabular}[c]{@{}l@{}}  Distance-majorization\\ based algorithm, AO \end{tabular}& \begin{tabular}[c]{@{}l@{}}  Signal waveform and peak-to-average-power ratio\\ constraints, amplitude on transmit and reflective \\RIS, constructive interference for communication \\users, security constraints for malicious radar targets\end{tabular} \\ \hline

%\cite{zhang2023irs}  &Continuous  &\begin{tabular}[c]{@{}l@{}} BS BF vector, BS AN covariance\\ matrix, RIS phase shift matrix,  \end{tabular}&\begin{tabular}[c]{@{}l@{}}Min. eavesdropper\\ data rate, Maxmin\\ communication SINR,\\ Max. communication\\ sum rate    \end{tabular}& \begin{tabular}[c]{@{}l@{}} Quadratic transformation,\\ Lagrangian dual transform.,\\ AO, Penalty-based method,\\Riemannian  gradient-\\based method, SDP  \end{tabular}& \begin{tabular}[c]{@{}l@{}} Radar similarity constraint,\\ RIS phase shifts constraint \end{tabular}\\ \hline

\cite{lyu2023robust}  &  Continous&\begin{tabular}[c]{@{}l@{}} BS BF vector, RIS phase shift \end{tabular}& \multirow{4}{*}{\begin{tabular}[c]{@{}l@{}} Min. BS \\ transmit power \end{tabular}} & \begin{tabular}[c]{@{}l@{}} AO, SDR, Sequential \\rank-one constraint relaxation \end{tabular} & \begin{tabular}[c]{@{}l@{}} Eve's worst-case SNR and SNR OP   \end{tabular}\\ \cline{1-3} \cline{5-6}

\cite{wang2023intelligent} & Continuous  & \begin{tabular}[c]{@{}l@{}} BS BF vector, RIS phase shift  \end{tabular}&  & BCD, SDP  &\begin{tabular}[c]{@{}l@{}} Minimum SINR requirements for legitimate\\ user, RIS phase shifts \end{tabular}\\ \hline

\cite{zhai2022improving} & Continuous  &\begin{tabular}[c]{@{}l@{}}Energy price, energy transfer time, energy\\ and information phases matrices, BF vector  \end{tabular} & Max. utility function  & \begin{tabular}[c]{@{}l@{}}Stackelberg game, AO,\\SDR, BCD, Golden\\ search method, SCA \end{tabular}  & \begin{tabular}[c]{@{}l@{}}Scheduling and RIS phase shifts constraints\\ and Power budget\end{tabular} \\ \hline

%\cite{liao2023intelligent} &Continuous  & \begin{tabular}[c]{@{}l@{}} BS BF vector, RIS phase shifts,\\ users' offloading time, users' transmit power,\\ users' local computation frequency  \end{tabular}& \begin{tabular}[c]{@{}l@{}}  Max. secure \\computation \end{tabular}&\begin{tabular}[c]{@{}l@{}} SDR, Lagrange duality theory   \end{tabular}& \begin{tabular}[c]{@{}l@{}} Energy harvesting constraint, RIS phase shifts,\\ constraint on energy harvesting and computation\\ time. \end{tabular}\\\hline

\cite{ngo2023physical} &Continuous  & \begin{tabular}[c]{@{}l@{}} Redundant rate, caching probability \end{tabular}& \begin{tabular}[c]{@{}l@{}}  Max. secure \\transmission\\ probability \end{tabular}&\begin{tabular}[c]{@{}l@{}} AO, Bisection search,\\ Exhaustive search   \end{tabular}& \begin{tabular}[c]{@{}l@{}}Caching probability and cache storage capacity \end{tabular}\\\hline

\cite{ge2022active} &Continuous  & \begin{tabular}[c]{@{}l@{}} Transmit power, RIS phase shifts \end{tabular} & \begin{tabular}[c]{@{}l@{}}Min. ground BS\\ transmit power  \end{tabular}&  \begin{tabular}[c]{@{}l@{}} AO, S-procedure \end{tabular}     & \begin{tabular}[c]{@{}l@{}} Min. secrecy rate requirements\end{tabular} \\ \hline

\cite{liu2022minimization} &Continuous  &\begin{tabular}[c]{@{}l@{}} BS BF vector, RIS phase shifts \end{tabular} & \begin{tabular}[c]{@{}l@{}}Min. SOP  \end{tabular} & \begin{tabular}[c]{@{}l@{}} AO, SDR, Manifold optimization \end{tabular}   & \begin{tabular}[c]{@{}l@{}} BS transmit power budget, RIS phase shifts \end{tabular} \\ \hline

\textcolor{black}{\cite{zhou2023secure}} & \textcolor{black}{Continuous} & \begin{tabular}[c]{@{}l@{}} \textcolor{black}{RIS phase shift, UAV
deployment,} \\ \textcolor{black}{power and computing resource allocation } \end{tabular} & \begin{tabular}[c]{@{}l@{}} \textcolor{black}{Max. Secure} \\ \textcolor{black}{computational tasks} \end{tabular} &\begin{tabular}[c]{@{}l@{}} \textcolor{black}{BCD method} \end{tabular} & \begin{tabular}[c]{@{}l@{}} \textcolor{black}{Non-negative SR, UAV and user power constraints,} \\ \textcolor{black}{UAV location, RIS phase shift} \end{tabular} \\\hline

\textcolor{black}{\cite{liu2024intelligent}} & \textcolor{black}{Continuous} &\begin{tabular}[c]{@{}l@{}}  \textcolor{black}{RIS phase shift} \end{tabular}& \textcolor{black}{Max. radar SINR} &\begin{tabular}[c]{@{}l@{}} \textcolor{black}{Manifold optimization} \end{tabular} & \begin{tabular}[c]{@{}l@{}} \textcolor{black}{RIS unit modulus constraints}\end{tabular}\\\hline

\textcolor{black}{\cite{xia2024joint}} & \textcolor{black}{Continuous} &\begin{tabular}[c]{@{}l@{}} \textcolor{black}{BS transmit waveform, RIS phase shifts}  \end{tabular}& 
\textcolor{black}{Max. sum rate} &\begin{tabular}[c]{@{}l@{}} \textcolor{black}{AO, SCA, manifold optimization} \end{tabular} & \begin{tabular}[c]{@{}l@{}} \textcolor{black}{Aerial Eve SNR and Cramer-Rao lower bound}\end{tabular}\\\hline

\textcolor{black}{\cite{de2024malicious}} & \textcolor{black}{Continuous} &\begin{tabular}[c]{@{}l@{}} \textcolor{black}{RIS phase shifts}  \end{tabular}& 
{\begin{tabular}[c]{@{}l@{}} \textcolor{black} {Max. inter-user}\\ \textcolor{black}{interference} \end{tabular}} &\begin{tabular}[c]{@{}l@{}} \textcolor{black}{Projected gradient-based algorithm} \end{tabular} & \begin{tabular}[c]{@{}l@{}} \textcolor{black}{RIS amplitude coefficients}\\ \textcolor{black} {RIS phase shifts}\end{tabular}\\\hline

\textcolor{black}{\cite{ye2023robust}} & \textcolor{black}{Continuous} &\begin{tabular}[c]{@{}l@{}} \textcolor{black}{UAV transmit power, UAV trajectory}\\ \textcolor{black}{RIS phase shifts}  \end{tabular}& 
{\begin{tabular}[c]{@{}l@{}} \textcolor{black} {Max. average}\\\textcolor{black} {secrecy rate} \end{tabular}} &\begin{tabular}[c]{@{}l@{}} \textcolor{black}{AO, Lagrangian dual transform,} \\ \textcolor{black}{Quadratic transform,S-procedure,}\\\textcolor{black} {SCA, SDP} \end{tabular} & \begin{tabular}[c]{@{}l@{}} \textcolor{black}{UAV's  mobility, UAV's power budget,}\\ \textcolor{black}{RIS amplitude constraint}\end{tabular}\\\hline

\end{tabular}
}
\vspace*{-3mm}
\end{table*}

\subsection{Other Secrecy Metrics for Optimization}
\subsubsection{Signal-to-Interference-Noise/ Signal-to-Noise Ratio Optimization}
In~\cite{cai2022symbol}, the authors studied \gls{SINR} maximization problem via \gls{RIS} phase shift and \gls{BS} precoding vector optimization. By utilizing the \gls{BCD} technique, a brute force-maximum likelihood detection method was proposed to obtain the BS precoding vector while a penalty method-based solution was developed for the \gls{RIS} phase shift optimization. Focusing on maximizing \gls{SNR} of the legitimate user, the authors in~\cite{wang2023zero} considered the joint optimization of BS \textcolor{black}{\gls{BF}} vector and \gls{RIS} phase shifts while considering constraints on the SNR of the eavesdropper, the \gls{RIS} phase shifts, and the maximum transmit power. An \gls{AO} solution method that uses the \gls{SDR} technique to solve the \gls{RIS} phase shift optimization subproblem and the \gls{SCA} and \gls{CCP} to solve the BS \textcolor{black}{\gls{BF}} optimization subproblem was developed. In~\cite{elhoushy2021exploiting}, the authors proposed  Dinkelbach-style algorithms to optimize the transmit power of \gls{BS}s and the \gls{RIS} phase shifts to minimize the \gls{SINR} of the eavesdroppers subject to constraints on the legitimate users' \gls{SINR}, \gls{BS} power, and \gls{RIS} phase shifts. In~\cite{ahmed2022joint}, the authors studied the optimization of \gls{RIS} phase shifts and sensor nodes precoding vector to maximize the \gls{SNR} at the fusion center under constraints on transmit power budgets for sensor nodes, \gls{RIS} phase shifts, and the \gls{SNR} at the eavesdropper.  

\subsubsection{Power Consumption Optimization}
In~\cite{cheng2023ris}, the authors proposed a \gls{BS} \textcolor{black}{\gls{BF}} vector and \gls{RIS} phase optimization scheme based on the Riemannian conjugate gradient method to minimize the total power consumed while considering the minimum \gls{SR} requirement. 
%Two AO algorithms based on SDR and SCA to obtain the minimum transmit power by optimizing the BS \textcolor{black}{\gls{BF}} vector and the RIS phase shift under SNR requirements and RIS phase constraints were developed in~\cite{han2021reconfigurable}.  
An alternating algorithm to optimize the \textcolor{black}{\gls{BF}} vectors for cognitive and primary \gls{BS}s and the phase shifts of an \gls{RIS} while considering rate requirements of primary and secondary users, interference thresholds, and power budgets of the BS was designed in~\cite{wu2022joint}. In~\cite{lv2022ris}, the authors proposed an alternating algorithm to optimize the beamformer vector at the BS and the \gls{RIS} phase shift under \gls{SR} constraints and power budget in order to minimize the total power consumption. A transmit power minimization algorithm based on \gls{AO} and \gls{SDR} techniques to optimize BS \textcolor{black}{\gls{BF}} vector and \gls{RIS} phase shifts subject to the \gls{SR} and \gls{RIS} phase shifts constraints was investigated in~\cite{chu2019intelligent}. In~\cite{niu2022efficient}, the authors proposed an alternating direction algorithm to optimize the \gls{BS} \textcolor{black}{\gls{BF}} vector and \gls{RIS} phase shifts under \gls{QoS}, transmit power, and \gls{RIS} phase shifts constraints. In~\cite{xiu2020secure}, an \gls{SCA}-based algorithm was proposed to maximize the \gls{AN} power by optimizing the \gls{BS} \textcolor{black}{\gls{BF}} vector, the \gls{RIS} phase shifts, and \gls{AN} power under transmit power and minimum user rate constraint. Focusing on minimizing the transmit power at the ground \gls{BS} while guaranteeing the achievable secrecy rate of the satellite user and the achievable rate of the terrestrial user, an \gls{AO}-based algorithm for optimizing BS precoding matrix and RIS phase matrix was proposed in~\cite{ge2022active}.

\subsubsection{Computation Efficiency Optimization}
In~\cite{mao2022reconfigurable}, the authors studied the optimization of time-slot assignment, \gls{RIS} phase shifts, local computing frequencies, and the transmit power of \gls{IoT} devices, with the objective of max-min computation efficiency, while guaranteeing the \gls{IoT} devices' secure computation rate and minimum computation requirements. A secure energy consumption minimization problem that optimizes the \gls{BS} receive \textcolor{black}{\gls{BF}} vectors, \gls{AN} covariance matrix, \gls{RIS} phase shifts, users' offloading time, transmit power, and local computation tasks was examined in~\cite{li2021intelligent}. To efficiently solve this non-convex problem, the authors decomposed the joint problem and adopted a \gls{SDR} algorithm to optimize the BS receive \textcolor{black}{\gls{BF}} vectors, AN covariance matrix, and \gls{RIS} phase shifts, and the Dinkelbach method to optimize the power and local computation tasks. The design constraints included task offloading time, users' maximum power constraint, maximum delay constraint, maximum local computation tasks constraint, \gls{BS} receive \textcolor{black}{\gls{BF}} vectors constraint, maximum power constraint for \gls{BS} transmitting \gls{AN}, and the phase shifts constraint of the \gls{RIS}. In~\cite{liao2023intelligent}, the design problem of maximizing the secure computation task bits of users by optimizing the \gls{AP} energy transmit \textcolor{black}{\gls{BF}}, the RIS phase shifts, users' uplink transmit power, users' offloading time, and the local computation frequency of users under energy harvesting, energy \textcolor{black}{\gls{BF}}, and RIS phase constraints and users' transmit power budget was examined. The authors proposed an alternating solution method in which \gls{SDP} and \gls{SDR} algorithms were used to optimize the energy transmit \textcolor{black}{\gls{BF}} of the \gls{AP} and the \gls{RIS} phase shifts, and the Lagrange duality method and Karush-Kuhn-Tucker condition were utilized to obtain the users' transmit power and computation time.

\subsubsection{Energy Efficiency Optimization}
An \textcolor{black}{\gls{EE}} maximization problem to optimize BS \textcolor{black}{\gls{BF}} vector, jamming \textcolor{black}{\gls{BF}} vector, and \gls{RIS} phase shifts under minimum \gls{SR} requirement, power budgets, and \gls{RIS} phase shift constraints was investigated in~\cite{wang2020energy}. To tackle the problem, the authors proposed an \gls{AO}-based algorithm that leverages the \gls{SDR} technique and Dinkelbach method. In~\cite{li2022secure}, the authors proposed an \gls{AO} algorithm that exploits a path-following procedure and quadratic transformation to maximize \textcolor{black}{\gls{EE}} by optimizing BS \textcolor{black}{\gls{BF}} vector, \gls{RIS} phase shift, and transmission rate under \gls{SOP}, \gls{BS} power and \gls{RIS} phase shifts constraint. In~\cite{li2023enhancing}, the authors studied secure \gls{EE} maximization by optimizing the \gls{BS} transmit covariance matrix, \gls{RIS} phase shifts, and the redundancy rate subject to \gls{SOP} constraint, \gls{RIS} phase shifts constraints, and transmit power constraint.

\subsubsection{Ergodic Secrecy Rate}
In~\cite{peng2023robust}, the authors investigated the optimization of the BS \textcolor{black}{\gls{BF}} vector and the phase shifts of the reflecting elements at the \gls{RIS} so as to maximize the weighted minimum approximate ergodic \gls{SR} subject to \gls{BS} transmit power constraint and \gls{RIS} phase shifts constraint. To solve this problem, the joint problem was first decoupled into two tractable subproblems, which were tackled alternatively via the \gls{BCD} method. Two different methods were proposed to solve the two subproblems. The first method exploited \gls{SOCP}, \gls{CCP}, and the quadratic transform while the second method was based on the \gls{MM} algorithm.

\subsubsection{Channel Gain}
An \gls{SDR} approach to maximize the variance of the Eve-\gls{RIS} generated deceiving channel by optimizing \gls{RIS} phase shift was investigated in~\cite{wei2023adversarial}.

\subsubsection{Secret Key Rate}
In~\cite{lu2023joint}, the authors investigated an \gls{RIS}-assisted key generation system by focusing on the design of precoding and phase shift matrices to fully exploit the randomness from the direct and cascaded channels. First, a water-filling algorithm was proposed to find the upper bound on the key rate of the system. In addition, the authors developed an algorithm by exploring the bisection search method and Grassmann manifold optimization to obtain the phase shift and precoding matrices such that the key rate approaches the upper bound. The authors in~\cite{li2021maximizing} proposed a multi-user secret key generation system that utilized RISs within the given environment. The channel model induced by RISs was characterized as the sum of products involving the channel from the transmitter to the RIS, the phase shifts introduced by the RIS, and the subsequent channel from the RIS to the receiver. A closed-form expression for the secret key rate was derived, providing a general analytical representation of the key generation process.

\subsubsection{Utility Function}
The study in~\cite{zhai2022improving} explored the integration of \gls{RIS} in a wireless-powered communication network, focusing on energy harvesting from a power station and secure information transmission to \gls{IoT} devices in the presence of eavesdroppers. In that paper, the authors proposed a \gls{RIS}-aided energy trading and secure communication scheme that employs a Stackelberg game model between the transmitter and power station where the transmitter optimizes the energy price, wireless energy transfer time, \gls{RIS} phase shift, and \textcolor{black}{\gls{BF}} vector to maximize its utility function. On the other hand, the power station adjusts the transmit power based on the energy price of the transmitter.

\subsubsection{Outage Probability}
In~\cite{liu2022minimization}, the authors first developed an expression of \textcolor{black}{\gls{SOP}} as a metric for RIS-assisted PLS system and then formulated a \textcolor{black}{\gls{SOP}} minimization problem that optimizes BS \textcolor{black}{\gls{BF}} vectors and RIS phase shift matrices under RIS phase constraints and BS transmit power budget. An AO-based procedure that involves a closed-form solution for the optimal \textcolor{black}{\gls{BF}} vector and two techniques for phase shifter matrix optimization (i.e., SDR-based and manifold-based methods) were proposed. An iterative scheme to maximize the secure transmission probability of an IRS-assisted cache-enabled space-terrestrial network by optimizing the transmission rates and caching probability under cache storage capacity limitations was explored in~\cite{ngo2023physical}. 

\subsection{Lessons Learnt}
This subsection has investigated various optimization frameworks for \gls{RIS}-assisted \gls{PLS} systems. A summary and the key points are presented below.
{\textcolor{black}{
\begin{itemize}
    \item  Most objective functions and constraint sets in \gls{RIS}-assisted \gls{PLS} systems are inherently non-convex and non-linear, involving highly coupled decision variables. This complexity is compounded when \gls{RIS} phase shift optimization is integrated with techniques like BS \textcolor{black}{\gls{BF}}, \gls{UAV} trajectory optimization, and \gls{RIS} location design. To address these challenges, various transformation and convexifying techniques are often employed. As an example, \gls{AO} and \gls{BCD} are used to decouple joint optimization variables into subproblems, \gls{SDR} is used to relax non-convex \gls{RIS} phase constraints, \gls{MM}, \gls{SCA}, and quadratic transform are used for non-convex and non-linear objective functions and constraints. Although these current approaches are effective, they often yield suboptimal solutions. In order to realize practical and optimal solutions, future research should focus on advanced global optimization techniques capable of handling coupled variables without decomposition, addressing non-convexity, and efficiently navigating large solution spaces. This shift could lead to more optimal and robust solutions for \gls{RIS}-assisted \gls{PLS} systems.
    \item Most studies in the literature focused on \gls{SR} optimization. However, 6G and beyond networks seek significant improvements in \gls{EE}, connection density, mobility management, transmission latency, and the support for sensing services. There is a pressing need to formulate novel objective functions and multi-objective optimization problems that concurrently address these diverse requirements. This holistic optimization strategy will ensure that \gls{RIS}-assisted \gls{PLS} systems align with the comprehensive performance targets of next-generation networks.
    \item It can be observed from Tables~IV and V that most existing works assume continuous, i.e., infinite resolution, phase shifts for \gls{RIS}, which is impractical for real-world implementations. There is a significant gap in understanding the impact of limited phase shifts on system performance, particularly on secrecy rates. Future studies should rigorously analyze the performance degradation due to discrete (finite-level) phase shifters and develop optimization techniques tailored to these realistic scenarios. This practical perspective is crucial for the deployment of \gls{RIS} in actual network environments. 
\end{itemize}
}}
%%%%%%%%%%%%%%%%%%%%%%%%%%%%%%%%%%%%%%%%%%%%%%%%%%%%%%%%%% 

\section{Machine Learning Techniques for RIS-Assisted PLS}
\label{Section: Machine Learning Techniques for RIS-Assisted PLS}

\begin{table*}[!t]
\centering
\vspace*{-6mm}
\caption{Summary of \gls{ML} related literature for \gls{RIS}-Assisted \gls{PLS} systems}
\vspace*{-3mm}
\label{Table:ML}
\resizebox{\textwidth}{!}{%
\begin{tabular}{|l|l|l|c|l|l|}
\hline
\textbf{[\#]} &
\textbf{\begin{tabular}[c]{@{}l@{}}Phase-shift\\ resolution\end{tabular}} &
  \textbf{Optimization variables} &
  \textbf{Objective} &
  \textbf{ML technique(s)} &
  \textbf{Constraints} 
\\ \hline

\cite{yang2020intelligent} & Continuous & \begin{tabular}[c]{@{}l@{}} BS transmit power, \\ RIS phase shifts\end{tabular} & \multirow{10}{*}{Max. SR} & \begin{tabular}[c]{@{}l@{}} Fast-policy hill-climbing RL algorithm \end{tabular} & \begin{tabular}[c]{@{}l@{}} Minimum SINR, transmit power  \end{tabular} \\ \cline{1-3} \cline{5-6} 

\cite{li2022reinforcement} & Continuous & \begin{tabular}[c]{@{}l@{}} BS beamforming vector, \\ RIS phase shifts  \end{tabular} &  & \begin{tabular}[c]{@{}l@{}} Deep Q-learning  algorithm \end{tabular} &  \begin{tabular}[c]{@{}l@{}} BS transmit power  \end{tabular}\\ \cline{1-3} \cline{5-6} 

\cite{peng2022deep} & Continuous & \begin{tabular}[c]{@{}l@{}} BS BF matrix, \\ RIS phase shifts \end{tabular} &  &\begin{tabular}[c]{@{}l@{}} DDPG RL algorithm \end{tabular}  & \begin{tabular}[c]{@{}l@{}} BS transmit power, RIS phase  \end{tabular} \\ \cline{1-3} \cline{5-6} 

\cite{saleem2022deep} & Continuous & \begin{tabular}[c]{@{}l@{}}  BS BF matrix, \\ RIS phase shifts\end{tabular} &  & \begin{tabular}[c]{@{}l@{}} DDPG RL algorithm \end{tabular} & \begin{tabular}[c]{@{}l@{}} Minimum SINR, BS transmit power, RIS phase  \end{tabular} \\ \cline{1-3} \cline{5-6} 

\cite{liu2023drl} & Continuous & \begin{tabular}[c]{@{}l@{}} BS BF, BS AN vector, \\ RIS phase shifts \end{tabular} &  & \begin{tabular}[c]{@{}l@{}} Soft actor-critic DRL algorithm \end{tabular} &  \begin{tabular}[c]{@{}l@{}} BS transmit power, maximum error \end{tabular}  \\ \cline{1-3} \cline{5-6} 

\cite{guo2021learning} & Continuous & \begin{tabular}[c]{@{}l@{}} 
UAV BF matrix, \\ UAV trajectory, \\ RIS phase shifts \end{tabular} &  & \begin{tabular}[c]{@{}l@{}} Twin-DDPG RL algorithm \end{tabular} & \begin{tabular}[c]{@{}l@{}} SOP  \end{tabular} \\ \cline{1-3} \cline{5-6} 

\cite{jiang2021aerial} & Continuous & \begin{tabular}[c]{@{}l@{}} ARIS deployment, \\ ARIS phase shifts\end{tabular} &  & \begin{tabular}[c]{@{}l@{}}Inner layer: semidefinite programming\\ and outer layer: deep  Q-learning algorithm \end{tabular} & \begin{tabular}[c]{@{}l@{}} ARIS location, ARIS phase   \end{tabular} \\ \cline{1-3} \cline{5-6} 

\cite{tang2023secure} & Continuous & \begin{tabular}[c]{@{}l@{}} RIS phase shifts, \\ UAV trajectory \end{tabular} & & \begin{tabular}[c]{@{}l@{}} Inner layer: manifold optimization and\\ outer layer: DRL algorithm \end{tabular} &  \begin{tabular}[c]{@{}l@{}} UAV flying waypoints and velocities, RIS phase  \end{tabular} \\ \cline{1-3} \cline{5-6} 

\cite{hoang2021ris} & Continuous & \begin{tabular}[c]{@{}l@{}} BS BF vector, \\ RIS phase shifts \end{tabular} &  & \begin{tabular}[c]{@{}l@{}} Unsupervised projection-based DNN algorithm \end{tabular} & \begin{tabular}[c]{@{}l@{}} BS transmit power, RIS phase  \end{tabular} \\ \cline{1-3} \cline{5-6} 

%\cite{thien2022secure} & Continuous & \begin{tabular}[c]{@{}l@{}} BS transmit power, \\ RIS phase shifts,\\ user power-splitting factors \end{tabular} & & \begin{tabular}[c]{@{}l@{}} DL-based\\ algorithm \end{tabular} &  \begin{tabular}[c]{@{}l@{}} BS transmit power,\\ required minimum \\ harvested energy \\ constraints \end{tabular}  \\ \cline{1-3} \cline{5-6}

\cite{saifaldeen2022dr} & N/A & \begin{tabular}[c]{@{}l@{}} AP BF vector, \\ RIS mirror orientations \end{tabular} &  & \begin{tabular}[c]{@{}l@{}} DDPG RL algorithm \end{tabular} & \begin{tabular}[c]{@{}l@{}}  BF weights,  orientation angles limits  \end{tabular} \\ \hline
% For ORIS in VLC, it is essential to control the reflection parameters to adjust the optical signals in the target directions, which is different from the IRS in RFs that focus on optimizing the phase of the reflected signals

\cite{yang2020deep} & Continuous & \begin{tabular}[c]{@{}l@{}} BS BF matrix, \\ RIS phase shifts \end{tabular} & Maxmin SR & \begin{tabular}[c]{@{}l@{}} Post-decision state and prioritized experience\\ replay DRL algorithm \end{tabular} & \begin{tabular}[c]{@{}l@{}} Minimum SR and data rate, power, RIS phase  \end{tabular} \\ \hline

\cite{huang2021multi} & Discrete & \begin{tabular}[c]{@{}l@{}} Relay selection, \\ RIS phase shifts \end{tabular} & \begin{tabular}[c]{@{}l@{}} Max. SR, \\ Max. throughput \end{tabular} & \begin{tabular}[c]{@{}l@{}} Distributed multi-agent RL algorithm \end{tabular} & \begin{tabular}[c]{@{}l@{}} delay, secrecy, RIS phase  \end{tabular} \\ \hline

\cite{zhang2023deep} & Continuous & \begin{tabular}[c]{@{}l@{}}  BS BF matrix, \\ Jammer BF matrix, \\ RIS phase shifts\end{tabular} & \begin{tabular}[c]{@{}l@{}} Max. SR, \\ Max. secure EE \end{tabular} & \begin{tabular}[c]{@{}l@{}} DDPG RL algorithm \end{tabular} & \begin{tabular}[c]{@{}l@{}} BS transmit power, user minimum SR, RIS phase  \end{tabular} \\ \hline

\cite{li2022irs} & Discrete & RIS phase shifts & Max. Eve rate & \begin{tabular}[c]{@{}l@{}} Double DQN algorithm \end{tabular} &  \begin{tabular}[c]{@{}l@{}} RIS phase, proactive eavesdropping   \end{tabular}\\ \hline

%\cite{xiao2022irs} & Continuous & \begin{tabular}[c]{@{}l@{}} Sensor encryption key, \\ sensor transmit power, \\ RIS phase shifts \end{tabular} & \begin{tabular}[c]{@{}l@{}}  Max. long-term\\ utility that\\ consists: the data\\ protection level,\\ the Eve rate,\\ the SINR,\\ the sensor energy \\ consumption,\\ the transmission\\ latency \end{tabular}  & \begin{tabular}[c]{@{}l@{}} Actor-critic-\\based \\DRL algorithm  \end{tabular} &  RIS phase constraint \\ \hline

\cite{sun2023leveraging} & Continuous & \begin{tabular}[c]{@{}l@{}} RIS phase shifts, \\ UAV trajectory \end{tabular} & \begin{tabular}[c]{@{}l@{}} Max. achievable rate \end{tabular} & \begin{tabular}[c]{@{}l@{}}  Q-learning and DQN algorithms \end{tabular} &  \begin{tabular}[c]{@{}l@{}} UAV horizontal coordinate, RIS phase  \end{tabular}  \\ \hline 

\cite{xu2023deep} & Continuous & \begin{tabular}[c]{@{}l@{}} RIS phase shift, \\ Device power control,\\ Device computation rate, \\  Device time allocation \end{tabular} & \begin{tabular}[c]{@{}l@{}} Weighted sum secrecy\\ computation efficiency \end{tabular} & \begin{tabular}[c]{@{}l@{}} DDPG  RL algorithm \end{tabular} &  \begin{tabular}[c]{@{}l@{}} Device transmit power, RIS phase  \end{tabular}  \\ \hline

\textcolor{black}{\cite{guo2024ris}} & \textcolor{black}{Continuous} & \begin{tabular}[c]{@{}l@{}} \textcolor{black}{UAV trajectories,} \\ \textcolor{black}{UAV active BF,} \\ \textcolor{black}{RIS passive BF} \end{tabular} & \textcolor{black}{Maxmin multicast rate} & \textcolor{black}{Multi-agent RL algorithm} &  \textcolor{black}{Flight duration, SOP} \\ \hline

\textcolor{black}{\cite{dong2024secure}} & \textcolor{black}{Discrete} & \begin{tabular}[c]{@{}l@{}} \textcolor{black}{UAV BF matrix,} \\ \textcolor{black}{UAV trajectory,} \\ \textcolor{black}{RIS phase shifts} \end{tabular} & \begin{tabular}[c]{@{}l@{}} \textcolor{black}{Max. ASR, Min. secrecy} \\ \textcolor{black}{outage duration} \end{tabular} & \begin{tabular}[c]{@{}l@{}} \textcolor{black}{Proximal policy optimization algorithm} \\ \textcolor{black}{and DRL algorithm} \end{tabular} &  \begin{tabular}[c]{@{}l@{}} \textcolor{black}{BS transmit power, RIS phase, UAV speed} \end{tabular}  \\ \hline

\textcolor{black}{\cite{li2024exploiting}} & \textcolor{black}{Continuous} & \begin{tabular}[c]{@{}l@{}} \textcolor{black}{BS power,} \\ \textcolor{black}{RIS phase shifts} \end{tabular} &  \textcolor{black}{Max. system EE} & \begin{tabular}[c]{@{}l@{}} \textcolor{black}{Batch normalization (BN),}\\ \textcolor{black}{prioritized experience replay (PER),} \\ \textcolor{black}{and DDPG  RL algorithm} \end{tabular} &  \begin{tabular}[c]{@{}l@{}} \textcolor{black}{Backhaul capacity limit,} \\ \textcolor{black}{minimum user SR, BS transmit power,} \\ \textcolor{black}{RIS phase} \end{tabular} \\ \hline

\textcolor{black}{\cite{saifaldeen2024drl}} & \textcolor{black}{Continuous} & \begin{tabular}[c]{@{}l@{}} \textcolor{black}{LED BF weights,} \\ \textcolor{black}{mmWave AP BF weights,} \\ \textcolor{black}{RIS mirror orientations,} \\ \textcolor{black}{mmWave RIS phase shifts} \end{tabular} & \textcolor{black}{Max. SC} & \begin{tabular}[c]{@{}l@{}} \textcolor{black}{DDPG  RL algorithm} \end{tabular} &  \begin{tabular}[c]{@{}l@{}} \textcolor{black}{Orientation angles limits, mmWave RIS phase,} \\ \textcolor{black}{LED and mmWave AP BF weights}\end{tabular}  \\ \hline

\end{tabular}%
}
\vspace*{-3mm}
\end{table*}

\gls{ML} has proven highly successful across diverse domains, including solving involved optimization problems of wireless communication systems. Integrating \textcolor{black}{\gls{ML}} techniques into enhancing \gls{PLS} within wireless communications shows significant potential across diverse characteristics. A notable advantage lies in the capability for proactive detection and prevention of threats. By 
using supervised learning algorithms trained on datasets, patterns indicative 
of malicious activities can be identified in real time, enabling the prompt recognition of eavesdropping attempts. On the other hand, unsupervised learning methods offer adaptability in revealing emerging threats without predefined training datasets. The applications of \textcolor{black}{\gls{ML}} in fortifying \gls{PLS} contain the mitigation of eavesdropping, jamming attacks, and unauthorized access. Advanced techniques demonstrate effectiveness in extracting complicated features from wireless signals, enhancing the precision of threat detection. Additionally, \gls{RL} can be utilized to optimize adaptive strategies for securing communication channels over time, establishing a responsive defense mechanism. The presented works in the literature that solve optimization problems resulting from the integration of \gls{RIS} technology in secure \gls{RF} and \gls{OWC} systems mainly adopted \gls{DRL} techniques %% Note: No need to list them here since they are all discussed in the following subsections%%%
%%~\cite{peng2022deep,saleem2022deep,saifaldeen2022dr,zhang2023deep,xu2023deep,guo2021learning,liu2023drl,xiao2022irs,tang2023secure,li2022reinforcement,li2022irs,sun2023leveraging,jiang2021aerial,huang2021multi,yang2020intelligent,yang2020deep}
and unsupervised learning. Data-intensive requirements of supervised learning, coupled with the potential unavailability of detailed datasets in real-world scenarios and the challenges associated with labeling, hinder the adoption of supervised learning algorithms. %Moreover, integrating \textcolor{black}{\gls{ML}} in the physical layer enhances security and optimizes network performance. Intelligent adaptation of modulation and coding schemes based on channel conditions by \textcolor{black}{\gls{ML}} algorithms improves wireless communication systems' overall efficiency and reliability.
Hence, this segment explores many \gls{ML} technique, involving \gls{DRL} and unsupervised learning techniques, that help in solving optimization problems resulting from the integration of \gls{RIS} technology in secure \gls{RF} and \gls{OWC} systems. \textcolor{black}{This section reviews state-of-the-art \gls{ML} techniques for RIS-assisted PLS and summarizes existing \gls{ML} schemes for \gls{SR} maximization and other secrecy metric optimization in Table~\ref{Table:ML}.}

\subsection{Deep Reinforcement Learning}
\gls{RL} is a potent \textcolor{black}{\gls{ML}} technique with significant implications for advancing \gls{AI}. In \gls{RL}, an autonomous agent undergoes a learning process, consistently making decisions by observing environmental states and autonomously adjusting its approach to attain an optimal policy~\cite{sutton2018reinforcement}. However, the time-intensive nature of this learning process renders \gls{RL} less suitable for large-scale networks, given the dynamic nature of wireless channels over time. Consequently, \gls{RL} applications face substantial limitations in real-world scenarios~\cite{luong2019applications}. Recently, the emergence of deep learning~\cite{goodfellow2016deep} has been identified as a transformative approach capable of addressing \gls{RL}'s shortcomings. In light of these considerations, \gls{DRL} is proposed, representing a revolutionary development in AI and a promising avenue for autonomous systems. Presently, deep learning facilitates the scalability of \gls{RL} to previously intractable problems. \Gls{DRL} algorithms find application in robotics, enabling real-time learning from inputs, such as camera data, by leveraging the advantages of \textcolor{black}{\glspl{DNN}} to enhance the learning speed in the \gls{DRL}~\cite{arulkumaran2017deep,goodfellow2016deep}. In~\cite{peng2022deep}, a secure communication system utilizing \gls{RIS} in a full-duplex setting was investigated, considering the impact of hardware impairment. An optimization problem was formulated to maximize the SSR through joint optimization of transmit \textcolor{black}{\gls{BF}} at the BS and phase shifts at the \gls{RIS}. Due to the mathematical intractability, a \gls{DRL}-based algorithm was proposed to obtain a solution. The efficacy of the developed \gls{DRL}-based algorithm in enhancing \gls{SSR} was validated through extensive simulation results. A \gls{RIS}-assisted secure \gls{ISAC} system was investigated in~\cite{liu2023drl}. The technique aimed to maximize the \gls{SR} through the joint optimization of transmit \textcolor{black}{\gls{BF}}, \gls{AN} matrix, and \gls{RIS} phase-shift matrix. An algorithm based on \gls{DRL} was introduced to tackle the optimization problem. The effectiveness of the proposed technique revealed higher secrecy performance. An energy-efficient and secure \gls{WPAN} transmission system, enhanced by \gls{RIS} and \gls{RL}, was introduced in~\cite{xiao2022irs}. The sensor encryption key, transmit power, and optimized \gls{RIS} phase shifts were coordinated to minimize eavesdropping rates. Secrecy performance improvement was achieved through a \gls{DRL} approach. \Gls{ARIS}-assisted secure communications were explored in~\cite{tang2023secure} to enhance security. The proposed approach involved trajectory design based on \gls{RL} and reflection optimization. The \gls{PLS} enhancement was validated through simulation results. The \gls{RL} was utilized to maximize the eavesdropping rate in~\cite{li2022irs } in the presence of legitimate \gls{RIS} to assess the eavesdropping performance. The authors in~\cite{huang2021multi } utilized the multi-agent \gls{DRL}-based technique to design \gls{RIS} reflection coefficients and relay selection to secure buffer-aided cooperative networks. To enhance the anti-jamming strategy of wireless communication systems, the \gls{RIS} was used in~\cite{yang2020intelligent}, where the fast \gls{RL} technique was presented to achieve the optimal anti-jamming design.

\subsubsection{Deep Deterministic Policy Gradient}
\gls{DDPG} is a specialized \textcolor{black}{\gls{RL}} algorithm for addressing challenges associated with continuous action spaces. Extending the foundations of the \gls{DPG}, \gls{DDPG} incorporates \textcolor{black}{\glspl{DNN}} for both the policy and value function~\cite{qiu2019deep}. Noteworthy is its adoption of a deterministic policy, directly mapping states to specific actions, offering a streamlined approach for learning in continuous action domains. The algorithm employs an actor-critic architecture, where the actor-network formulates the optimal action policy, and the critic assesses the selected actions. Integration of experience replay involves storing past experiences in a buffer, mitigating temporal correlations in training sequences~\cite{ning2021deep}. 

Utilizing the \gls{DDPG} in \gls{PLS} introduces an innovative approach to addressing security challenges within wireless communication networks. The \gls{DDPG}'s ability to handle continuous action spaces makes it a strategic choice for optimizing decision-making processes about \gls{PLS}. The deterministic policy inherent in \gls{DDPG} facilitates precise and strategic decision-making for secure wireless communication systems. The \gls{DDPG} provides a sophisticated framework capable of adapting security strategies in response to the dynamic nature of wireless channels by integrating \textcolor{black}{\glspl{DNN}} for policy formulation and value function evaluation. The advantages of implementing \gls{RIS} in \gls{MEC} 
to secure the \gls{IIoT} networks were investigated in~\cite{xu2023deep}. The optimization of \gls{RIS} phase shift, power control, local computation rate, and time slot was collectively undertaken to maximize the weighted sum secrecy computation efficiency, in which a \gls{DDPG}-based algorithm was developed to address the intricate optimization problem. Simulation results demonstrated that deploying \gls{RIS} in \gls{MEC}-enabled \gls{IIoT} networks enhanced task offloading security, and the proposed \gls{DDPG}-based algorithm surpassed baseline methods in terms of \gls{WSSCE}. A cooperative jamming technique, incorporating DRL for \gls{RIS}-enhanced secure cooperative network in the \gls{IoT}, was introduced in~\cite{zhang2023deep}, considering the presence of eavesdroppers. The primary objective was maximizing the ASR by jointly optimizing transmit and jamming \textcolor{black}{\gls{BF}} matrices and the \gls{RIS} phase-shift matrix, where the \gls{DDPG} optimization algorithm was proposed. The obtained results show improvements in secrecy rate compared to benchmark schemes. Secure communication within diverse multi-device \gls{IoT} environments was explored in~\cite{saleem2022deep}. The classification of legitimate devices into trusted and untrusted categories addressed varying levels of network security in the presence of potential eavesdroppers. A joint optimization problem involving active and passive \textcolor{black}{\gls{BF}} was formulated to maximize the \gls{SSR} of trusted devices while ensuring performance guarantees for all trusted and untrusted devices. An algorithm based on the \gls{DDPG} was introduced to obtain the optimal phases for \gls{RIS} and the transmit \textcolor{black}{\gls{BF}} matrix. The results demonstrated a significant increase in the \gls{SR} of trusted devices. The authors of~\cite{saifaldeen2022dr} introduced a \gls{DRL}-based approach for a secure \gls{VLC} system utilizing \gls{RIS}. The adjustment of \gls{BF} weights at light fixtures and mirror orientations within the mirror array sheet to optimize \gls{SR} was controlled by the DDPG-based algorithm. The adaptability of the \gls{DDPG}-based algorithm in managing the system's high complexity and user mobility was demonstrated. Findings indicated enhancing the security of the \gls{VLC} communication system. In~\cite{guo2021learning}, the robust and secure transmission of \gls{RIS}-assisted mmWave \gls{UAV} communications was explored. A \gls{DDPG} algorithm was introduced to maximize the \gls{SR} for all authorized users. Simulation results confirmed that superior performance could be attained compared to several benchmark methods through the joint optimization of \gls{UAV} trajectory and active (passive) \textcolor{black}{\gls{BF}}.

\subsubsection{Deep Q-learning Algorithm}
\gls{DQN} extends the capabilities of the Q-learning algorithm by incorporating \textcolor{black}{\glspl{DNN}} to manage intricate and high-dimensional state spaces effectively. In \gls{DQN}, the conventional Q-table is replaced with a neural network called the Q-network, which approximates Q-values for each action within a given state. This adaptation equips \gls{DQN} to navigate scenarios with continuous or extensive state spaces proficiently. The neural network undergoes training to minimize the disparity between its predicted and target Q-values, computed using the Bellman equation. DQN has demonstrated notable success in tackling demanding problems within \textcolor{black}{\gls{RL}}, especially in domains like playing intricate video games, where the conventional Q-learning approach encounters challenges due to the expansive and continuous state space. Integrating \textcolor{black}{\glspl{DNN}} enables \gls{DQN} to generalize and acquire insights into intricate patterns, establishing it as a robust tool in \textcolor{black}{\gls{AI}}.

Secure communications enabled by \gls{ARIS} were explored in~\cite{jiang2021aerial}, employing \gls{ARIS} deployment and passive \textcolor{black}{\gls{BF}}, with the strategic design leveraging \gls{DQN}. Simulation results revealed that the proposed \gls{DQN} successfully approximated optimal deployment, resulting in significant improvements in security performance compared to baseline methods. Moreover, a better performance could be achieved with more reflecting elements. The authors of~\cite{sun2023leveraging} presented \gls{ARIS}, wherein \gls{UAV} and \gls{RIS} capabilities were utilized to enhance average downlink achievable rates in wireless networks. The implementation used Q-learning and \gls{DQN} algorithms to evaluate the approach's effectiveness in enhancing security. The findings indicated that the \gls{DQN} algorithm was more suitable than the Q-learning algorithm for systems with huge action and state spaces. In~\cite{li2022reinforcement}, the security of \gls{RIS}-enabled wireless networks was investigated in the presence of intelligent attackers. The \gls{DQN} was introduced to enable adaptive modulation of BS and \gls{RIS} reflection \textcolor{black}{\gls{BF}}. When an eavesdropping attempt was detected, the \textcolor{black}{\gls{BS}} allocated a portion of transmit power to emit AN, disrupting eavesdroppers. The provided results demonstrated the efficacy of the proposed strategy, contributing to enhancing the \gls{SR} in \gls{RIS}-assisted wireless communication systems.

% Omar: I will consider writing about them later, this is if we need to elaborate more on the DRL variants.

%\subsubsection{Multi-agent Reinforcement Learning Algorithm}

%In multi-agent \gls{RL} algorithms, individual agents independently employ either \gls{RL} or \gls{DRL} to enhance their performance or accomplish a collective objective. However, the coordination mechanism for multiple agents needs to be carefully designed.

%\cite{huang2021multi}

%\subsubsection{Fast-policy Hill-climbing Reinforcement Learning Algorithm}

%\cite{yang2020intelligent}

%\subsubsection{Post-decision State and Prioritized Experience Replay Deep Reinforcement Learning Algorithm}

%\cite{yang2020deep}

\subsection{Unsupervised Learning}

Unsupervised learning-based algorithms do not rely on labeled training data to enhance and optimize processes. Instead, these algorithms autonomously uncover patterns and structures within the data, making them particularly valuable in situations where labeled data is limited or costly to obtain. In~\cite{hoang2021ris}, an \textcolor{black}{\gls{ASR}} maximization for secure \gls{RIS}-aided \gls{AANET} has been designed through a deep unsupervised learning algorithm. Specifically, the proposed algorithm relied on a projection-based \gls{DNN} to get suitable solutions. The \gls{DNN} addresses the given optimization problem without constraints. Simultaneously, the projection method ensures that the \gls{DNN}'s output is subsequently mapped to the constrained domain. 

\textcolor{black}{
\subsection{Lessons Learnt}
This subsection investigates the promising ML techniques for RIS-assisted PLS. Here are the key takeaways:
\begin{itemize}
    \item Integrating ML techniques into RIS-assisted wireless communications to enhance PLS holds significant promise across various dimensions. Supervised and unsupervised ML algorithms offer robust frameworks for proactive threat detection and prevention, addressing attacks like eavesdropping, jamming attacks, and unauthorized access. Specifically, DDPG algorithms, known for managing continuous action spaces effectively, are pivotal in optimizing decision-making processes related to PLS. Supervised learning algorithms enable proactive threat detection by learning from labeled datasets to recognize security breach patterns, while DDPG's deterministic policy enhances decision precision in establishing secure communication channels and optimizing parameters like RIS phase shifts and beamforming matrices. Additionally, algorithms like DQN enhance security protocols in RIS-enabled systems by managing complex state spaces and optimizing deployment strategies to mitigate evolving security risks without relying on labeled data, making them cost-effective solutions for optimizing PLS in wireless networks.
    \item From Table VI, it is evident that much of the existing literature on RIS-enabled PLS  in wireless networks assumes continuous phase shifts, implying infinite resolution, whereas practical implementations necessitate discrete phase shifts. This contrast highlights the gap between theoretical assumptions and real-world application scenarios. Furthermore, the majority of research efforts have focused on leveraging DRL to enhance secrecy performance in RIS-assisted PLS for wireless networks. However, there remains potential in exploring advanced ML techniques such as federated, graph-based, transfer, and quantum machine learning for optimizing RIS functionalities. Incorporating these advanced ML methods offers novel insights and potentially superior performance in practical deployments, addressing existing constraints and paving the way for more robust and efficient RIS-assisted wireless communication systems.
\end{itemize}}

\section{Performance Analysis for RIS-Assisted PLS}
\label{Section: Performance Analysis for RIS-Assisted PLS}

The performance analysis for RIS-assisted PLS represents an essential aspect of evaluating the potential of RIS technology in enhancing the security of wireless communication systems. 
In PLS, RIS holds promise for strengthening security measures 
by selectively adjusting signal characteristics, managing interference, and mitigating eavesdropping attempts. \textcolor{black}{This section focuses on conducting a comprehensive performance analysis to assess the impact 
of RIS-assisted PLS in terms of SNR improvement, SC enhancement, and SOP reduction. By examining the performance of RIS-enabled systems under different scenarios, valuable insights can be gained into the significance and feasibility of leveraging RIS to enhance wireless communication systems' PLS.}

\begin{table*}[!t]
\centering
\vspace*{-6mm}
\caption{Summary of performance analysis related literature for \gls{RIS}-Assisted \gls{PLS} systems}
\vspace*{-3mm}
\label{Table:PerformanceAnalysis}
\resizebox{\textwidth}{!}{%
\begin{tabular}{|c|c|c|c|c|l|}
\hline
\textbf{{[}\#{]}} & {\textbf{\begin{tabular}[c]{@{}c@{}}Phase-shift \\ \textbf{resolution} \end{tabular}}} & \textbf{\begin{tabular}[c]{@{}c@{}}Security \\ attack\end{tabular}}  & {\textbf{\begin{tabular}[c]{@{}c@{}}Eavesdropper \\ \textbf{type}\end{tabular}}} & {\textbf{\begin{tabular}[c]{@{}c@{}}Performance \\ \textbf{metric(s)}\end{tabular}}} & \textbf{Main Findings}\\ \hline

\cite{vega2022physical} & Continuous & Jamming & Passive & \multirow{10}{*}{SOP} & \begin{tabular}[c]{@{}l@{}} Impacting of electromagnetic interference's on secrecy performance\end{tabular}  \\ \cline{1-4} \cline{6-6}

\cite{yang2020secrecyGeneral} & Continuous & Eavesdropping & Passive &  & \begin{tabular}[c]{@{}l@{}} Enhancing PLS using RIS \end{tabular}  \\ \cline{1-4} \cline{6-6}

\cite{wang2023uplink} & Continuous &  Eavesdropping & Passive  &  & \begin{tabular}[c]{@{}l@{}}  Analyzing secure RIS-assisted HAP-UAV  joint MU mixed RF/FSO system  \end{tabular}\\ \cline{1-4} \cline{6-6}

\cite{jiang2021robust}  & Continuous & Eavesdropping & Passive &  & \begin{tabular}[c]{@{}l@{}} Utilizing cooperative jamming and BF design to improve PLS of NOMA \end{tabular}  \\ \cline{1-4} \cline{6-6}

\cite{pei2023secrecy} & Continuous & Eavesdropping & Passive &  & \begin{tabular}[c]{@{}l@{}} Utilizing cooperative jamming and BF design to improve PLS of NOMA \end{tabular}  \\ \cline{1-4} \cline{6-6}

% Removed, as we agreed that we would not discuss covert communications in this survey paper
%\cite{yang2023covert} & Continuous & Eavesdropping & Passive &  & \begin{tabular}[c]{@{}l@{}} Improving the performance of covert  \\communication by using  RS, NOMA, \\and RIS  \end{tabular}  \\ \cline{1-4} \cline{6-6}

\cite{shi2022secure} & Discrete & Eavesdropping & Passive &  & \begin{tabular}[c]{@{}l@{}} Revealing the SOP's scaling law for RIS's elements and quantization's bits \end{tabular}  \\ \cline{1-4} \cline{6-6}

\cite{li2022enhancing} & Continuous & Eavesdropping & Passive &  & \begin{tabular}[c]{@{}l@{}} Reducing the secrecy performance due      to residual hardware impairments  \end{tabular}  \\ \cline{1-4} \cline{6-6}

\cite{ai2021secure}  &Continuous&Eavesdropping & Passive &  & \begin{tabular}[c]{@{}l@{}} Enhancing secrecy performance of  V2V   communications with the aid of RIS  \end{tabular}  \\ \cline{1-4} \cline{6-6}

\textcolor{black}{\cite{pei2024secrecy}}  & \textcolor{black}{Discrete} & \textcolor{black}{Eavesdropping}   & \textcolor{black}{Passive}  & & \begin{tabular}[c]{@{}l@{}} \textcolor{black}{Investigating an RIS-aided ambient \gls{BackCom} with} \\ \textcolor{black}{perfect and imperfect CSI and showing the superior secrecy outage behaviours} \\ \textcolor{black}{compared to the conventional BackCom networks} \end{tabular}\\ \hline

\cite{khoshafa2020reconfigurable}  & Continuous&Eavesdropping & Passive &  \multirow{4}{*}{SOP, PNSC} & \begin{tabular}[c]{@{}l@{}} Friendly Jamming Eve using RIS to improve the security performance  \end{tabular}  \\ \cline{1-4} \cline{6-6}

\cite{khoshafa2023ris}& Continuous&Eavesdropping & Passive &  & \begin{tabular}[c]{@{}l@{}} Improving the secrecy performance of primary network using secondary RIS  \end{tabular}  \\ \cline{1-4} \cline{6-6}

\cite{cao2022ergodic}& Continuous& Eavesdropping & Passive &  & \begin{tabular}[c]{@{}l@{}} Designing phase shift to maximize  information at the legitimate destination  \end{tabular}  \\ \cline{1-4} \cline{6-6}

\cite{khoshafa2021active} & Continuous & Eavesdropping & Passive &  & \begin{tabular}[c]{@{}l@{}} Enhancing secrecy performance of wireless network using active RIS \end{tabular}  \\ \hline

\cite{khoshafa2023securing}& Continuous&Eavesdropping & Passive & \multirow{2}{*}{\begin{tabular}[c]{@{}l@{}} SOP, PNSC, ASR \end{tabular}}   & \begin{tabular}[c]{@{}l@{}} Improving a secure data transmission of LPWAN IoT sensor  \end{tabular}  \\ \cline{1-4} \cline{6-6}

\cite{zhang2021physical} & Continuous & Jamming & Active & & \begin{tabular}[c]{@{}l@{}} Reducing the secrecy performance due to residual hardware impairments  \end{tabular}  \\ \hline

\cite{trigui2021secrecy} & Discrete & Eavesdropping & Passive & SOP, ASR & \begin{tabular}[c]{@{}l@{}} Unveiling the secrecy loss due to phase resolution \end{tabular}  \\ \hline

\cite{tang2022physical} & Continuous & Eavesdropping & Passive & SOP, ASC & \begin{tabular}[c]{@{}l@{}} Investigating the effect of REs on the secrecy diversity orders \end{tabular}  \\ \hline

\cite{shi2023secrecy} & Continuous & Eavesdropping & Passive & SOP, ESC & \begin{tabular}[c]{@{}l@{}} Showing the effects of RE on the SOP and ESC \end{tabular}  \\ \hline

\cite{sheng2023performance} & Continuous & Eavesdropping & Passive & SOP, ANT & \begin{tabular}[c]{@{}l@{}} HARQ is beneficial to improve system security-reliability balance \end{tabular}  \\ \hline

\textcolor{black}{\cite{zhang2023secrecy}}  & \begin{tabular}[c]{@{}l@{}} \textcolor{black}{Continuous/} \\ \textcolor{black}{Discrete} \end{tabular} &  \textcolor{black}{Eavesdropping}  & \textcolor{black}{Passive}  & \textcolor{black}{SOP, ESR} & \begin{tabular}[c]{@{}l@{}} \textcolor{black}{Investigating the fundamental limits of RIS-aided wiretap MIMO communications} \end{tabular}\\ \hline

\textcolor{black}{\cite{yuan2023security}}  & \textcolor{black}{Continuous}  & \textcolor{black}{Eavesdropping}   & \textcolor{black}{Passive}  & \textcolor{black}{SOP, IP} & \begin{tabular}[c]{@{}l@{}} \textcolor{black}{Showing that the active RIS improves the UAV system security and reliability}\\ \textcolor{black}{tradeoff (SRT) more than passive and hybrid RIS} \end{tabular}\\ \hline

\cite{rahman2023ris} &Continuous  & Eavesdropping & Passive &\begin{tabular}[c]{@{}l@{}}  SOP, ASC, SPSC, EST \end{tabular}  & \begin{tabular}[c]{@{}l@{}}  Exploring secure RIS-assisted relaying system for mixed RF/FSO system  \end{tabular} \\ \hline

\cite{gu2022physical} & Discrete & Eavesdropping & Passive &  PNSC, ESC & \begin{tabular}[c]{@{}l@{}} Investigating the influence of RIS's size and location on the secrecy performance  \end{tabular}  \\ \hline

\cite{luo2021reconfigurable} & Continuous & Eavesdropping & Passive & \multirow{3}{*}{SR}  & {{\begin{tabular}[c]{@{}c@{}}Enhancing the security by using RIS as {a source of multiplicative randomness}\end{tabular}}}  \\ \cline{1-4} \cline{6-6}

\textcolor{black}{\cite{shang2024secure}}  & \textcolor{black}{Continuous}  &  \textcolor{black}{Eavesdropping}  &  \textcolor{black}{Passive} &  & \begin{tabular}[c]{@{}l@{}} \textcolor{black}{Investigating a secure RIS-aided vehicle road cooperation (VRC) system with} \\ \textcolor{black}{consecutive artificial noise and index modulation} \end{tabular}\\ \hline

\cite{chai2023secure} & Continuous & Spoofing & Active & Sum SR & \begin{tabular}[c]{@{}l@{}} Detecting spoofing and designing BF to improve secrecy performance  \end{tabular}  \\ \hline 

%\cite{ren2022performance} & \begin{tabular}[c]{@{}l@{}} Continuous/   \\ Discrete \end{tabular} & Eavesdropping & Passive & \begin{tabular}[c]{@{}l@{}} Sum achievable\\ security data\\ rate\end{tabular} & \begin{tabular}[c]{@{}l@{}} Showing the significance of utilizing \\ an RIS in enhancing the security \\performance \end{tabular}  \\ \hline

\cite{ye2022secure} & Continuous & Eavesdropping & Passive & \multirow{2}{*}{SER} & Reducing the side-lobe energy toward Eve  \\ \cline{1-4} \cline{6-6}

\cite{wei2022secrecy} & Continuous & Eavesdropping & Passive &  & {{\begin{tabular}[c]{@{}c@{}}Enhancing the secrecy performance by   {integrating multiple antennas with RIS}\end{tabular}}}  \\ \hline

\cite{lin2022security} & Continuous & Eavesdropping & Passive &  \begin{tabular}[c]{@{}l@{}} \gls{IP} \end{tabular} & \begin{tabular}[c]{@{}l@{}} Improving secrecy performance by using friendly jamming and RIS \end{tabular}  \\ \hline

\cite{elhoushy2021exploiting}& Continuous&Spoofing & Active & \begin{tabular}[c]{@{}l@{}}   ASR \end{tabular} & \begin{tabular}[c]{@{}l@{}} Utilizing RIS in increasing the    reliability of CF mMIMO systems  \end{tabular}  \\ \hline

\cite{wei2022multi} & Continuous & Eavesdropping & Passive &  \begin{tabular}[c]{@{}l@{}} Secret key rates \end{tabular}  & \begin{tabular}[c]{@{}l@{}} Analyzing the impact of cooperative multi-Eve  scheme against the RIS-secured  \end{tabular}  \\ \hline

\cite{yuan2022secure} \ &  Discrete & Eavesdropping & Passive & \multirow{3}{*}{ESR} & \begin{tabular}[c]{@{}l@{}} Investigating secure transmission of a LEO satellite using ARIS-equipped HAP  \end{tabular}  \\ \cline{1-4} \cline{6-6}

\textcolor{black}{\cite{yang2024spatially}}  & \textcolor{black}{Continuous} &  \textcolor{black}{Eavesdropping}  &  \textcolor{black}{Passive} &  & \begin{tabular}[c]{@{}l@{}} \textcolor{black}{Investigating secure massive MIMO RIS-aided system with the CSI imperfections,} \\ \textcolor{black}{transceiver hardware impairments, RIS phase noise, and correlated Rayleigh fading} \end{tabular}\\ \hline

\cite{cao2023physical}  & Continuous &  Eavesdropping & Passive  & ESC & \begin{tabular}[c]{@{}l@{}}  Enhancing the PLS of  WPC  system vis RIS \end{tabular}\\ \hline

%\textcolor{black}{\cite{wang2024channel}}  & \textcolor{black}{Continuous} &    &   &  & \begin{tabular}[c]{@{}l@{}} \end{tabular}\\ \hline  %Channel Reciprocity Attacks

\textcolor{black}{\cite{diao2024secure}}  & \textcolor{black}{Continuous} &  \textcolor{black}{Eavesdropping}  & \textcolor{black}{Passive}  & \textcolor{black}{EC, ECP, ESC} & \begin{tabular}[c]{@{}l@{}} \textcolor{black}{Investigating three secure UAV wireless-powered NOMA transmission modes} \\ \textcolor{black}{under Nakagmi-$m$ channels and phase compensation errors} \end{tabular}\\ \hline

\textcolor{black}{\cite{li2024secure}}  & \textcolor{black}{Continuous} & \textcolor{black}{Eavesdropping}   &  \textcolor{black}{Passive} & \textcolor{black}{OP, IP} & \begin{tabular}[c]{@{}l@{}} \textcolor{black}{Investigating secure RIS-Aided cognitive V2X networks with imperfect CSI over} \\ \textcolor{black}{double Rayleigh fading channels} \end{tabular}\\ \hline

%\cite{tang2021novel} &  &  &  &  &  \\ \hline

%&   & \begin{tabular}[c]{@{}c@{}} Secrecy Rate \\ SOP \\ etc.\end{tabular} & \begin{tabular}[c]{@{}c@{}} MUD \\ MMSE-SIC \\ etc.\end{tabular} &  &  \\ \hline

%\cite{kong2021effective} &  &  &  &  &  \\ \hline

%\cite{adam2023intelligent} &  &  &  &  &  \\ \hline %Note(Omar): this reference does not contain PLS [ignored]

%\cite{wijewardena2021physical} &Continuous&Eavesdropping & Passive &   Sum secrecy rate & \begin{tabular}[c]{@{}l@{}} Enhancing secrecy performance of  V2V  \\  communications with the aid of RIS  \end{tabular}  \\ \hline

\end{tabular}%
}
\vspace*{-3mm}
\end{table*}

\subsection{SNR Improvement with RIS}
This subsection explores the \gls{SNR} enhancement via \gls{RIS} within the paradigm of \gls{PLS}, highlighting its crucial role in mitigating noise and optimizing signal strength. Significantly, the adaptive nature of \gls{RIS} facilitates dynamic adjustments, a feature of paramount significance in addressing varying interference patterns and ensuring high \gls{SNR}. The high \gls{SNR} extends to fortifying communication links, particularly in challenging urban settings. This increases communication reliability and directly impacts security 
by strengthening communication links against eavesdropping and other security threats. The secrecy capabilities of wireless communication systems enhanced 
by using the \gls{RIS} were examined in the presence of a potential eavesdropper, where the \gls{RIS} was strategically placed between the source and the intended user, creating an intelligent environment to maintain link security. The results confirmed the positive influence of \gls{RIS} utilization on improving the secrecy performance of wireless systems~\cite{khoshafa2020reconfigurable, yang2020secrecy}. The performance analysis of secure communications assisted by \gls{RIS} for discrete phase shifts was investigated in~\cite{trigui2021secrecy, shi2022secure, xu2020ergodic}. The security performance of a communication system aided by \gls{RIS} was examined in~\cite{wei2022secrecy}, with spatially random \glspl{UAV} serving as eavesdroppers. The obtained results highlighted the security enhancements achieved through the deployment of \gls{RIS}. The secrecy performance of a relay communication system utilizing an \gls{RIS} on a \gls{UAV} was investigated in~\cite{wang2021secrecy11} in the presence of multiple ground eavesdroppers. Specifically, a \gls{UAV} equipped with a \gls{RIS} was a passive relay, forwarding signals from the \textcolor{black}{\gls{BS}} to users. 
Secrecy performance for integrated satellite and \gls{UAV} relay networks with multiple vehicle eavesdroppers was investigated in~\cite{zhou2022secrecy} by employing \gls{RIS}. The \gls{UAV} was used to relay the legitimate signal to the destination user, facilitated by \gls{RIS}. The authors in~\cite{khoshafa2021active}, proposed a novel technique by designing active elements in \gls{RIS} to overcome the double-fading problem introduced in the \gls{RIS}-aided link in a wireless communications system to secure the transmission data. In this approach, the reflected incident signal was amplified by \gls{RIS}'s active elements, departing from the sole reflective function observed in passive \gls{RIS} modules. The analysis and simulation results revealed a significant reduction in the required size of the \gls{RIS} to achieve a specified performance level when active elements were employed. Additionally, a practical design for active \gls{RIS} was proposed. 

\subsection{Secrecy Capacity Enhancement}
This subsection focuses on enhancing \gls{SC} through using \gls{RIS} within the context of PLS. Recent research conveys how the reconfigurable nature of \gls{RIS} strategically influences channel dynamics, strengthening the system's capacity for secure data transmission through adaptive configurations that dynamically manipulate signal propagation, offering insights into potential gains in secure communication. The results illustrate how the \gls{RIS} contributes to heightened SC by dynamically adapting to the communication environment and strategically configuring itself to minimize the impact of potential eavesdropping attempts. Comparative analyses against scenarios without \gls{RIS} intervention further highlight the high improvements in \gls{SC} obtained by \gls{RIS}. This integrated approach clarifies how \gls{RIS} enhances \gls{SC}, establishing its role as a promising technology in improving \gls{PLS}. The effectiveness of \gls{STAR}-\gls{RIS} in enhancing security within the \gls{MISO} network was explored in~\cite{niu2021simultaneous}. Three transmission protocols, energy splitting, mode selection, and time splitting, were examined to maximize the weighted sum secrecy rate. In~\cite{zhao2022secrecy}, high passive \textcolor{black}{\gls{BF}} gains for signal enhancement were achieved by \gls{RIS}s through the dynamic adjustment of their reflection coefficients, improving wireless security and \gls{RF}-based wireless power transfer efficiency. A \gls{RIS}-assisted secure \gls{SWIPT} system was investigated for transferring information and power. The worst-case downlink secure communication scenario was addressed in~\cite{liu2022ris}, applying \gls{RIS}-\gls{UAV} as an aerial passive relay to enhance communication. The objective was to maximize the worst-case downlink secrecy rate by optimizing power allocation, \gls{RIS} passive \textcolor{black}{\gls{BF}}, and UAV trajectory. The results demonstrated the effectiveness of the proposed scheme compared to other benchmarks. The impact of hardware impairments on the secrecy performance of a \gls{mmWave} system assisted by an \gls{RIS} was investigated in~\cite{ragheb2023ris}. In this regard, a system model was constructed without considering hardware impairments, and optimal solutions were presented for signal and \gls{AN} powers, \textcolor{black}{\gls{BF}} design, and phase shifts of the \gls{RIS} elements. The authors in~\cite{li2023secure} explored the \gls{PLS} of active \gls{RIS}-assisted \gls{NOMA} networks in the presence of external and internal eavesdroppers. Specifically, closed-form expressions for \gls{SOP} and secrecy system throughput were derived, considering both imperfect \gls{SIC} and perfect \gls{SIC}. An innovative approach to enhancing the secrecy performance of a \gls{LPWAN} by integrating an \gls{RIS} with a \gls{UAV} was introduced in~\cite{khoshafa2023securing}. The primary goal was to improve the secure data transmission between an \gls{IoT} sensor and a gateway in \gls{LPWAN} applications. Closed-form expressions for the \gls{SOP}, \gls{PNSC}, and \gls{ASR} were derived for the proposed network operating over Nakagami-$m$ fading channels. Additionally, the practical impact of the eavesdropper's location was explored. The obtained results illustrated that integrating a \gls{RIS}-\gls{UAV} significantly improves the secrecy performance of the \gls{LPWAN}, enabling reliable long-range transmission.

\subsection{Secrecy Outage Probability Reduction}

By employing thorough techniques, \gls{RIS} has great promise in reducing the \gls{SOP} in wireless communication. Through selective signal reflection and \textcolor{black}{\gls{BF}}, \gls{RIS} directs signals precisely to the intended receiver, minimizing the risk of information leakage. The dynamic adaptability of \gls{RIS} allows real-time adjustments in response to changing wireless conditions, enhancing secrecy against evolving threats. Intelligent jamming approaches strengthen the communication link, while controlled \gls{AN} and collaborative \gls{RIS} networks add complexity to intercepted signals, maintaining confidentiality and reducing the \gls{SOP}. Utilizing \gls{RIS} presents an opportunity to increase \gls{PLS} in wireless communication systems, as highlighted in~\cite{zhang2021physical}. One strategy involves leveraging \gls{RIS} to implement virtual \textcolor{black}{\gls{BF}}, enabling the product of multiple directed beams toward different users. This can potentially enhance the secrecy performance of cellular networks by reducing the signal strength at unintended receivers, thereby confounding the eavesdroppers and decreasing the SOP. %The authors in~\cite{khoshafa2021active} proposed an active \gls{RIS} as a novel approach to improve the secrecy performance by using active elements to amplify the reflected signal, leading to the secrecy performance enhancement.
Also, \gls{RIS} can establish a secure communication zone by directing the signal towards the intended receiver while diverting it away from potential eavesdroppers~\cite{wei2022secrecy}. Additionally, \gls{RIS} can serve as a friendly jammer to disrupt eavesdropper signals~\cite{tang2021securing}. The unique capabilities of \gls{RIS} extend to detecting and identifying eavesdropper locations. By doing so, \gls{RIS} can assess the proximity of unintended receivers and adjust \textcolor{black}{\gls{BF}} to minimize signal strength at those locations. To secure the transmission link, the \gls{RIS} was strategically placed near the eavesdropper, effectively shortening information disclosure and heightening the confidentiality of the wireless network~\cite{zhang2021improving}. The secrecy performance of a communication system assisted by \gls{RIS} was investigated in~\cite{shi2023secrecy} in the presence of spatially random eavesdroppers, where closed-form expressions for the \gls{SOP} and the eavesdropper \textcolor{black}{\gls{SC}} were derived. To evaluate the \gls{SOP}, the \gls{STAR}-\gls{RIS}-assisted cognitive \gls{NOMA}-\gls{HARQ} network was investigated in~\cite{sheng2023performance}. The reviewed performance analysis works on RIS-assisted PLS are summarized in Table~\ref{Table:PerformanceAnalysis}.

{\color{black}{
\subsection{Lessons Learnt}
This subsection analyzes how the RIS technology improves the performance of the PLS in wireless communication systems. Here are the key takeaways:
\begin{itemize}
%    \item Integrating the \gls{RIS} technology within \gls{PLS} frameworks allows for a dynamic signal optimization, significantly enhancing SNR and \gls{SC}. The reconfigurable nature of \gls{RIS} facilitates adaptive configurations and \textcolor{black}{\gls{BF}} capabilities, mitigating noise, managing interference, and influencing channel dynamics. This adaptability ensures high \gls{SNR} and secure data transmission, particularly in challenging environments, directly strengthening the security of communication links against eavesdropping.
    \item An \gls{RIS} effectively reduces the \gls{SOP} by precisely directing signals to intended receivers and minimizing leakage. The strategic placement of \gls{RIS} creates intelligent environments that enhance link security. Techniques such as intelligent jamming and controlled \textcolor{black}{\gls{AN}} add complexity to intercepted signals, further improving the secrecy performance. This setup is especially effective in maintaining secure communication links in complex scenarios, including urban settings and diverse network configurations.
    \item Deploying \gls{RIS} in relay communication systems, including satellite and \gls{UAV} networks, enhances security against multiple ground eavesdroppers. Incorporating active elements in \gls{RIS} design addresses the double-fading issues, resulting in smaller \gls{RIS} sizes without compromising performance. This innovation highlights the potential for more compact and efficient \gls{RIS} deployments that maintain high security and signal quality, showcasing the versatility of the \gls{RIS} technology in various applications.
    \item Conducting thorough performance analyses of \gls{RIS}-assisted \gls{PLS} provides valuable insights into \gls{SNR} improvement, \gls{SC} enhancement, and the \gls{SOP} reduction. These analyses demonstrate the feasibility and significance of leveraging the \gls{RIS} technology 
    to bolster the security of wireless communication systems. Investigating real-world scenarios, such as hardware impairments and spatially random eavesdroppers, emphasizes the practical applicability of the \gls{RIS} technology. This ensures that theoretical benefits translate into tangible security improvements in real-world deployments, highlighting the crucial role of \gls{RIS} in enhancing wireless communication security.
    \item From Table VII, it is clear that most existing research on RIS-assisted PLS performance primarily focuses on passive eavesdropping, where attackers only intercept communication. While mathematically tractable, this approach overlooks the growing threat of active eavesdropping. Active attackers can disrupt communication or manipulate signals, making passive-only analysis insufficient. Moreover, the RIS technology itself can be exploited 
    for jamming by active attackers. 
    Future research needs to prioritize active eavesdropping scenarios to ensure robust security, including developing analytical models for 
    these attacks and designing secure signal processing techniques to 
    detect and mitigate them.
\end{itemize}}}
\section{Open Research Challenges and Future Directions}
\label{Section: Open Research Challenges and Future Directions}

\textcolor{black}{This section highlights key challenges and future research directions for \gls{RIS}-assisted \gls{PLS} in emerging \gls{RF} and \gls{OWC} networks aiming at improving wireless network security.} 
\subsection{RIS-Assisted Secure Wireless Communications From Optimization Techniques Perspective}
{\color{black}{The optimization of \gls{RIS}-assisted secure wireless communications usually involves challenging mathematical problems due to the multiple and typically coupled variables, non-convex objective functions and constraints, the presence of discrete and continuous decision variables, and the high dimensionality of the optimization problems. As a result, most proposed approaches adopt relaxation, approximation and decomposition techniques which lead to suboptimal performance or result in heuristics without any performance guarantee. This calls for studies on the application of advanced optimization techniques such as global optimization solution methods, i.e.,~\cite{nerini2023closed}, and non-convex non-smooth optimization such as~\cite{cutkosky2023optimal, jordan2023deterministic}. Metaheuristics on the other hand can tackle various optimization problems without any stringent requirements (e.g., convexity, continuity, and differentiability) for objective functions and constraints, and can cope with high dimensional optimization problems. It is therefore important to examine the application of some recently proposed metaheuristics~\cite{alorf2023survey} in optimizing \gls{RIS}-assisted secure wireless systems.}}

%%%%%%%%%%%%%%%%%%%%%%%%%%%%%%%%%%%%%%%%%%%%%%%%%%%%%
\subsection{RIS-Assisted Secure Wireless Communications via Advanced ML Techniques}
Utilizing advanced \gls{ML} techniques with RIS (see~\cite{zhou2023survey} and the references therein) offers a compelling avenue to increase \gls{PLS} within wireless communication systems. Federated learning, characterized by decentralized model training while preserving data privacy, could be utilized to collectively optimize \gls{RIS} configurations across distributed network nodes, protecting sensitive information. Graph learning techniques have the potential to facilitate the analysis of complex interaction patterns between \gls{RIS} components and wireless channels, thereby enabling more accurate prediction and mitigation of potential security vulnerabilities. Transfer learning, involving knowledge transfer from related tasks to enhance \gls{RIS} configuration optimization for \gls{PLS}, could expedite the convergence of security-aware \gls{RIS} designs. Quantum learning techniques, leveraging the computational power of quantum computing, could offer unprecedented capabilities for optimizing \gls{RIS} configurations and cryptographic protocols, strengthening \gls{PLS} against quantum-based attacks. Additionally, meta-learning strategies could empower \gls{RIS} to autonomously adapt and refine their security mechanisms based on past experiences and emerging threats, strengthening the resilience of wireless communication systems against security challenges. Exploring the integrated deployment of these advanced \gls{ML} techniques with \gls{RIS} represents a promising future research direction to advance \gls{PLS} within wireless communication systems.
%\textcolor{black}{Employ newly developed \gls{ML} techniques, such as federated learning, graph learning, transfer learning, hierarchical learning, Quantum learning, and meta-learning. \textcolor{black}{Mentioned in table XI~\cite{zhou2023survey}}.}

\subsection{\textcolor{black}{RIS-Assisted Secure Multiband Networks}}
\textcolor{black}{The \gls{6G} technology will enable the standalone and integrated deployment of various transmission frequencies (including conventional \gls{RF}, \gls{mmWave}, \gls{THz}, and \gls{VLC}) to offer a trade-off between capacity, latency, network coverage and mobility. However, this comes at the expense of some drawbacks among which wireless security. Malicious users can exploit the different signal propagation characteristics of the various frequency bands for enhanced eavesdropping. A promising avenue for future research studies is to leverage the \gls{RIS} technology for the security improvement of \glspl{MWN} which can be intricate. This is due to the fact that with the coexistence of multiple spectrum transmissions, employing \gls{RIS} for security improvement to operate in one frequency band may not necessarily work as well on another spectrum band. Moreover, \glspl{RIS} have different capabilities in various frequency bands. These capabilities affect the role of \gls{RIS} and the decision variables that need to be optimized in enhancing \glspl{PLS}. It is therefore important to investigate the role of multiband \glspl{RIS} in \gls{PLS}.}

\subsection{\textcolor{black}{RIS-Assisted Secure Wireless Communications using a Movable Antenna Technology}}
\textcolor{black}{\Gls{MA} is a novel antenna architecture (see~\cite{hu2024movable} and the references therein) that enables the performance improvement of wireless communication systems by fully leveraging the dynamic transmit/receive antenna placement at positions with more favorable channel conditions, in a confined region\footnote{\textcolor{black}{A similar approach to \gls{MA} is the \gls{FA} technology wherein an antenna can freely switch positions in a fixed-length linear space in search of the strongest signal~\cite{xlai2024performance,new2024fluid,xu2024channel}.}}. 
This architecture can be integrated into the \gls{RIS} technology to ensure the dynamic nature of the latter with a non-uniform activity across all its elements, and thus facilitate more real-time adjustments to wireless communications properties than the \gls{FPA}-\gls{RIS} setup. However, obtaining precise \gls{CSI} at the transmitting/receiving ends or optimizing antenna positions that yields the best communication performance can be very challenging. There is limited research on the adoption of \gls{RIS} in \gls{MA}-based  wireless systems, let alone in the context of \gls{PLS}. Exciting areas worth investigating include jointly optimizing \gls{MA} position, power allocation, and \gls{RIS} deployment and parameter(s), such as phase shift for RF-based \gls{RIS} and orientation angles for optical \gls{RIS} tuning to maximize \gls{PLS}. Exploring this research direction has the potential to bring significant advancements in future wireless networks while enabling secure and reliable networks 
for practical scenarios.}

%\Gls{RIS} can be adopted into \gls{BackCom} systems to provide secure transmissions as attested by some recent research efforts~\cite{wang2022multicast,han2023broadcast,wang2023intelligent}. Nonetheless, the research on \gls{RIS}-assisted secure \gls{BackCom} systems is still in its infancy and numerous research directions are worthwhile to be explored in future studies. One potential avenue warrants the investigation of the secrecy performance of \gls{RIS}-assisted symbiotic \gls{BackCom} 

\subsection{\textcolor{black}{RIS-Assisted Secure Systems in Generalized Fading}}
\textcolor{black}{Exploring the potential of \gls{RIS} to improve the information-theoretic security of wireless systems in generalized fading scenarios represents a promising direction for future studies. The generalized fading models such as $\alpha$-$\mu$, $\kappa$-$\mu$ 
and $\eta$-$\mu$, adequately fit real-time experimental data and can be used to characterize small-scale fading in the \gls{mmWave} and \gls{THz} bands where a large contiguous bandwidth is available and can be exploited to support the demand for bandwidth-hungry applications in ultra-dense networks. As in the case of classical fading models (Rayleigh, Rician, Nakagami-$m$, etc...), wireless transmissions over generalized fading channels are susceptible to security vulnerabilities. In this context, the deployment of \gls{RIS} wireless systems subject to generalized fading emerges as an appealing direction for future investigations. However, it is worth noting the analytical intricacies of obtaining the exact statistics of \gls{RIS}-aided secure systems in such fading models, let alone the various exact performance measures. Addressing these challenges is crucial and will therefore constitute an advancement to the current state-of-the-art.}

\subsection{RIS-Assisted Secure Communications With Other Emerging Wireless Technologies}
\textcolor{black}{This subsection highlights the key challenges and future research directions for emerging applications that have not yet been explored in terms of information-theoretic security through the \gls{RIS} technology.}

\subsubsection{RIS-Assisted Secure CoMP Communications}
RIS-assisted secure \gls{CoMP} Communications are considered an open research area that aims to enhance the performance and security of wireless networks. By integrating \gls{RIS} within the \gls{CoMP} framework, the technology endeavors to optimize parameters such as signal strength, coverage, and overall system capacity while simultaneously addressing issues such as interference mitigation and signal confidentiality enhancement. Key areas of research focus include the development of efficient strategies for \gls{RIS} deployment and joint optimization of \gls{CoMP} and \gls{RIS} to increase the security level. Despite the complexity of such challenges, \gls{RIS}-assisted secure \gls{CoMP} communications maintain significant promise in securing wireless communication networks by offering enhanced performance and security capabilities that align with the evolving requirements of next-generation applications and services.
%%%%%%%%%%%%%%%%%%%%%%%%%%%%%%%%%%%%%%%%%%%%%%%%%%%%%%%%%%
\subsubsection{RIS-Assisted Secure Blockchain Technology}
Integrating \gls{RIS} into secure blockchain technology is an innovative strategy for reinforcing the security and confidentiality of distributed ledgers~\cite{liu2022energy}. \gls{RIS} actively plays a role in mitigating potential attacks, such as jamming from malicious nodes, by intelligently redirecting and optimizing communication links. Consequently, the security of blockchain technology is improved, establishing a more robust and reliable framework. Moreover, the adaptability and flexibility of \gls{RIS} align with the decentralized nature of blockchain, offering an additional layer of security. Despite existing challenges in deployment and integration, the prospective advantages of \gls{RIS}-assisted secure blockchain technology lie in its ability to elevate the reliability and robustness of decentralized systems, resulting in secure and efficient blockchain applications.

%%%%%%%%%%%%%%%%%%%%%%%%%%%%%%%%%%%%%%%%%%%%%%%%%%%%%

%\subsubsection{RIS-Assisted Secure Quantum-enabled Systems}

%%%%%%%%%%%%%%%%%%%%%%%%%%%%%%%%%%%%%%%%%%%%%%%%%%%%%

\subsubsection{\textcolor{black}{RIS-Assisted Secure RSMA-based Systems}}
\textcolor{black}{\Gls{NOMA} has emerged as a key-enabling technology for \gls{5G} networks~\cite{maraqa2020survey}. However, major challenges such as hardware complexity, receiver design, error propagation in \gls{SIC} just to name a few, have hampered its development for \gls{5G} and beyond wireless systems. To this end, \gls{RSMA} is envisioned as a key-enabling multiple-access scheme that addresses the massive connectivity problem by accommodating different users in future heterogeneous networks. Incorporating \gls{RIS} into \gls{RSMA}-based systems may be used to improve the network performance in the context of \gls{PLS}. By taking advantage of the reflective elements in the \gls{RIS} technology, it is feasible to establish a highly directional and controllable wireless transmission environment. This capability may facilitate the improvement of the legitimate user's signal strength or the attenuation of the the malicious' signal, thereby enhancing the overall wireless security of practical \gls{RSMA}-based systems without the need for extra processing, and signaling. This is worthwhile to be explored as a potential direction for future research studies.}

%%%%%%%%%%%%%%%%%%%%%%%%%%%%%%%%%%%%%%%%%%%%%%%%%%%%%
\subsubsection{\textcolor{black}{RIS-Assisted Secure AR and VR Communications}}
\textcolor{black}{\Gls{AR} and \gls{VR} applications require seamless communications with negligible latency and ultra-high bandwidth that can be achieved via abundant spectrum bands, such as \gls{mmWave} and \gls{THz} bands, at the expense of limited range and high blockage by physical barriers. To this end, ensuring the wireless security of such applications is a challenging task. A potential research direction on \gls{RIS}-aided secure \gls{AR} and \gls{VR} communications may focus on leveraging \gls{RIS} to enhance the security of communication channels in \gls{AR} and \gls{VR} applications. The goal is to strategically deploy \gls{RIS} elements to ensure the confidentiality of sensitive data transmitted in immersive environments. The investigation will address challenges such as adapting \gls{RIS} configurations to the dynamic nature of \gls{AR} and \gls{VR} and balancing security needs with low-latency communication.}

%%%%%%%%%%%%%%%%%%%%%%%%%%%%%%%%%%%%%%%%%%%%%%%%%%%%%
\subsubsection{RIS-Assisted Secure Under-water Communications}
Achieving secure wireless communications in underwater is a necessity for governments, commercial companies, and the military. This is because underwater wireless communications play an important role in both civil applications such as oil and gas exploration and military applications. Exploring the potential of securing both underwater \textcolor{black}{\glspl{OWC}} that has a medium communication range that spans between $50$-$100$ meters and underwater \gls{RF} wireless communications that is suitable for short-distance propagation of around $10$ meters can be an attractive area for future research. One possible way to attain this is by mounting \glspl{RIS} on underwater entities (e.g., autonomous underwater vehicles, submarines, divers, etc.), at the sea level, or at the ground level under the sea while employing one of \gls{PLS} techniques. 

\subsubsection{\textcolor{black}{Mixed RIS/FSS-Assisted Secure Wireless Networks}}
\textcolor{black}{Despite providing more security and interference immunity, the users' confidentiality of a \gls{FSS}-aided wireless communication system may still be compromised through the \gls{FSS} filters which may expose users' sensitive data including location information, behavior patterns, etc., by revealing the users' communication frequency usage. In this context, the deployment of \gls{RIS} in \gls{FSS}-assisted communication networks emerges as an appealing research direction for future studies to mitigate users' security. Although both the \gls{RIS} and \gls{FSS} belong to the family of intelligent surfaces, they differ in their structure and working principle. To this end, the adoption of \gls{RIS} in the present context is undeniably of utmost interest for the development of \gls{6G} communication networks since it is likely to bring forth design issues as well as challenges on the optimization of the signal quality and coverage.}

%%%%%%%%%%%%%%%
%\begin{figure}[!ht]
%\centering
%\includegraphics[width=3.4in]{Figures/Omar/4586312.png}
%\caption{Dummy Chart: Future work's section division.}
%\label{Future work}
%\end{figure}
%%%%%%%%%%%%%%%%%%%%%%%%%%

\section{Conclusion}
\label{Section: Conclusion}

This paper has provided a comprehensive review on the up-to-date research for the information-theoretic security of \gls{RIS}-enabled future wireless systems. To begin with, an overview of the \gls{PLS} fundamentals has been presented by focusing on the various types of attacks in \gls{PLS}, the techniques used to achieve \gls{PLS}, and the secrecy performance metrics. Moreover, the working principles and architecture of the \gls{RIS} technology have been explored followed by its corresponding deployment strategies for future wireless communication systems. Besides, the \gls{RIS} technology suited for \gls{OWC} has also been presented. Subsequently, the \gls{RIS}-enabled \gls{PLS} scenarios and techniques in both \gls{RF} and \gls{OWC} systems have been discussed. Furthermore, the state-of-the-art optimization techniques for \gls{PLS} have been reviewed and some existing optimization methods for secrecy performance maximization have been summarized. To address the complexity issues associated with the rapid increase in the number of interactions between users and infrastructures in future wireless networks, the adoption of \gls{ML} in RIS-assisted PLS-based systems has been well investigated. A comprehensive discussion on the performance analysis of \gls{RIS}-assisted \gls{PLS}-based wireless systems has been presented, and useful insights into the adoption of the \gls{RIS} technology for performance improvement have been provided. Finally, some insightful open research challenges have been proposed and thoroughly discussed, and corresponding future research directions have ensued to further close the knowledge gap towards the development of forthcoming wireless technologies. We hope that this survey paper will serve as a valuable resource for researchers and practitioners in an effort to bring the deployment of \gls{RIS}-assisted \gls{PLS} in emerging wireless systems closer to reality.

%\clearpage
 %\vspace
 \begingroup
 	\bibliographystyle{IEEEtran}
 	\bibliography{References/Majid,References/Jules, References/Omar, References/Sylvester}
  
 \endgroup

\ifCLASSOPTIONcaptionsoff
  \newpage
\fi

\end{document}